\documentclass[a4paper,12pt]{report}
\usepackage[latin1]{inputenc}
\usepackage{graphicx}
\usepackage{amsmath,amssymb,euscript,array,mathrsfs}
\usepackage{epstopdf}
\usepackage{setspace}
\usepackage[left=4cm, right=2.5cm]{geometry}
\usepackage{cite}

\numberwithin{equation}{section}

\newcommand{\Tr}{\textnormal{Tr}}
\newcommand{\Ncal}{\mathcal{N}}
\newcommand{\EQ}[1]{\begin{equation} #1 \end{equation}}
\newcommand{\AL}[1]{\begin{subequations} \begin{align} #1
\end{align}\end{subequations}}
\newcommand{\ALlabel}[2]{\begin{subequations} \label{#2} \begin{align} #1
\end{align}\end{subequations}}
\newcommand{\SP}[1]{\begin{equation}\begin{split} #1
\end{split}\end{equation}}

\allowdisplaybreaks

\begin{document}

\begin{titlepage}

\center

{\Huge Aspects of Gauge-Gravity Duality}\\[1.5cm]

\textbf{Per Albert Daniel Elander}
\bigskip
\\
{\it Department of Physics}, \\ {\it Swansea University}, \\ {\it Swansea, SA2 8PP, UK}
\bigskip
\bigskip
\bigskip
\bigskip
\\
Submitted to the University of Wales in fulfilment of the requirements for the Degree of Doctor of Philosophy.
\bigskip
\bigskip
\\
Swansea University, 2010.

\end{titlepage}

\begin{spacing}{1}

\abstract{
In this Ph.D. thesis, we study various backgrounds in Type IIB supergravity which admit interpretations in terms a dual field theory, and compute properties such as effective potentials and spectra, using both holographic and field theoretic methods. This thesis is based on the papers \cite{Elander:2008vw, Elander:2009bm, Elander:2009pk}.

First, we study the phase structure of $\beta$-deformed $\mathcal N = 4$ SYM on $S^3$ at weak and strong 't Hooft coupling. We compute the one-loop effective potential, and find that at near critical chemical potential and small finite temperature, there is a metastable state at the origin of moduli space. We derive the gravitational background describing the theory at strong coupling, and by performing a probe-brane calculation, we find qualitative agreement between the weak and strong coupling results.

Next, we study gravitational backgrounds obtained by wrapping $N_c$ D5 color branes on an $S^2$ inside a CY3-fold, and $N_f$ D5 backreacting flavor branes on a non-compact two-cycle inside the same CY3-fold. These backgrounds are believed to be dual to certain SQCD-like theories. We compute how the spectrum depends on the number of flavors, and find that the mass of the lightest scalar glueball increases with the number of flavors until the point $N_f = 2 N_c$ is reached after which the opposite behaviour is observed.

Finally, we consider a class of backgrounds that exhibit walking behaviour, i.e. a suitably defined four-dimensional gauge coupling stays nearly constant in an intermediate energy regime. The breaking of approximate scale invariance has been conjectured to lead to the existence of a light scalar in the spectrum. This so-called dilaton would be the pseudo-Goldstone boson of dilatations. Using holographic techniques, we compute the spectrum and find a light state whose mass is suppressed by the length of the walking region, suggesting that this might be the dilaton.
}

\end{spacing}

\newpage
\tableofcontents
\newpage

\section*{Acknowledgements}

I would like to thank Asad~Naqvi and Maurizio~Piai, who have both acted as my Ph.D. advisor. I would also like to thank S.~Prem~Kumar, whose help and supervision were integral to the project about the phase structure of $\beta$-deformed $\mathcal N = 4$, which comprises Chapter~\ref{ch:2} and Chapter~\ref{ch:3} of this thesis. In the completion of the project described in Chapter~\ref{ch:5}, on the spectrum of glueballs in SQCD-like theories, I benefitted from many discussions with Carlos~N\'u\~nez, whom I thank here. Carlos~N\'u\~nez also collaborated with Maurizio~Piai and me on the walking backgrounds discussed in Chapter~\ref{ch:6}. It has been a pleasure working with you. You have all taught me a lot about physics. Throughout my studies, I have also benefitted from discussions with Timothy~Hollowood, Emiliano~Imeroni, Johannes~Schmude, Adi~Armoni, and Ioannis Papadimitriou. This work was in part supported by a grant from Stiftelsen Karl och Annie Leons Minnesfond för Vetenskaplig forskning. Finally, I would like to thank my parents, Arne and Eva Elander for their love and support.

\newpage

\chapter{Introduction}
\label{ch:1}

Gauge-gravity duality has provided us with invaluable insights into the dynamics of quantum field theories at strong coupling. The AdS/CFT conjecture was originally proposed in \cite{Maldacena:1997re} and subsequently refined in \cite{Witten:1998qj, Gubser:1998bc}. In its original form, it relates Type IIB string theory on $AdS_5 \times S^5$ to $\mathcal N = 4$ SYM in four dimensions. However, it has since been extended to apply to more generalized settings with less supersymmetry and also to backgrounds where conformal symmetry is broken. In this Ph.D. thesis, we study various different backgrounds for which there exists a dual interpretation in terms of a quantum field theory. Using field theoretic, as well as holographic methods, we compute properties such as spectra and effective potentials. This thesis is based on the papers \cite{Elander:2008vw, Elander:2009bm, Elander:2009pk}.

In Chapter~\ref{ch:2}, we study the phase structure of $\beta$-deformed $\mathcal N = 4$ SYM at weak 't Hooft coupling. The $\beta$ deformation is an exactly marginal deformation that breaks the amount of supersymmetry from $\mathcal N = 4$ to $\mathcal N =1$ \cite{Leigh:1995ep}. Furthermore, the global $SO(6)$ R-symmetry of $\mathcal N = 4$ SYM gets broken to $U(1)^3$. We add chemical potentials for these $U(1)$s, and study the field theory at finite temperature. The addition of chemical potentials produces an instability in the theory in the sense that it generates negative mass squared terms for the scalars charged under the associated symmetries. For this reason, we define the theory on $S^3$. This generates positive mass squared terms for the scalars since they couple to the curvature through the conformal coupling. For critical values of the chemical potentials (in particular, those for which the aforementioned negative and positive contributions to the mass squared terms of the scalars cancel), the theory has a non-trivial moduli space: there is a Coulomb branch, and also, for special values of the deformation parameter $\beta$, additional Higgs branches open up \cite{Berenstein:2000ux, Dorey:2004xm, Berenstein:2000hy}. We compute the one-loop effective potential and find among other things that at finite temperature and near critical chemical potential, there is a metastable state at the origin of moduli space, which decays through thermal activation and quantum tunnelling. On the Higgs branch, this has the interpretation in terms of deconstruction as an extra-dimensional torus whose volume decays from infinite to zero size.

In Chapter~\ref{ch:3}, we perform the analogous study as in Chapter~\ref{ch:2}, but at strong 't Hooft coupling. In order to obtain the gravity dual of $\beta$-deformed $\mathcal N = 4$ SYM with chemical potentials at finite temperature, we start with a solution in Type IIB supergravity that is known to describe the corresponding $\mathcal N = 4$ case, then apply a solution generating technique called a TsT-transformation to obtain the $\beta$-deformed background. The resulting solution describes a black hole rotating in an internal (deformed) $S^5$. Compactifying to five dimensions, one obtains a solution in $\mathcal N = 2$ $U(1)^3$ gauged supergravity that describes a Reissner-Nordström black hole carrying charges with respect to the three $U(1)$s. The boundary values of the gauge fields of the $U(1)$s correspond to the values of the chemical potentials in the dual field theory. In order to compute the effective potential at strong coupling, we use D3 probes branes on the Coulomb branch and D5 probe branes on the Higgs branch. We find qualitatively the same results as at weak coupling, i.e. there is a metastable state at the origin of moduli space.

In Chapter~\ref{ch:4}, we review and develop holographic techniques for the computation of spectra of strongly coupled quantum field theories. We will then apply these techniques to specific examples in Chapter~\ref{ch:5} and Chapter~\ref{ch:6}. In order to compute spectra holographically, one studies fluctuations around the particular background one is interested in. The fluctuations satisfy linearized equations of motion, and, in general, solutions with the correct IR and UV behaviour only exist for specific values of $K^2$, where $K$ is the four-momentum. From these values of $-K^2 = M^2$, the spectrum is obtained. In \cite{Berg:2005pd}, a gauge-invariant formalism was developed for this purpose. Given a five-dimensional non-linear sigma model consisting of a number of scalars coupled to gravity, and with a potential that can be derived from a superpotential $W$, general formulas were given for the linearized equations of motion that the fluctuations satisfy. Furthermore, this formalism leads to a simplification in the sense that even though the metric is allowed to fluctuate, the gravitational modes effectively decouple, so that in the end one obtains expressions that involve only the scalar fluctuations. We generalize these methods to the case where the potential for the scalars cannot be derived from a superpotential. This will be needed for the model that we study in Chapter~\ref{ch:6}.

In Chapter~\ref{ch:5}, we study the spectrum of glueballs in SQCD-like theories whose Type IIB supergravity description is in terms of $N_c$ D5 color branes wrapped on an $S^2$ inside a CY3-fold, and $N_f$ backreacting D5 flavor branes wrapped on a non-compact two-cycle inside the same CY3-fold \cite{Casero:2006pt}. The D5 flavor branes are smeared along the transverse angular coordinates, breaking the $SU(N_f)$ global symmetry to $U(1)^{N_f}$. The dual field theory is believed to be similar in the IR to $\mathcal N = 1$ SQCD with a quartic superpotential for the quark superfields. However, the full theory cannot be dual to SQCD for a number of reasons. It does not have an $SU(N_f) \times SU(N_f) \times U(1)_R$ global symmetry as SQCD does, but instead only one $SU(N_f)$ (broken further to $U(1)^{N_f}$ by the smearing). The backgrounds correponding to this setup have been to found fall into two categories known as Type A and Type N \cite{HoyosBadajoz:2008fw, Casero:2007jj}. Type A backgrounds are special cases of Type N backgrounds for which the VEV of the gaugino condensate as well as the mesons are zero. In this chapter, we study the spectrum of a few backgrounds of Type A for which the dilaton grows linearly in the UV. In the IR, there are different possible behaviours for the background (known as Type I, II and III \cite{Casero:2007jj}) corresponding to different vacua in the dual field theory. These backgrounds have a singularity in the IR which is ``good'' according to the criterion given in \cite{Maldacena:2000mw}, and are believed to capture the non-perturbative physics of the dual field theory.

Technically, it is difficult to compute the spectrum while working in ten dimensions. However, we show that there exists a consistent truncation to a five-dimensional non-linear sigma model consisting of four scalars coupled to gravity, so that the methods of Chapter~\ref{ch:4} can be applied. We find that the mass of the lightest scalar glueball increases as the number of flavors is increased, until the point $N_f = 2 N_c$ is reached after which the opposite behaviour is observed. For a particular class of backgrounds that are Seiberg dual to themselves, we demonstrate explicitly that the spectrum obeys Seiberg duality. In the gravity picture, Seiberg duality is realized for these theories as a diffeomorphism, i.e. just a change of variables \cite{Casero:2006pt, HoyosBadajoz:2008fw}. Therefore, the background itself does not change under Seiberg duality, but since we have changed variables, the dictionary interpretation of the dual field theory is changed. We show that for the five-dimensional model, Seiberg duality corresponds to a set of transformations of the scalar fields and $N_c \rightarrow N_f - N_c$. The five-dimensional Lagrangian is invariant under these transformations, and therefore anything that can be computed within this framework obeys Seiberg duality.

In Chapter~\ref{ch:6}, we study models for which a suitably defined gauge coupling exhibits walking behaviour, i.e. it stays nearly constant in an intermediate energy-regime. These models are obtained by wrapping $N_c$ number of D5-branes on an $S^2$, and are of Type N, according to the classification mentioned above. They contain no flavors, and can be thought of as deformations of the background known as non-singular Maldacena-Nunez \cite{Maldacena:2000yy}. Although explicit examples have proved difficult to find, strongly coupled systems with walking behaviour have for a long time been considered as viable candidates for physics beyond the Standard Model. This idea is known as Walking Technicolor \cite{Technicolor}. While the models we study share certain qualitative features of Walking Technicolor, i.e. the walking behaviour, we do not couple them to the Standard Model, and therefore they are not in their present form to be thought of as phenomenological models. Nevertheless, we are in the position to ask questions regarding the effect of the walking behaviour on physical quantities.

Simply plotting the gauge coupling as a function of energy scale does not conclusively establish that we are dealing with a walking theory, since such a plot could potentially look very different in another regularization scheme. From the gravity point of view, this corresponds to the fact that we can always choose a different radial coordinate. We need to compute something that is actually physical, such as the spectrum. It has been argued that in theories with walking behaviour there should exist a light state, corresponding to the spontaneous breaking of approximate scale invariance. This pseudo-Goldstone boson of dilatations is referred to as the dilaton, and its presence in phenomenological models would be dramatic. Using the five-dimensional methods described in Chapter~\ref{ch:4}, we show that, in addition to two towers of states, the spectrum indeed contains such a light state for the theories that we study. Furthermore, its mass is suppressed by the length of the walking region, suggesting that it might be interpreted as a dilaton.

\newpage

\chapter{Phase Structure of $\beta$-deformed $\Ncal=4$ SYM at Weak Coupling}
\label{ch:2}
One of the many exciting results to have come out of the AdS/CFT correspondence \cite{Maldacena:1997re, Witten:1998qj, Gubser:1998bc} is that $\Ncal=4$ supersymmetric Yang-Mills at finite temperature is related to black holes in $AdS_5$. For example, the Hawking-Page phase transition \cite{Hawking:1982dh}, in which a black hole forms above a critical value of the temperature, turns out to be dual, by the correspondence, to a confinement-deconfinement phase transition in the quantum field theory on the boundary \cite{Witten:1998zw}. The link between the thermodynamics of black holes and that of $\Ncal=4$ SYM makes it an interesting project to map out the phase structure, and compare the results at strong and weak 't~Hooft coupling. Much effort has been devoted towards this goal \cite{Sundborg:1999ue, Aharony:2003sx, Aharony:2005bq, Liu:2004vy, Yamada:2006rx, Harmark:2006di, Hollowood:2006xb, Harmark:2006ta, Harmark:2006ie, Harmark:2007px, AlvarezGaume:2005fv, Basu:2005pj, AlvarezGaume:2006jg, Azuma:2007fj, Dutta:2007ws, Yamada:2008em, Hollowood:2008gp, Murata:2008bg, Hawking:1999dp, Cvetic:1999ne, Yamada:2007gb, Hawking:1998kw}.

In this chapter, we will derive weak coupling results about the phase structure of $\beta$-deformed $\Ncal=4$ SYM. The $\beta$-deformation is a marginal deformation of $\Ncal=4$ SYM, which changes the superpotential of $\Ncal=4$ SYM to
\EQ{
	W = i 2 \sqrt{2} \Tr \left( e^{i\pi\beta} \Phi_1 \Phi_2 \Phi_3 - e^{-i\pi\beta} \Phi_1 \Phi_3 \Phi_2 \right),
}
where $\beta$ is the deformation parameter. While the $\beta$-deformation breaks the amount of supersymmetry to $\Ncal=1$, it is interesting in that it preserves the conformal invariance of the original theory \cite{Leigh:1995ep}. The global $SO(6)$ R-symmetry of $\Ncal=4$ SYM is broken to $U(1)^3$, and we can add chemical potentials $\mu_i$ for these three $U(1)$s. The addition of chemical potentials breaks the conformal invariance, as well as all the supersymmetry of the theory. Furthermore, a negative mass squared term $-\mu_i^2$ gets generated for the scalars charged under the associated symmetry, and therefore the theory becomes unstable, unless it is defined at finite volume where the scalars also couple to the curvature through the conformal coupling, thus generating positive mass squared terms. We will define the $\beta$-deformed theory on $S^1 \times S^3$, where $S^1$ is the compactified time direction. In particular, we will be interested in chemical potentials which are close to critical, meaning that the negative mass squared terms that they generate almost cancel the ones from the conformal coupling. Classically, it is only for critical chemical potentials that there are flat directions and a non-trivial moduli space. The moduli space of the $\beta$-deformed theory has a Coulomb branch, and also, for special values of the deformation parameter $\beta$, additional Higgs branches open up \cite{Berenstein:2000ux, Dorey:2004xm, Berenstein:2000hy}. On these branches the theory is equivalent at low energies to $\Ncal = 4$ SYM. At intermediate energies, it can be viewed as the deconstruction of $\Ncal = (1,1)$ SYM in six dimensions, with the two extra dimensions forming a latticized torus \cite{Dorey:2003pp, Dorey:2004iq}. In essence, the torus forms because we can reinterpret the two gauge group indices of the adjoint scalars as discretized extra dimensions.

It was found in \cite{Hollowood:2008gp} that, at zero temperature, $\Ncal=4$ SYM on $S^1 \times S^3$ with critical chemical potentials has a one-loop effective action that is independent of the scalar VEVs. In this article, we repeat the calculation for the $\beta$-deformed theory and find the same result for $SU(N)$ gauge group, but a different one for $U(N)$. Since, for gauge group $U(N)$, the overall $U(1)$ decouples for $\Ncal=4$ SYM, this could not have happened in that case. However, in the $\beta$-deformed theory, it is no longer true that the overall $U(1)$ decouples. At finite temperature, and near critical chemical potential, $\Ncal=4$ SYM has a metastable state at the origin of moduli space, which decays through thermal activation or quantum tunnelling due to the runaway behaviour of the potential for large values of the scalar VEVs \cite{Hollowood:2008gp}. This is also true for the Coulomb branch of the $\beta$-deformed theory. We perform calculations which show that the same is true for the Higgs branch, where an interpretation can be made in which the extra-dimensional torus has a metastable state when its volume is infinite, that then decays to zero volume.

The structure of this chapter is as follows. In section~2.1, we review how to add chemical potentials to the theory, and the moduli space of $\beta$-deformed $\Ncal=4$ SYM. In section~2.2, we compute the one-loop effective action for the theory on the Coulomb branch, whereas in section~2.3 we do the same for the Higgs branch. Section~2.4 covers the metastable phases that occur at finite temperature and near critical chemical potentials. Finally, in section~2.5 we summarize our results.

\section{The $\beta$-deformation of $\mathcal{N}=4$ SYM}

\subsection{Lagrangian}

The $\beta$-deformation of $\mathcal{N}=4$ SYM is obtained by deforming the $\mathcal{N}=4$ superpotential to
\EQ{
	W = i 2 \sqrt{2} \Tr \, \Phi_1 [\Phi_2,\Phi_3]_\beta,
}
where
\EQ{
	[A,B]_\beta \equiv e^{i\pi\beta} AB - e^{-i\pi\beta} BA,
}
and $\mathcal{N}=4$ SYM corresponds to $\beta=0$. Here, we have used the following conventions. The generators of the $SU(N)$ ($U(N)$) Lie algebra are normalized as follows:
\EQ{
	\Tr \, T^a T^b = \frac{1}{2} \delta_{ab}.
}
This implies (for U(N) the second term on the right hand side is not present)
\EQ{
	T^a_{ij} T^a_{kl} = \frac{1}{2} \delta_{il} \delta_{jk} - \frac{1}{2N} \delta_{ij} \delta_{kl},
}
which in turn implies that (again, the second term is not present for $U(N)$)
\EQ{
	\label{eq:superderivation}
	\left( \Tr \, X T^a \right) \left( \Tr \, T^a Y \right) = \frac{1}{2} \Tr \, XY - \frac{1}{2N} \left( \Tr \, X \right) \left( \Tr \, Y \right).
}
In the following, we will take $\beta$ to be real, and focus on the $U(N)$ case (we will explain how the results differ in the SU(N) case as we go along). Using \eqref{eq:superderivation}, the potential for the scalars coming from the superpotential is
\EQ{
	V_W = \frac{4}{g^2} \, \Tr \left( | [\phi_1,\phi_2]_\beta |^2 + | [\phi_2,\phi_3]_\beta |^2 + | [\phi_3,\phi_1]_\beta |^2 \right).
}
The potential due to the D-term is
\EQ{
	V_D = \frac{1}{g^2} \Tr \left( [\phi_1^\dagger,\phi_1] + [\phi_2^\dagger,\phi_2] + [\phi_3^\dagger,\phi_3] \right)^2
}

A non-zero $\beta$ breaks the original $SU(4)$ R-symmetry to $U(1)^3$, where each of the $\Phi_i$ is charged under a
different $U(1)$.\footnote{We note that we can also use a basis with one $U(1)_R$ and two global $U(1)$, where the global $U(1)$s are linear combinations of the original three $U(1)_R$s. It is therefore not unreasonable to expect that the results will be qualitatively different when one turns on two of the chemical potentials $\mu_i$ from when one only turns on one.} For the complex scalars $\phi_i$, we write this as
\SP{
	\hat Q_1 (\phi_1,\phi_2,\phi_3) &= (1,0,0), \\
	\hat Q_2 (\phi_1,\phi_2,\phi_3) &= (0,1,0), \\
	\hat Q_3 (\phi_1,\phi_2,\phi_3) &= (0,0,1),
}
and similarly for the fermions:
\SP{
	\hat Q_1 (\lambda,\chi_1,\chi_2,\chi_3) &= \tfrac{1}{2} (1,1,-1,-1), \\
	\hat Q_2 (\lambda,\chi_1,\chi_2,\chi_3) &= \tfrac{1}{2} (1,-1,1,-1), \\
	\hat Q_3 (\lambda,\chi_1,\chi_2,\chi_3) &= \tfrac{1}{2} (1,-1,-1,1).
}
The grand canonical partition function is
\EQ{
	Z(T,\mu_i) = \Tr \, e^{-\frac{1}{T} (\hat H - \sum_i \mu_i \hat Q_i ) },
}
where $\hat H$ is the Hamiltonian and $\mu_i$ are the chemical potentials. Viewed as a Euclidean path integral with time compactified on $S^1$, adding chemical potentials to the theory is equivalent to letting \cite{Yamada:2006rx}
\EQ{
	D_\mu \rightarrow D_\mu - \delta_{\mu,0} \sum_i \mu_i \hat Q_i.
}
Hence, the kinetic terms for the complex scalars have the form
\SP{
	\label{eq:kineticscalars}
	2 \Tr \left( (D_\mu + \mu_i \delta_{0,\mu}) \phi_i \right)^\dagger (D_\mu - \mu_i \delta_{0,\mu}) \phi_i = \\ =
	2 \Tr \left( (D_\mu \phi_i)^\dagger D_\mu \phi + 2 \mu_i \phi_i^\dagger \tilde{D}_0 \phi_i - \mu_i^2 \phi_i^\dagger \phi_i \right)
}
whereas for the fermions the kinetic terms are
\EQ{
	2 \Tr \, \bar\chi_i ( i \sigma_\mu D^\mu - i \bar\mu_i ) \chi_i,
}
where
\SP{
	\bar\mu_0 &= \tfrac{1}{2} (\mu_1 + \mu_2 + \mu_3), \\
	\bar\mu_1 &= \tfrac{1}{2} (\mu_1 - \mu_2 - \mu_3), \\
	\bar\mu_2 &= \tfrac{1}{2} (-\mu_1 + \mu_2 - \mu_3), \\
	\bar\mu_3 &= \tfrac{1}{2} (-\mu_1 - \mu_2 + \mu_3),
}
and we have made the definition $\chi_0 \equiv \lambda$.

We will give VEVs to the scalars as
\SP{
	\phi_i \rightarrow \frac{\varphi_i}{\sqrt{2}} + \phi_i,
}
where $\varphi_i$ is the background value of the field. In order to fix the gauge, we add a term
\EQ{
	\mathcal{L}_{gf} = \frac{1}{g^2} \Tr \left( \nabla_i A^i + \tilde{D}_0 A^0 - \frac{i}{\sqrt{2}} \sum_{i=1}^3 \left( [\varphi_i^\dagger,\phi_i] + [\varphi_i,\phi_i^\dagger] \right) \right)^2
}
to the Lagrangian, corresponding to $R_\xi$-gauge with Feynman parameter $\xi=1$. This cancels cross terms of the form
\EQ{
	\frac{1}{g^2 \sqrt{2}} \left( - i [A_\mu,\varphi_i^\dagger] \partial_\mu \phi_i
	- i \partial_\mu \phi_i^\dagger [A_\mu,\varphi_i] \right).
}
The kinetic terms for the fields with a critical chemical potential have the following form:
\EQ{
	2 \Tr \left( (D_\mu (\frac{\varphi_i}{\sqrt{2}} + \phi_i))^\dagger D_\mu (\frac{\varphi_i}{\sqrt{2}} + \phi_i) + 2 \mu_i (\frac{\varphi_i}{\sqrt{2}} + \phi_i)^\dagger \tilde{D}_0 (\frac{\varphi_i}{\sqrt{2}} + \phi_i) \right)
}
Taking care to cancel the cross terms from the gauge fixing, the first term contributes
\EQ{
	2 \Tr \left( \phi_i^\dagger (-D^2) \phi_i + \frac{1}{2} A_\mu (\varphi_i^\dagger \varphi_i) A_\mu \right)
}
at second order, whereas the second term contributes
\EQ{
	2 \Tr \left( 2 \mu_i \left[ \phi_i^\dagger \tilde{D}_0 \phi_i
	+ \frac{i}{\sqrt{2}} \phi_i^\dagger \varphi_i A_0 + \frac{i}{\sqrt{2}} A_0 \varphi_i^\dagger \phi_i \right] \right),
}
where we have used the following notation for the commutator action:
\AL{
	\varphi &\equiv [\varphi, \cdot] \\
 	\varphi_\beta &\equiv [\varphi, \cdot]_\beta, \\
 	\varphi_\beta^\dagger &\equiv [\varphi^\dagger, \cdot]_{-\beta}.
}
Let us note that some useful relations are
\AL{
	\Tr \, [X,A] B &= - \Tr \, A [X,B], \\
	[A,B]_\beta &= - [B,A]_{-\beta}, \\
	[A,B]_\beta^\dagger &= - [A^\dagger,B^\dagger]_\beta, \\
	\Tr \, [X,A]_\beta^\dagger B &= - \Tr \, A^\dagger [X^\dagger,B]_{-\beta}.
}

\subsection{Classical Moduli Space}
Since the theory is defined on $S^3$, there is a conformal coupling of the scalars to the curvature which takes the form
\EQ{
	2 \Tr \, R^{-2} \phi_i^\dagger \phi_i,
}
where $R$ is the radius of the $S^3$. Furthermore, from \eqref{eq:kineticscalars} we get a similar term but with opposite sign:
\EQ{
	- 2 \Tr \, \mu_i^2 \phi_i^\dagger \phi_i.
}
Only when at least one of the $\mu_i$ has the critical value $\mu_i = R^{-1}$ is there any possibility of flat directions and a non-trivial moduli space. However, the F- and D-flatness conditions also need to be satisfied:
\EQ{
	[\phi_1,\phi_2]_\beta = [\phi_2,\phi_3]_\beta = [\phi_3,\phi_1]_\beta = 0,
}
\EQ{
	\sum_{i=1}^3 [\phi_i^\dagger,\phi_i] = 0.
}
These are solved by giving each $\phi_i$ a diagonal VEV, while imposing the restriction that for each row (equivalently column), no more than one of $\phi_i$ is allowed to have a non-zero entry. We also have to mod out by the Weyl group. This defines the Coulomb branch, where, for generic VEVs, the original $U(N)$ ($SU(N)$) gauge symmetry is broken down to $U(1)^N$ ($U(1)^{N-1}$).

For rational values of $\beta$, there are additional Higgs branches. For example, we can take $\beta = 1/N$ and give VEVs to the scalars as
\AL{
	\left\langle \phi_1 \right\rangle &= \lambda^{(1)} U_{(N)}, \\
	\left\langle \phi_2 \right\rangle &= \lambda^{(2)} V_{(N)}, \\
	\left\langle \phi_3 \right\rangle &= \lambda^{(3)} V_{(N)}^\dagger U_{(N)}^\dagger,
}
where $\lambda^{(1)}$, $\lambda^{(2)}$, and $\lambda^{(3)}$ are complex numbers, and
\AL{
	U_{(N)} &= \textnormal{diag} \left( \omega, \omega^2, \ldots, \omega^N \right) \\
	\left(V_{(N)}\right)_{ab} &=
	\begin{cases}
	1 \textnormal{	if } b=a+1 \textnormal{ mod } N, \\
	0 \textnormal{ otherwise}
	\end{cases},
}
with $\omega = e^{2\pi i \beta}$. This breaks $U(N)$ to $U(1)$, while in the $SU(N)$ case the gauge group is completely broken. To obtain the right moduli space, we also have to mod out by the discrete gauge transformations
\EQ{
	\phi_i \rightarrow \Gamma_j \phi_i \Gamma_j^{\dagger},
}
where $\Gamma_1 = U_{(N)}$ and $\Gamma_2 = V_{(N)}$. These rotate $\lambda^{(i)}$ by discrete phases $\omega$ \cite{Dorey:2003pp}. After taking this to account, the moduli space is ${\mathbb C}^3/({\mathbb Z}^N \times {\mathbb Z}^N)$.

More generally, we have the solution
\AL{
	\left\langle \phi_1 \right\rangle &= \Lambda^{(1)} \otimes U_{(n)}, \\
	\left\langle \phi_2 \right\rangle &= \Lambda^{(2)} \otimes V_{(n)}, \\
	\left\langle \phi_3 \right\rangle &= \Lambda^{(3)} \otimes V_{(n)}^\dagger U_{(n)}^\dagger,
}
with
\EQ{
	\Lambda^{(i)} = \textnormal{diag} \left( \lambda^{(i)}_1, \lambda^{(i)}_2, \ldots, \lambda^{(i)}_m \right)
}
and $N=nm$, $\beta=1/n$. For generic $\Lambda^{(i)}$, this breaks $U(N)$ to $U(1)^m$, and $SU(N)$ to $U(1)^{m-1}$. The low energy theory turns out to be $\mathcal{N}=4$ on the Coulomb branch \cite{Dorey:2003pp, Dorey:2004iq}.

\section{One-Loop Effective Potential for the Coulomb Branch}

\subsection{General Considerations}
We will now compute the Wilsonian one-loop effective potential by integrating out all but the lightest fields of the theory. The fields can be expanded on $S^3 \times S^1$ in terms of spherical harmonics and Matsubara modes. The analysis is similar to that in \cite{Hollowood:2008gp}. We will turn on one critical chemical potential $\mu_1 = R^{-1}$ and give a background VEV to the mode constant on $S^3 \times S^1$ of the associated complex scalar
\EQ{
	\phi_1 \rightarrow \frac{\varphi}{\sqrt{2}} + \phi_1.
}
In addition, there will be a background value for the spatial zero mode of the holonomy of the time component of the gauge field around the thermal circle:
\EQ{
	A_0 \rightarrow \alpha + A_0.
}
The effective action is parametrized by $\varphi$, and $\alpha$, which in this section we shall take to both be diagonal.

The Lagrangian for the bosons and the ghosts $(\bar c,c)$ at second order is
\SP{
	\mathcal{L}_b^{(2)} = \frac{1}{g^2} 2 \Tr \bigg( \frac{1}{2} A_0 ( - \tilde{D}_0^2 - \Delta^{(s)} + \varphi^\dagger \varphi ) A_0 + \\
	+ \frac{1}{2} A_i ( - \tilde{D}_0^2 - \Delta^{(v)} + \varphi^\dagger \varphi ) A_i + \\
	+ \bar{c} ( - \tilde{D}_0^2 - \Delta^{(s)} + \varphi^\dagger \varphi ) c + \\
	+ \phi_1^\dagger (-\tilde{D}_0^2 - \Delta^{(s)} + \varphi^\dagger \varphi ) \phi_1 + \\
	+ \mu_1 \left[ 2 \phi_1^\dagger \tilde{D}_0 \phi_1 + i \sqrt{2} \phi_1^\dagger \varphi A_0 + i \sqrt{2} A_0 \varphi^\dagger \phi_1 \right] + \\
	+ \phi_2^\dagger ( - \tilde{D}_0^2 - \Delta^{(s)} + \varphi_\beta^\dagger \varphi_\beta + R^{-2} ) \phi_2 + \\
	+ \phi_3^\dagger ( - \tilde{D}_0^2 - \Delta^{(s)} + \varphi_{-\beta}^\dagger \varphi_{-\beta} + R^{-2} ) \phi_3
	\bigg),
}
where $\Delta^{(s)}$ and $\Delta^{(v)}$ are the scalar and vector Laplacians on $S^3$ respectively. For the fermions, we have
\SP{
		\mathcal{L}_f^{(2)} = 2 \Tr \bigg( \sum_{i=0}^3 \bar{\chi_i} ( i \sigma_\mu D^\mu - i \bar{\mu_i} ) \chi_i -
	\chi_0 (i \varphi^\dagger) \chi_1 - \bar{\chi}_1 ( -i \varphi) \bar{\chi}_0 - \\
	- \chi_3 (i \varphi_\beta) \chi_2 - \bar{\chi}_2 ( -i \varphi_\beta^{\dagger}) \bar{\chi}_3 \bigg)
}

The one-loop correction to the effective potential is given by
\EQ{
	\label{eq:logdeteff}
	V_1 = \frac{T}{2\pi^2R^3} \frac{1}{2} \sum_{\textnormal{species}} \sum_{ij}^N \sum_{\ell = \ell_0}^\infty d_\ell^{B(F)} \log \det \left( - \tilde D_0^2 + \varepsilon_\ell \left( \varphi \right)  \right),
}
where $\ell$ is the angular momentum quantum number of the mode with $\ell_0$ its lowest value, $d_\ell^{B(F)}$ is the degeneracy including differing signs for bosons and fermions, and finally $\varepsilon_\ell$ is the energy of the mode. After a Poisson resummation over the Matsubara frequencies, \eqref{eq:logdeteff} can be recast as a sum over species \cite{Hollowood:2008gp} with bosons contributing
\EQ{
	\label{eq:bpoissonsum}
	\frac{1}{\textnormal{Vol}(S^3)} \frac{1}{2} \sum_{i,j=1}^N \sum_{\ell = \ell_0}^\infty d_\ell^B \left( |\varepsilon_\ell(\varphi)| - T \sum_{k=1}^\infty \frac{1}{k} e^{-\frac{k}{T} |\varepsilon_\ell(\varphi)|} \cos(k \alpha_{ij} / T) \right),
}
and fermions contributing
\EQ{
	\label{eq:fpoissonsum}
	\frac{1}{\textnormal{Vol}(S^3)} \frac{1}{2} \sum_{i,j=1}^N \sum_{\ell = \ell_0}^\infty d_\ell^F \left( |\varepsilon_\ell(\varphi)| - T \sum_{k=1}^\infty \frac{(-1)^k }{k} e^{-\frac{k}{T} |\varepsilon_\ell(\varphi)|} \cos(k \alpha_{ij} / T) \right),
}
where $\alpha_{ij} = \alpha_i - \alpha_j$ ($\alpha_i$ refers to the $i$th diagonal component of $\alpha$).

\subsection{Energy Levels}

We will now compute what the energy levels are. Consider first the scalar fields. We expand in spherical harmonics and use that
\EQ{
	\Delta^{(s)} Y_{\ell} = R^{-2} \ell(\ell+2) Y_\ell \ \ \ \ (\ell=0,1,\ldots).
}
The only non-trivial case concerns the fields $(A_0,\phi_1,\phi_1^\dagger)$, whose fluctuation matrix is equal to
\EQ{
	{\small
	\begin{pmatrix}
		- \tilde{D}_0^2 - \nabla^{(s)} + \varphi^\dagger \varphi & i \sqrt{2} R^{-1} \varphi^\dagger & - i \sqrt{2} R^{-1} \varphi \\
		i \sqrt{2} R^{-1} \varphi & - \tilde{D}_0^2 - \nabla^{(s)} + \varphi^\dagger \varphi + 2 R^{-1} \tilde{D}_0 & 0 \\
		- i \sqrt{2} R^{-1} \varphi^\dagger & 0 & - \tilde{D}_0^2 - \nabla^{(s)} + \varphi^\dagger \varphi - 2 R^{-1} \tilde{D}_0 
	\end{pmatrix}}.
}
Putting its determinant
\EQ{
	(- \tilde{D}_0^2 + \ell(\ell+2) R^{-2} + \varphi^\dagger \varphi)
	(- \tilde{D}_0^2 + \ell^2 R^{-2} + \varphi^\dagger \varphi)
	(- \tilde{D}_0^2 + (\ell+2)^2 R^{-2} + \varphi^\dagger \varphi)
}
equal to zero and solving for $\tilde{D}_0$ gives three energy levels associated with $(A_0,\phi_1,\phi_1^\dagger)$. These are
\EQ{
	\sqrt{R^{-2}\ell(\ell+2) + \varphi^\dagger \varphi}
}
and
\EQ{
	\sqrt{R^{-2}(\ell+1 \pm R \mu_1)^2 + \varphi^\dagger \varphi},
}
with degeneracies $d_\ell = (\ell+1)^2$.

Calculating the energy levels for the fermions is more involved. We will evaluate the determinant of the fluctuation matrix by brute force. In the path integral, we will expand $e^{-S}$ to the power that saturates the measure
\EQ{
	\int \prod_{i=0}^3 D \bar{\chi}_i(p) D \bar{\chi}_i(-p) D \chi_i(p) D \chi_i(-p).
}
This happens for $S$ to the 16th power. However, matters simplify, because of the block diagonal form of the fluctuation matrix, and also because of how the $\chi_i(p)$s and $\chi_i(-p)$s have to combine. Each fermionic variable can only appear once. We can represent the way the terms in the Lagrangian combine as four graphs that look like:
\SP{
	\begin{matrix}
		\chi_0(-p) & \leftarrow (-i \varphi) \rightarrow & \chi_1(p) \\
		\uparrow & & \uparrow \\
		(i \sigma_\mu D^\mu(-p) - i \bar{\mu_0}) & & (i \sigma_\mu D^\mu(p) - i \bar{\mu_1}) \\
		\downarrow & & \downarrow \\
		\bar{\chi}_0(p) & \leftarrow (i \varphi) \rightarrow & \bar{\chi}_1(-p)
	\end{matrix}
}
The idea is to trace out closed paths in this graph. Consider the example of a closed path travelling through all corners of this graph. Starting in the upper left corner and going to the upper right corner, we are instructed to write down a factor $\chi_0(-p) (-i \varphi) \chi_1(p)$. Then, continuing down to the lower right corner, we pick up a factor $\chi_1(p) (i \sigma_\mu D^\mu(p) - i \bar{\mu_1}) \bar{\chi}_1(-p)$. When we have travelled around the four corners and back to the original one, all the fermionic variables $\chi_{0,1}^\alpha$ and their complex conjugates have occured exactly once. There is another graph which is the same as the one above, but with $p \rightarrow -p$, and similarly for $\chi_2$ and $\chi_3$ (but with $\varphi \rightarrow \varphi_\beta$).\footnote{When all three VEVs are turned on, there is a single graph which is a four-dimensional hypercube. When two VEVs are turned on, this gets cut into two cubes, which then get cut into the four squares for one VEV.} Since the four graphs do not connect, we can consider them separately. Finally, we sum up all the different ways to form closed paths.

Following closed paths around the graphs, there are three ways to saturate the measure; the two closed paths
\SP{
	\begin{matrix}
		\chi_0(-p) & & \chi_1(p) \\
		\uparrow \downarrow & & \uparrow \downarrow \\
		\bar{\chi}_0(p) &  & \bar{\chi}_1(-p)
	\end{matrix}
}
pick up a term
\SP{
	\frac{1}{4} (i \sigma_\mu D^\mu(-p) - i \bar{\mu_0})^{\dot{\alpha}\alpha} (i \sigma_\mu D^\mu(-p) - i \bar{\mu_0})_{\dot{\alpha}\alpha} \\
	(i \sigma_\nu D^\nu(p) - i \bar{\mu_1})^{\dot{\beta}\beta} (i \sigma_\nu D^\nu(p) - i \bar{\mu_1})_{\dot{\beta}\beta} = \\
	= ( \tilde{D}_0^2 + \nabla^{(f)} + 2 \bar{\mu}_0 \tilde{D}_0 + \bar{\mu}_0^2) ( \tilde{D}_0^2 + \nabla^{(f)} - 2 \bar{\mu}_1 \tilde{D}_0 + \bar{\mu}_1^2),
}
while the two closed paths
\SP{
	\begin{matrix}
		\chi_0(-p) & \rightleftharpoons & \chi_1(p) \\
		& & \\
		\bar{\chi}_0(p) & \rightleftharpoons & \bar{\chi}_1(-p)
	\end{matrix}
}
pick up
\EQ{
	(\varphi^\dagger \varphi)^2,
}
and, finally, the one closed path that travels around the whole graph
\SP{
	\begin{matrix}
		\chi_0(-p) & \rightarrow & \chi_1(p) \\
		\uparrow & & \downarrow \\
		\bar{\chi}_0(p) & \leftarrow & \bar{\chi}_1(-p)
	\end{matrix}
}
picks up
\SP{
	\frac{1}{4} (i \sigma_\mu D^\mu(-p) - i \bar{\mu_0})^{\dot{\alpha}\alpha} (i \sigma_\nu D^\nu(p) - i \bar{\mu_1})^{\dot{\beta}\beta} (i \varphi^\dagger \epsilon_{\dot{\alpha}\dot{\beta}}) (-i \varphi \epsilon_{\alpha\beta}) = \\
	= 2 \varphi^\dagger \varphi ( - \tilde{D}_0^2 - \nabla^{(f)} + (\tilde{\mu_1} - \tilde{\mu_0}) \tilde{D}_0 + \tilde{\mu_0} \tilde{\mu_1}).
}
Together, we have
\SP{
	( \tilde{D}_0^2 + \nabla^{(f)} + 2 \bar{\mu}_0 \tilde{D}_0 + \bar{\mu}_0^2) ( \tilde{D}_0^2 + \nabla^{(f)} - 2 \bar{\mu}_1 \tilde{D}_0 + \bar{\mu}_1^2) - \\
	+ 2 \varphi^\dagger \varphi ( - \tilde{D}_0^2 - \nabla^{(f)} + (\tilde{\mu_1} - \tilde{\mu_0}) \tilde{D}_0 + \tilde{\mu_0} \tilde{\mu_1}) + (\varphi^\dagger \varphi)^2.
}
Putting this equal to zero and solving for $\tilde{D}_0$ yields the following energies ($\nabla^{(f)} = - (l + 1/2)^2 R^{-2}$):
\EQ{
	\frac{\tilde{\mu_1} - \tilde{\mu_0}}{2} \pm \sqrt{  R^{-2} \left( l + \frac{1}{2} \pm \frac{R \tilde{\mu_0} + R \tilde{\mu_1}}{2} \right)^2 + \varphi^\dagger \varphi}.
}
The graph with $p \rightarrow -p$ just exchanges the roles of $\tilde{\mu_0}$ and $\tilde{\mu_1}$, so that together these two graphs yield
\EQ{
	\sqrt{  R^{-2} \left( l + \frac{1}{2} \pm \frac{R \tilde{\mu_0} + R \tilde{\mu_1}}{2} \right)^2 + \varphi^\dagger \varphi} \pm \frac{\tilde{\mu_1} - \tilde{\mu_0}}{2},
}
which becomes
\EQ{
	\sqrt{  R^{-2} \left( l + \frac{1}{2} \pm \frac{R \mu_1}{2} \right)^2 + \varphi^\dagger \varphi} \pm \frac{\mu_2 + \mu_3}{2}.
}
The two remaining graphs exchange $\varphi \rightarrow \varphi_\beta$ (for diagonal VEVs $\varphi_\beta^\dagger \varphi_\beta = \varphi_\beta \varphi_\beta^\dagger $), $\tilde{\mu}_0 \rightarrow \tilde{\mu}_3$, $\tilde{\mu}_1 \rightarrow \tilde{\mu}_2$, thus yielding:
\EQ{
	\sqrt{  R^{-2} \left( l + \frac{1}{2} \pm \frac{R \mu_1}{2} \right)^2 + \varphi_\beta^\dagger \varphi_\beta} \pm \frac{\mu_2 - \mu_3}{2}.
}

We summarize these results in Table~2.1. All possible sign combinations are allowed. Note that the expressions are only valid at vanishing or critical chemical potentials.\footnote{More precisely, the expressions for the fermions are valid for any values of the chemical potentials, but the chemical potentials need to be vanishing or critical in order for the expressions for the complex scalars and $A_0$ to take the simple form of Table 2.1.}

\begin{table}
\begin{center}
\begin{tabular}{c | c | c | c}
  Field & $d_\ell$ & $|\varepsilon_\ell|$ & $\ell_0$ \\ \hline
 	$B_i$ & $2\ell(\ell+2)$ & $\sqrt{R^{-2}(\ell+1)^2 + \varphi^\dagger \varphi}$ & 1 \\
 	$C_i$ & $(\ell+1)^2$ & $\sqrt{R^{-2}\ell(\ell+2) + \varphi^\dagger \varphi}$ & 1 \\
 	$(c,\bar{c})$ & $-2(\ell+1)^2$ & $\sqrt{R^{-2}\ell(\ell+2) + \varphi^\dagger \varphi}$ & 0 \\
 	$(A_0,\phi_1,\phi_1^\dagger)_1$ & $(\ell+1)^2$ & $\sqrt{R^{-2}\ell(\ell+2) + \varphi^\dagger \varphi}$ & 0 \\
 	$(A_0,\phi_1,\phi_1^\dagger)_{2,3}$ & $(\ell+1)^2$ & $\sqrt{R^{-2}(\ell+1 \pm R \mu_1)^2 + \varphi^\dagger \varphi}$ & 0 \\
 	$\phi_2$ & $(\ell+1)^2$ & $\sqrt{R^{-2}(\ell+1)^2 + \varphi_\beta^\dagger \varphi_\beta} \pm \mu_2$ & 0 \\
 	$\phi_3$ & $(\ell+1)^2$ & $\sqrt{R^{-2}(\ell+1)^2 + \varphi_{-\beta}^\dagger \varphi_{-\beta}} \pm \mu_3$ & 0 \\
 	$(\lambda,\chi_1)$ & $-\ell(\ell+1)$ & $	\sqrt{  R^{-2} \left( \ell + \frac{1}{2} \pm \frac{R \mu_1}{2} \right)^2 + \varphi^\dagger \varphi} \pm \frac{\mu_2 + \mu_3}{2}$ & 1 \\
 	$(\chi_2,\chi_3)$ & $-\ell(\ell+1)$ & $	\sqrt{  R^{-2} \left( \ell + \frac{1}{2} \pm \frac{R \mu_1}{2} \right)^2 + \varphi_\beta^\dagger \varphi_\beta} \pm \frac{\mu_2 - \mu_3}{2}$ & 1
\end{tabular}
\end{center}
\caption{\footnotesize The energies $\varepsilon_\ell$ for the Coulomb branch and gauge group $U(N)$, together with their degeneracies $d_\ell$ for the various fields. $\ell_0$ is the minimum value of the angular momentum quantum number $\ell$. The expressions are valid for vanishing or critical ($\mu_i = R^{-1}$) chemical potentials. All possible sign combinations are allowed.}
\end{table}

\subsection{Zero Temperature}
At zero temperature, only the Casimir energy parts of \eqref{eq:bpoissonsum} and \eqref{eq:fpoissonsum} contribute:
\EQ{
	V_1(T=0) = \frac{1}{\textnormal{Vol} (S^3)} \frac{1}{2} \sum_{\textnormal{species}} \sum_{i,j=1}^N \sum_{\ell = \ell_0}^\infty d_\ell^{B(F)} |\varepsilon_\ell(\varphi)|.
}
We regularize this expression by introducing a cut-off that does not depend on the chemical potentials, as follows \cite{Hollowood:2008gp}:
\EQ{
	\frac{1}{2} \sum_{i,j=1}^N \sum_{l=l_0}^\infty d_l^{B(F)} |\epsilon_\ell(\varphi)| f(\left|\epsilon_\ell(\varphi)|_{\mu_i=0}\right|/\Lambda),
}
where $\Lambda$ is the cut-off, and $f(x)$ is a function that is equal to 1 for $x \leq 1$ and zero for $x>1$.

Since $(\varphi \phi)_{ij} = (\varphi_i - \varphi_j ) \phi_{ij}$, it makes sense to define
\EQ{
	\varphi_{ij} \equiv \varphi_i - \varphi_j,
}
where $\varphi_i$ is the $i$th diagonal component of $\varphi$,
and similarly for the beta-commutator
\EQ{
	{\varphi_\beta}_{ij} \equiv e^{i\pi\beta} \varphi_i - e^{-i\pi\beta} \varphi_j.
}
$\varepsilon_\ell(\varphi)$ should then be thought of as a function of $\varphi_{ij}$ and ${\varphi_\beta}_{ij}$. We then have that
\EQ{
	(\varphi_\beta^\dagger \varphi_\beta \phi)_{ij} = |e^{i\pi\beta} \varphi_i - e^{-i\pi\beta} \varphi_j|^2 \phi_{ij},
}
\EQ{
	(\varphi_{-\beta}^\dagger \varphi_{-\beta} \phi)_{ij} = |e^{i\pi\beta} \varphi_j - e^{-i\pi\beta} \varphi_i|^2 \phi_{ij},
}
so that, as we sum over $i$ and $j$, there is no need to distinguish between the two in the calculation.

Furthermore, we note that $C_i$, $(c,\bar{c})$, $(A_0,\phi_1,\phi_1^\dagger)_1$, and the contribution given by $(A_0,\phi_1,\phi_1^\dagger)_{2,3}$ with a minus sign and $\ell=0$ cancel against each other. Converting the sum over $\ell$ into an integral for the remaining fields by using the Abel-Plana formula \cite{Mostepanenko:1997sw}
\EQ{
	\sum_{n=0}^\infty F(n) = \int_0^\infty dx \, F(x) + \frac{1}{2} F(0) - 2 \int_0^\infty dx \frac{\textnormal{Im} F(ix)}{e^{2\pi x} - 1},
}
and summing over the species, we find the zero temperature effective potential
\SP{
	\label{eq:zeroTeffpot}
	\textnormal{Vol}(S^3) \, V_1(T=0) = \frac{3N^2}{16 R} + \frac{R}{8} \Tr  \left( \varphi_\beta^\dagger \varphi_\beta - \varphi^\dagger \varphi \right) = \\
	= \frac{3N^2}{16 R} + \frac{R}{8} \sum_{i,j=1}^N \left( |e^{i\pi\beta} \varphi_i - e^{-i\pi\beta} \varphi_j|^2 - |\varphi_i - \varphi_j|^2 \right) = \\
	= \frac{3N^2}{16R} + \frac{R}{2} \sin^2(\pi \beta) \left| \sum_{i=1}^N \varphi_i \right| ^2.
}
Although we have used the expressions of the energies for gauge group $U(N)$, in the large $N$ limit this expression is valid for $SU(N)$ also. The reason is that the only energies which are affected in going from $U(N)$ to $SU(N)$ are those for the diagonal fluctuations. For gauge group $SU(N)$, \eqref{eq:zeroTeffpot} reduces to the same expression as in the $\mathcal{N}=4$ case. For $U(N)$, the result is sensitive to the overall $U(1)$ which, unlike in the $\mathcal{N}=4$ theory, does not decouple from the dynamics for generic $\beta$. We note that \eqref{eq:zeroTeffpot} also is valid when $\mu_2$ or $\mu_3$ are critical, since they appear outside the square roots with plus or minus signs in the expressions for the energies and therefore cancel against each other when we sum all modes. Therefore, at zero temperature there is no difference between turning on a chemical potential for a $U(1)_R$ or a global $U(1)$. (We note that even though there is a positive mass squared for the traceful part of $\phi$, there is no metastable phase for near (and above) critical chemical potential due to the fact that the traceless modes still have negative masses squared.)

We also note briefly that using another Abel-Plana formula \cite{Mostepanenko:1997sw}
\EQ{
	\sum_{n=0}^\infty F(n+1/2) = \int_0^\infty dx \, F(x) + 2 \int_0^\infty dx \frac{\textnormal{Im} F(ix)}{e^{2\pi x} + 1},
}
we can derive an expression for the off-shell effective action without chemical potentials:
\SP{
	\label{eq:offshelleffaction}
	\textnormal{Vol}(S^3) V = \frac{R}{8} \left( |\varphi|^2 - |\varphi_\beta|^2 \right) + \\ +
	\frac{R}{4} \left( |\varphi_\beta|^2 - |\varphi|^2 \right) \log \left( 2 R \Lambda \right) + \\ +
	\frac{R}{4} |\varphi|^2 \log \left( R |\varphi| \right) - \frac{R}{4} |\varphi_\beta|^2 \log \left( R |\varphi_\beta| \right) + \\ +
	R^{-1} \int_{R|\varphi|}^\infty dl \, \frac{\left(4l^2 + \frac{3}{2} + \frac{1}{2} e^{-2\pi l} \right) \sqrt{l^2 - R^2 |\varphi|^2}}{\sinh(2\pi l)} + \\
	+ R^{-1} \int_{R|\varphi_\beta|}^\infty dl \, \frac{\left(4l^2 + \frac{1}{2} - \frac{1}{2} e^{-2\pi l} \right) \sqrt{l^2 - R^2 |\varphi_\beta|^2}}{\sinh(2\pi l)}.
}
In the above expression, the sum over $i$ and $j$ is implicit. Since
\EQ{
 \sum_{ij} \left( |\varphi_\beta|^2 - |\varphi|^2 \right) = 4 \sin^2 \left( \pi \beta \right) \left| \sum_{i} \varphi_i \right|^2,
}
there is no ultraviolet divergence in the $SU(N)$ theory. Note also, that the ultraviolet divergence for the $U(N)$ theory is a finite volume effect. For $\beta = 0$ and $\mathcal{N}=4$ SYM, \eqref{eq:offshelleffaction} reduces to
\EQ{
	2 R^{-1} \int_{R|\varphi|}^\infty dl \, \frac{\left(4l^2 + 1 \right) \sqrt{l^2 - R^2 |\varphi|^2}}{\sinh(2\pi l)},
}
first computed in \cite{Hollowood:2006xb}.

\subsection{Finite Temperature}
Consider
\EQ{
		\sum_{l = l_0}^\infty d_l^{B(F)} e^{-\frac{k}{T} |\varepsilon_l(\varphi)|}
}
appearing in the temperature dependent part of the expression \eqref{eq:bpoissonsum} and \eqref{eq:fpoissonsum} for the bosonic and fermionic contributions to the one-loop effective potential. Summing over all bosonic modes, we obtain
\SP{
	\sum_{l = 1}^\infty 4l^2 \bigg( e^{-\frac{k}{T} \sqrt{R^{-2} l^2 + |\varphi|^2}} +
	\frac{1}{2} \left[ \cosh \left( \frac{k \mu_2}{T} \right) + \cosh \left( \frac{k \mu_3}{T} \right) \right] e^{-\frac{k}{T} \sqrt{R^{-2} l^2 + |\varphi_\beta|^2}}  \bigg).
}
Similarly for the fermionic modes, we have
\SP{
	- \sum_{l = 1}^\infty 4l^2 \bigg(
	\cosh \left( \frac{k (\mu_2 +\mu_3)}{2T} \right)	e^{-\frac{k}{T} \sqrt{R^{-2} l^2 + |\varphi|^2}} + \\ + \cosh \left( \frac{k (\mu_2 - \mu_3)}{2T} \right) e^{-\frac{k}{T} \sqrt{R^{-2} l^2 + |\varphi_\beta|^2}}  \bigg).
}
Hence, the full expression for the one-loop effective potential at finite temperature is
\SP{
	\label{eq:finitetempeffact}
	V_0 + V_1 = \frac{1}{\textnormal{Vol}(S^3)} \bigg\{
	\frac{3N^2}{16R} + \frac{R}{2} \sin^2(\pi \beta) \left| \sum_{i=1}^N \varphi_i \right| ^2  - \\ -
	2T \sum_{i,j=1}^N
		\sum_{k=1}^\infty \frac{\cos(k \alpha_{ij} / T)}{k}
		\sum_{l = 1}^\infty l^2 \bigg( \left[ 1 - (-1)^k \cosh \left( \frac{k (\mu_2 +\mu_3)}{2T} \right) \right] \\ e^{-\frac{k}{T} \sqrt{R^{-2} l^2 + |\varphi_{ij}|^2}} +
	\frac{1}{2} \bigg[ \cosh \left( \frac{k \mu_2}{T} \right) + \cosh \left( \frac{k \mu_3}{T} \right) - \\ - 2(-1)^k \cosh \left( \frac{k (\mu_2 - \mu_3)}{2T} \right) \bigg] e^{-\frac{k}{T} \sqrt{R^{-2} l^2 + |{\varphi_\beta}_{ij}|^2}}  \bigg)
	\bigg\}
}
Since there is an attractive potential for the $\alpha_i$, we can put $\alpha_{ij}=0$, which means that the theory is in the deconfined phase. We see that unlike in the zero temperature case, because $\varphi_\beta$ appears in the exponential, there is now a non-trivial dependence on $\beta$ not just for the overall $U(1)$, but also for $SU(N)$.

\section{One-Loop Effective Potential for the Higgs Branch}

Let us first work out the case $n=N$, $\beta = 1/N$. To simplify matters, we will only give VEVs to two of the complex scalars
\AL{
	\phi_1 \rightarrow \frac{\lambda^{(1)} U_{(N)}}{\sqrt{2}} + \phi_1, \\
	\phi_2 \rightarrow \frac{\lambda^{(2)} V_{(N)}}{\sqrt{2}} + \phi_2.
}
In order to be able to do this, we need to (at least) turn on the two chemical potentials $\mu_1 = \mu_2 = R^{-1}$. Although technically more involved, conceptually the calculation of the one-loop effective potential for the Higgs branch proceeds in the same way as that for the Coulomb branch. We expand the fluctuations as
\EQ{
	\phi = \frac{1}{\sqrt{2N}} \sum_{i,j=1}^N \phi_{i,j} J_{i,j},
}
where
\EQ{
	J_{a,b} \equiv V_{(N)}^a U_{(N)}^{-b} \omega^{\frac{ab}{2}}
}
is a basis for $N \times N$ matrices \cite{Dorey:2003pp}, and $a$ and $b$ are integers defined modulo $N$. In general, $\phi_{i,j}$ is complex. For hermitian $\phi$, we have
\EQ{
	\phi_{i,j}^\dagger = \phi_{-i,-j}.
}
Furthermore,
\AL{
	U_{(N)} &= J_{0,-1} \\
	V_{(N)} &= J_{1,0}.
}
The $J_{a,b}$ satisfy the commutation relations
\AL{
	[J_{a,b},J_{c,d}] &= 2 \sin \left( \frac{(bc-ad)\pi}{N} \right) J_{a+c,b+d}, \\
	[J_{a,b},J_{c,d}]_{\pm \beta} &= 2 \sin \left( \frac{(bc-ad \pm 1)\pi}{N} \right) J_{a+c,b+d},
}
and
\SP{
	J_{a,b}^\dagger &= J_{-a,-b} \\
	\Tr \left( J_{a,b}^\dagger J_{c,d} \right) &= n \delta_{ac} \delta_{bd}.
}

\subsection{Energy Levels for Scalars}

At second order, the contribution from the D-term comes from (to simplify expressions we have rescaled $\varphi_i \rightarrow \sqrt{2} \varphi_i$ in this section)
\EQ{
	V_D^{(2)} = \frac{1}{g^2} \Tr \left( [\varphi_1^\dagger,\phi_1] - [\varphi_1,\phi_1^\dagger] + [\varphi_2^\dagger,\phi_2] - [\varphi_2,\phi_2^\dagger] \right)^2,
}
while the gauge fixing contributes
\EQ{
	V_{gf}^{(2)} = - \frac{1}{g^2} \Tr \left( [\varphi_1^\dagger,\phi_1] + [\varphi_1,\phi_1^\dagger] + [\varphi_2^\dagger,\phi_2] + [\varphi_2,\phi_2^\dagger] \right)^2.
}
Together, this becomes
\SP{
	V_D^{(2)} + V_{gf}^{(2)} =
	\frac{1}{g^2} 2 \Tr \bigg( \phi_1^\dagger ( 2 \varphi_1 \varphi_1^\dagger ) \phi_1 +
	\phi_2^\dagger ( 2 \varphi_2 \varphi_2^\dagger ) \phi_2 + \\ +
	\phi_1^\dagger ( 2 \varphi_1 \varphi_2^\dagger ) \phi_2 +
	\phi_2^\dagger ( 2 \varphi_2 \varphi_1^\dagger ) \phi_1 \bigg)
}
At second order, the superpotential contributes
\SP{
	V_W^{(2)} = \frac{1}{g^2} 2 \Tr \bigg( \phi_1^\dagger (2 (\varphi_2)_{-\beta}^\dagger (\varphi_2)_{-\beta} ) \phi_1 + \phi_2^\dagger (2 (\varphi_1)_\beta^\dagger (\varphi_1)_\beta ) \phi_2 + \\
	+ \phi_2^\dagger (2 (\varphi_1)_\beta^\dagger (\varphi_2)_{-\beta} ) \phi_1 
	+ \phi_1^\dagger (2 (\varphi_2)_{-\beta}^\dagger (\varphi_1)_\beta ) \phi_2 + \\
	+ \phi_3^\dagger ( 2 (\varphi_1)_{-\beta}^\dagger (\varphi_1)_{-\beta} ) \phi_3 + \phi_3^\dagger ( 2 (\varphi_2)_\beta^\dagger (\varphi_2)_\beta ) \phi_3 \bigg).
}
We will now use the following relations:
\AL{
	&[J_{a,b},\phi] = \frac{1}{\sqrt{2N}} \sum_{i,j=1}^N 2 \sin \left( \frac{(bi-aj)\pi}{n} \right) \phi_{i,j} J_{a+i,b+j} \\
	&[J_{a,b},\phi]_{\pm \beta} = \frac{1}{\sqrt{2N}} \sum_{i,j=1}^N 2 \sin \left( \frac{(bi-aj \pm 1)\pi}{n} \right) \phi_{i,j} J_{a+i,b+j} \\
	&[J_{a,b},V^{k} \phi U^{l}] = \nonumber \\ &= \frac{1}{\sqrt{2N}} \sum_{i,j=1}^N 2 \sin \left( \frac{(b(i+k)-a(j-l))\pi}{n} \right) \phi_{i,j} J_{a+i+k,b+j-l} \\
	&[J_{a,b},V^{k} \phi U^{l}]_{\pm \beta} = \nonumber \\ &= \frac{1}{\sqrt{2N}} \sum_{i,j=1}^N 2 \sin \left( \frac{(b(i+k)-a(j-l) \pm 1)\pi}{n} \right) \phi_{i,j} J_{a+i+k,b+j-l}
}
After a change of basis
\AL{
	\phi_1' \equiv \phi_1 U^\dagger, \\
	\phi_2' \equiv V^\dagger \phi_2 \\
	\phi_3' \equiv V \phi_3 U,
}
matters simplify, and we will see that the cross terms between $\phi_1$ and $\phi_2$ cancel. We have that
\SP{
	(\varphi_2)_{-\beta} \phi_1 &= \lambda^{(2)} [J_{1,0}, \phi_1' U]_{-\beta} = \\ &= - \frac{1}{\sqrt{2N}} \sum_{i,j=1}^N 2\lambda^{(2)} \sin \left( \frac{j\pi}{n} \right) \phi'^{(1)}_{i,j} J_{i+1,j-1}
}
\SP{
	(\varphi_1)_\beta \phi_2 &= \lambda^{(1)} [J_{0,-1}, V \phi_1']_\beta = \\ &= - \frac{1}{\sqrt{2N}} \sum_{i,j=1}^N 2\lambda^{(1)} \sin \left( \frac{i\pi}{n} \right) \phi'^{(2)}_{i,j} J_{i+1,j-1}
}
\SP{
	\varphi_1^\dagger \phi_1 &= \lambda^{(1)} [J_{0,1}, \phi_1' U] = \\ &= \frac{1}{\sqrt{2N}} \sum_{i,j=1}^N 2\lambda^{(1)} \sin \left( \frac{i\pi}{n} \right) \phi'^{(1)}_{i,j} J_{i,j}
}
\SP{
	\varphi_2^\dagger \phi_2 &= \lambda^{(2)} [J_{-1,0}, V \phi_1'] = \\ &= \frac{1}{\sqrt{2N}} \sum_{i,j=1}^N 2\lambda^{(1)} \sin \left( \frac{j\pi}{n} \right) \phi'^{(2)}_{i,j} J_{i,j}
}
\SP{
	(\varphi_1)_{-\beta} \phi_3 &= \lambda^{(1)} [J_{0,-1}, V^\dagger \phi_2 U^\dagger]_{-\beta} = \\ &= - \frac{1}{\sqrt{2N}} \sum_{i,j=1}^N 2 \lambda^{(1)} \sin \left( \frac{i\pi}{n} \right) \phi'^{(3)}_{i,j} J_{i-1,j}
}
\SP{
	(\varphi_2)_\beta \phi_3 &= \lambda^{(2)} [J_{1,0}, V^\dagger \phi_2 U^\dagger]_\beta = \\ &= - \frac{1}{\sqrt{2N}} \sum_{i,j=1}^N 2 \lambda^{(2)} \sin \left( \frac{j\pi}{n} \right) \phi'^{(3)}_{i,j} J_{i,j+1}
}
After rescaling $\lambda^{(1)} \rightarrow \lambda^{(1)} / \sqrt{2}$, $\lambda^{(2)} \rightarrow \lambda^{(2)} / \sqrt{2}$ and writing $\phi' \rightarrow \phi$, we finally get
\EQ{
	V^{(2)}_b = \sum_{a=1}^3 \sum_{i,j=1}^N \left( 4 \left[ |\lambda^{(1)}|^2 \sin^2 \left( \frac{i\pi}{n} \right) + |\lambda^{(2)}|^2 \sin^2 \left( \frac{j\pi}{n} \right) \right] {\phi^{(a)}_{i,j}}^\dagger \phi^{(a)}_{i,j} \right)
}
The rest of the analysis is analogous to the case with one VEV. (Indeed, it is completely the same as for $\mathcal{N}=4$ SYM.)

\subsection{Energy Levels for Fermions}

The fermions get masses from coupling to gauginos
\EQ{
	\frac{1}{g^2} 2 \Tr \left( - i \sqrt{2} \lambda [\varphi_1^\dagger,\chi_1] - i \sqrt{2} \lambda [\varphi_2^\dagger,\chi_2] \right) + c.c.,
}
and from the superpotential
\EQ{
	\frac{1}{g^2} 2 \Tr \left( - i \sqrt{2} \chi_1 [\varphi_2,\chi_3]_\beta - i \sqrt{2} \chi_3 [\varphi_1,\chi_2]_\beta  \right) + c.c.
}
We make the following change of basis:
\AL{
	\chi_1' \equiv \chi_1 U^\dagger \\
	\chi_2' \equiv V^\dagger \chi_2 \\
	\chi_3' \equiv V \chi_3 U.
}
Then, we have that
\EQ{
	[\varphi_1^\dagger,\chi_1] = \lambda^{(1)} [J_{0,-1}, \chi_1 U] = \frac{1}{\sqrt{2N}} \sum_{i,j=1}^N 2 \sin \left( \frac{i\pi}{n} \right) \chi^{(1)}_{i,j} J_{i,j}
}
\EQ{
	[\varphi_2^\dagger,\chi_2] = \lambda^{(2)} [J_{1,0}, V \chi_2] = \frac{1}{\sqrt{2N}} \sum_{i,j=1}^N 2 \sin \left( \frac{j\pi}{n} \right) \chi^{(2)}_{i,j} J_{i,j}
}
\EQ{
	[\varphi_2,\chi_3]_\beta = \lambda^{(2)} [J_{1,0}, V^\dagger \chi_3 U^\dagger]_\beta = - \frac{1}{\sqrt{2N}} \sum_{i,j=1}^N 2 \sin \left( \frac{j\pi}{n} \right) \chi^{(3)}_{i,j} J_{i,j+1}
}
\EQ{
	[\varphi_1,\chi_2]_\beta = \lambda^{(1)} [J_{0,-1}, V \chi_2]_\beta = - \frac{1}{\sqrt{2N}} \sum_{i,j=1}^N 2 \sin \left( \frac{i\pi}{n} \right) \chi^{(2)}_{i,j} J_{i+1,j-1}
}
Similar calculations give the expected masses for the gauge bosons. The results for the energy levels are summarized in Table~2.2. As can be seen, everything is the same as $\mathcal{N}=4$ SYM, in the sense that $X_{ij}$ appears presicely where $\varphi^\dagger \varphi$ appears for $\mathcal{N}=4$ SYM.

\begin{table}
\begin{center}
\begin{tabular}{c | c | c | c}
  Field & $d_\ell$ & $|\varepsilon_\ell|$ & $\ell_0$ \\ \hline
 	$B_i$ & $2\ell(\ell+2)$ & $\sqrt{R^{-2}(\ell+1)^2 + X_{ij}(\lambda^{(1,2)})}$ & 1 \\
 	$C_i$ & $(\ell+1)^2$ & $\sqrt{R^{-2}\ell(\ell+2) + X_{ij}(\lambda^{(1,2)})}$ & 1 \\
 	$(c,\bar{c})$ & $-2(\ell+1)^2$ & $\sqrt{R^{-2}\ell(\ell+2) + X_{ij}(\lambda^{(1,2)})}$ & 0 \\
 	$(A_0,\phi_1,\phi_1^\dagger)_1$ & $(\ell+1)^2$ & $\sqrt{R^{-2}\ell(\ell+2) + X_{ij}(\lambda^{(1,2)})}$ & 0 \\
 	$(A_0,\phi_1,\phi_1^\dagger)_{2,3}$ & $(\ell+1)^2$ & $\sqrt{R^{-2}(\ell+1 \pm R \mu_1)^2 + X_{ij}(\lambda^{(1,2)})}$ & 0 \\
 	$\phi_2$ & $(\ell+1)^2$ & $\sqrt{R^{-2}(\ell+1)^2 + X_{ij}(\lambda^{(1,2)})} \pm \mu_2$ & 0 \\
 	$\phi_3$ & $(\ell+1)^2$ & $\sqrt{R^{-2}(\ell+1)^2 + X_{ij}(\lambda^{(1,2)})} \pm \mu_3$ & 0 \\
 	$(\lambda,\chi_1)$ & $-\ell(\ell+1)$ & $	\sqrt{R^{-2} \left( \ell + \frac{1}{2} \pm \frac{R \mu_1}{2} \right)^2 + X_{ij}(\lambda^{(1,2)})} \pm \frac{\mu_2 + \mu_3}{2}$ & 1 \\
 	$(\chi_2,\chi_3)$ & $-\ell(\ell+1)$ & $	\sqrt{R^{-2} \left( \ell + \frac{1}{2} \pm \frac{R \mu_1}{2} \right)^2 + X_{ij}(\lambda^{(1,2)})} \pm \frac{\mu_2 - \mu_3}{2}$ & 1
\end{tabular}\end{center}
\caption{\footnotesize The energies $\varepsilon_\ell$ for the Higgs branch and gauge group $U(N)$, together with their degeneracies $d_\ell$ for the various fields. $\ell_0$ is the minimum value of the angular momentum quantum number $\ell$. The expressions are valid for vanishing or critical ($\mu_i = R^{-1}$) chemical potentials. All possible sign combinations are allowed. $X_{ij}(\lambda^{(1,2)}) = 4 |\lambda^{(1)}|^2 \sin^2 \left( \frac{i\pi}{n} \right) + 4 |\lambda^{(2)}|^2 \sin^2 \left( \frac{j\pi}{n} \right)$.}
\end{table}

\subsection{One-Loop Effective Potential}

Inspecting Table~2.2, we see that the form of the spectrum has the interpretation as the appearance of two extra compact dimensions forming a discretized torus \cite{Dorey:2003pp} with radii given by
\ALlabel{
	R_1 &= \frac{N}{2\pi|\lambda^{(1)}|}, \\
	R_2 &= \frac{N}{2\pi|\lambda^{(2)}|},
}{eq:torusradii}
and lattice spacings given by
\AL{
	\epsilon_1 &= \frac{2 \pi R_1}{N} = \frac{1}{|\lambda^{(1)}|}, \\
	\epsilon_2 &= \frac{2 \pi R_2}{N} = \frac{1}{|\lambda^{(2)}|}.
}

Since
\EQ{
	X_{ij}(\lambda^{(1,2)}) = 4 |\lambda^{(1)}|^2 \sin^2 \left( \frac{i\pi}{N} \right) + 4 |\lambda^{(2)}|^2 \sin^2 \left( \frac{j\pi}{N} \right)
}
appears precisely where $|\varphi|^2$ would appear for $\beta = 0$ and $\mathcal{N}=4$ SYM, we see immediately from \eqref{eq:zeroTeffpot} that at zero temperature the effective potential on the Higgs branch must be independent of $\lambda^{(1)}$ and $\lambda^{(2)}$ and equal to
\EQ{
	V_0 + V_1 = \frac{1}{\textnormal{Vol}(S^3)} \frac{3N^2}{16 R}.
}
At finite temperature, we have
\SP{
	\label{eq:finitetempeffactHiggs}
	V_0 + V_1 = \frac{1}{\textnormal{Vol}(S^3)} \bigg\{
	\frac{3N^2}{16R} -	2T \sum_{i,j=1}^N
		\sum_{k=1}^\infty \frac{\cos(k \alpha_{ij} / T)}{k} \\
		\sum_{l = 1}^\infty l^2 e^{-\frac{k}{T} \sqrt{R^{-2} l^2 + X_{ij}(\lambda^{(1,2)})}} \\ \bigg( 1 + \frac{1}{2} \left[ \cosh \left( \frac{k \mu_2}{T} \right) + \cosh \left( \frac{k \mu_3}{T} \right) \right] - \\ - (-1)^k \left[ \cosh \left( \frac{k (\mu_2 +\mu_3)}{2T} \right) + \cosh \left( \frac{k (\mu_2 - \mu_3)}{2T} \right) \right] \bigg)
	\bigg\}.
}
Again, because of the attractive potential, we can put $\alpha_{ij}=0$ in the above expression, which shows that the large $N$ theory is in the deconfined phase.

Now, let us move on to the more general case when $n$ does not necessarily equal $N$. Again, we will only give VEVs to two of the complex scalar fields:
\EQ{
	\left\langle \phi_1 \right\rangle = \Lambda^{(1)} \otimes U_{(n)}, \ \left\langle \phi_2 \right\rangle = \Lambda^{(2)} \otimes V_{(n)},
}
with
\EQ{
	\Lambda^{(1,2)} = \textnormal{diag} \left( \lambda^{(1,2)}_1, \lambda^{(1,2)}_2, \ldots, \lambda^{(1,2)}_m, \right).
}
We can expand the fluctuations as
\EQ{
	\phi = \frac{1}{\sqrt{2n}} \sum_{a,b=1}^m \sum_{i,j=1}^n \phi_{i,j}^{a,b} M_{a,b} \otimes J_{i,j},
}
where
\EQ{
	\left( M_{a,b} \right)_{de} = \delta_{ad} \delta_{be}
}
is an $m \times m$ matrix.
Then, when the VEVs act on the fluctuations in commutators such as $[\phi_i,\phi]$, instead of getting expressions involving $\sin \theta$ with $\theta = \frac{j\pi}{n}$ or $\theta = \frac{i\pi}{n}$, we will now get expressions of the form
\EQ{
	\lambda^{(i)}_a e^{i\theta} - \lambda^{(i)}_b e^{-i\theta}.
}
It is the absolute value squared which will appear in the expressions for the energy levels. We have
\SP{
	&\left| \lambda^{(i)}_a e^{i\theta} - \lambda^{(i)}_b e^{-i\theta} \right|^2 = \\
	&= \left( \left| \lambda^{(i)}_a \right| - \left| \lambda^{(i)}_b \right| \right)^2 + 4 \left| \lambda^{(i)}_a \right| \left| \lambda^{(i)}_b \right| \sin^2 \theta',
}
with
\EQ{
	\theta' = \frac{1}{2} \left\{ \arg \lambda^{(i)}_a - \arg \lambda^{(i)}_b \right\} + \theta.
}
In other words, nothing is different from the case $n=N$ considered before (and summarized in Table~2.2) other than that $X$ now takes the form
\SP{
	X_{abij}(\Lambda^{(1,2)}) = \left( \left| \lambda^{(1)}_a \right| - \left| \lambda^{(1)}_b \right| \right)^2 + \left( \left| \lambda^{(2)}_a \right| - \left| \lambda^{(2)}_b \right| \right)^2 + \\ +
	4 \left| \lambda^{(1)}_a \right| \left| \lambda^{(1)}_b \right| \sin^2 \left( \frac{1}{2} \left\{ \arg \lambda^{(1)}_a - \arg \lambda^{(1)}_b \right\} + \frac{i\pi}{n} \right) + \\ +
	4 \left| \lambda^{(2)}_a \right| \left| \lambda^{(2)}_b \right| \sin^2 \left( \frac{1}{2} \left\{ \arg \lambda^{(2)}_a - \arg \lambda^{(2)}_b \right\} + \frac{j\pi}{n} \right).
}
At zero temperature, the one-loop effective action remains the same as for $n=N$ (i.~e. flat), while at finite temperature all that changes in the expression for the one-loop effective potential, equation \eqref{eq:finitetempeffactHiggs}, is the form of $X$ and that we now also have to sum over $a$ and $b$:
\SP{
	\label{eq:generalHiggs}
	V_0 + V_1 = \frac{1}{\textnormal{Vol}(S^3)} \bigg\{
	\frac{3N^2}{16R} -	2T \sum_{a,b=1}^m \sum_{i,j=1}^n
		\sum_{k=1}^\infty \frac{\cos(k \alpha_{ij} / T)}{k} \\
		\sum_{l=1}^\infty l^2 e^{-\frac{k}{T} \sqrt{R^{-2} l^2 + X_{abij}(\Lambda^{(1,2)})}} \\ \bigg( 1 + \frac{1}{2} \left[ \cosh \left( \frac{k \mu_2}{T} \right) + \cosh \left( \frac{k \mu_3}{T} \right) \right] - \\ - (-1)^k \left[ \cosh \left( \frac{k (\mu_2 +\mu_3)}{2T} \right) + \cosh \left( \frac{k (\mu_2 - \mu_3)}{2T} \right) \right] \bigg)
	\bigg\}.
}
The same remarks regarding the differences between gauge group $U(N)$ and $SU(N)$ remain true for the Higgs branch, with the only difference being that in order to use the same expressions for the one-loop effective potential in the two cases, we now need to take the large $m$ limit.

\section{Metastable Phases}
In this section, we will take one or more of the chemical potentials to be near critical, which we define as
\EQ{
	\mu_i = R^{-1} + \mathcal{O}(\lambda),
}
where $\lambda = g^2 N$ is the 't~Hooft coupling. In particular, this means that corrections to the preceding results appear at higher orders in perturbation theory. We will see that even though at a classical level this choice of chemical potential causes an instability, when we take into account the quantum corrections, there are metastable phases at small finite temperature $RT \ll 1$.

First, consider the Coulomb branch at small finite temperature and close to the origin of the moduli space, so that
\EQ{
	R^2 |\varphi_{ij}|^2, R^2 |{\varphi_\beta}_{ij}|^2 \ll RT \ll 1.
}
For $\mu_2 = \mu_3 = 0$, we can put $l=k=1$ in \eqref{eq:finitetempeffact} after which we get that the one-loop quantum correction to the effective potential is given by
\SP{
	V_1 = \frac{1}{\textnormal{Vol}(S^3)} \bigg\{
	\frac{3N^2}{16R} + \frac{R}{2} \sin^2(\pi \beta) \left| \sum_{i=1}^N \varphi_i \right| ^2 - \\ -
	4T \sum_{i,j=1}^N
		\bigg( e^{-\frac{1}{RT} \sqrt{1 + R^2 |\varphi_{ij}|^2}} +
	e^{-\frac{1}{RT} \sqrt{1 + R^2 |{\varphi_\beta}_{ij}|^2}} \bigg)
	\bigg\}.
}
Expanding in $\varphi$ and $\varphi_\beta$, we obtain
\SP{
	V_1 = \frac{1}{\textnormal{Vol}(S^3)} \bigg\{
	\frac{3N^2}{16R} + \frac{R}{2} \sin^2(\pi \beta) \left| \sum_{i=1}^N \varphi_i \right| ^2 - 8T N^2 e^{-\frac{1}{RT}} + \\
	+ 8RN e^{-\frac{1}{RT}} \sum_i |\varphi_i|^2 - 8R e^{-\frac{1}{RT}} \cos^2(\pi \beta) \left| \sum_{i=1}^N \varphi_i \right| ^2 \bigg\},
}
which again is the same result as for the $\mathcal{N}=4$ case in the case of gauge group $SU(N)$, but different for gauge group $U(N)$ \cite{Hollowood:2008gp}.
The tree level term is equal to
\EQ{
	V_0 = \frac{N}{\lambda} (R^{-2} - \mu_1^2) \sum_i |\varphi_i|^2,
}
so we see that we have a metastable state at the origin if
\EQ{
	0 < \mu_1 - R^{-1} < \frac{2 \lambda}{\pi^2 R} e^{-\frac{1}{RT}}.
}
This holds true for gauge group $U(N)$ also, since the only potentially negative contribution to the mass of the new field is suppressed exponentially for $RT \ll 1$. In the large $N$ limit, the decay rate, through tunnelling and thermal activation, becomes zero \cite{Hollowood:2008gp}.

Moving on to the Higgs branch and the case $n=N$, let us put $\mu_3 = 0$, $\alpha_{ij} = 0$. Again, we consider small temperature and VEVs:
\EQ{
	R^2 |\lambda^{(i)}|^2 \ll RT \ll 1.
}
The sum over $k$ in \eqref{eq:finitetempeffactHiggs} contains a piece equal to
\EQ{
	\label{eq:finitetempHiggssum}
	\sum_{k=1}^\infty \frac{1}{k} e^{-\frac{k}{RT} \left( \sqrt{l^2 + 4 R^2 |\lambda^{(1)}|^2 \sin^2 \left( \frac{i\pi}{N} \right) + 4 R^2 |\lambda^{(2)}|^2 \sin^2 \left( \frac{j\pi}{N} \right)} - 1 \right)},
}
which for $l = 1$ clearly leads to a logarithmic divergence for small VEVs. When more than one critical chemical potential is turned on, extra zero modes appear. The sum \eqref{eq:finitetempHiggssum} corresponds precisely to integrating out these zero modes, which really should have been kept in the effective action, and this is what causes the logarithmic divergence near the origin of the moduli space. This is analogous to what happens for $\Ncal = 4$ with two or three critical chemical potentials \cite{Hollowood:2008gp}. The next to leading contribution to the one-loop effective action \eqref{eq:finitetempeffactHiggs} comes from a term which is similar to \eqref{eq:finitetempHiggssum}, with $k = l = 1$ and a $\frac{1}{2}$ instead of a $1$ outside the square root in the exponent. Expanding in $|\lambda^{(i)}|$, we obtain
\SP{
	V_0 + V_1 = \frac{1}{\textnormal{Vol}(S^3)} \bigg\{ \frac{3N^2}{16R} - 4T N^2 e^{-\frac{1}{2RT}} + \\ +
	8 N R \left( \sum_{i=1}^N \sin^2 \left( \frac{i\pi}{N} \right) \right) e^{-\frac{1}{2RT}} \left( |\lambda^{(1)}|^2 + |\lambda^{(2)}|^2 \right) \bigg\}.
}
The only gauge invariant operator consistent with the symmetries of the theory, which would reproduce the same mass squared as above, is proportional to $\Tr ( \phi_1^\dagger \phi_1 + \phi_2^\dagger \phi_2 )$. Therefore, the extra zero modes in the effective action must also have a positive mass squared at the origin of moduli space. This shows that for near (and above) critical chemical potentials there is a metastable state at the origin. In terms of the radii \eqref{eq:torusradii} of the extra-dimensional torus, the torus is metastable at infinite volume, and decays to zero size.

For general $N = n m$, similar considerations lead to a one-loop effective action near the origin of the form
\SP{
	V_0 + V_1 = \frac{1}{\textnormal{Vol}(S^3)} \bigg\{ \frac{3N^2}{16R} - 4T N^2 e^{-\frac{1}{2RT}} + \\ +
	2 N R \sum_{a,b=1}^m \sum_{i,j=1}^n X_{abij}(\Lambda^{(1,2)}) \bigg\},
}
which also has a minimum with positive curvature for zero VEVs, showing that for near (and above) critical chemical potentials there is a metastable state at the origin of moduli space.

\section{Summary}
We have studied the $\beta$-deformation of $\mathcal{N}=4$ SYM on $S^3$ with chemical potentials. On the Coulomb branch, the one-loop effective potential at zero temperature and critical chemical potentials is flat for gauge group $SU(N)$, but for $U(N)$, there is a dependence on the overall $U(1)$ traceful part of the VEV. On the Higgs branch, the zero temperature one-loop effective action is flat both for $SU(N)$ and $U(N)$. This is expected since on the Higgs branch, the low energy theory is $\Ncal = 4$, and can be viewed as a six-dimensional theory with 16 supercharges compactified on a torus. At near critical chemical potential and small finite temperature, there is a metastable state at the origin of moduli space for both the Coulomb branch and Higgs branch. On the Higgs branch, this has the interpretation as an extra-dimensional torus which becomes metastable for infinite size and decays to zero size through quantum tunnelling and thermal activation.

\newpage

\chapter{Phase Structure of $\beta$-deformed $\Ncal=4$ SYM at Strong Coupling}
\label{ch:3}

At finite temperature, the gravitational dual of $\Ncal=4$ SYM with chemical potentials and gauge group $SU(N)$ is a solution of $\Ncal=2$ five-dimensional $U(1)^3$ gauged supergravity, which describes a Reissner-Nordström black hole carrying charges with respect to the three $U(1)$s \cite{Behrndt:1998ns, Behrndt:1998jd}. There are three background gauge fields $A_{i\mu}^{(1)}$ whose values at the boundary correspond to the value of the chemical potentials of the boundary quantum field theory. The five-dimensional charged black hole solution can be embedded in ten dimensional Type IIB supergravity compactified on $S^5$ \cite{Cvetic:1999xp}. The resulting Type IIB supergravity solution describes an (uncharged) $AdS_5$ black hole rotating in $S^5$. In \cite{Lunin:2005jy}, it was described how to, in general, generate the ten-dimensional solution describing the $\beta$-deformed theory by performing a TsT-transformation, a T-duality followed by a shift of variables and then another T-duality, on the solution describing $\Ncal=4$ SYM. Applying this method to the ten-dimensional rotating black hole solution, we obtain the Type IIB supergravity solution which is the gravity dual of finite temperature $\beta$-deformed $\Ncal=4$ SYM with chemical potentials.

In order to see if the picture remains qualitatively the same at strong 't~Hooft coupling as the weak coupling results derived in the previous chapter, we perform a probe-brane calculation in the dual gravitational background. This was done for finite temperature $\Ncal=4$ SYM in \cite{Yamada:2008em}, where it was found that for near critical chemical potentials, there is a metastable phase at strong coupling. On the Coulomb branch, the probe-branes we will use are D3-brane giant gravitons, which extend along the three non-radial spatial coordinates of $AdS_5$, whereas on the Higgs branch, the probe-branes are D5-brane giant gravitons, which, in addition to extending along the same coordinates in $AdS_5$ as the D3-branes, also wrap around the torus formed by the two directions in $S^5$ which involve the TsT-transformation. Giant gravitons were studied in, for example, \cite{McGreevy:2000cw, Grisaru:2000zn, Hashimoto:2000zp, Gubser:1998jb}, and in \cite{Imeroni:2006rb, Pirrone:2006iq} they were studied in the Lunin-Maldacena background. We show that for near critical chemical potentials, the metastable phases of $\beta$-deformed $\Ncal=4$ SYM at finite temperature and weak coupling persist at strong coupling as well.

The structure of this chapter is as follows. In section~3.1, we find the gravity dual describing the $\beta$-deformed theory, and in section~3.2, we carry out the probe-brane calculation which establishes the existence of a metastable phase at strong 't~Hooft coupling. Finally, we summarize our results in section~3.3.

\section{Gravity Dual}
\subsection{$AdS_5$ Black Hole Spinning in $S^5$}
Let us first review the Type IIB supergravity solution dual to finite temperature $\mathcal{N}=4$ SYM with chemical potentials. The solution describes an $AdS_5$ black hole spinning in $S^5$. The ten-dimensional background metric is given by \cite{Cvetic:1999xp}
\EQ{
	\label{eq:metric}
	ds^2_{10} = \tilde{\Delta}^{1/2} ds_5^2 + R^2\tilde{\Delta}^{-1/2} \sum_{i=1}^3 X_i^{-1} \left\{ dr_i^2 + r_i^2 \left( d\phi_i + R^{-1} A_i^{(1)} \right)^2 \right\},
}
where
\EQ{
	ds_5^2 = - H(r)^{-2/3} f(r) dt^2 + H(r)^{1/3} [ f(r)^{-1} dr^2 + r^2 d\Omega_{3,1}^2 ]
}
is the metric of the $AdS_5$ black hole, and $d\Omega_{3,1}$ is the volume element of the $S^3$. We have
\AL{
	H_i(r) &= 1 + \frac{q_i}{r^2}, \\
	H(r) &= H_1(r) H_2(r) H_3(r), \\
	f(r) &= 1 - \left( \frac{r_0}{r} \right)^2 + \left( \frac{r}{R} \right)^2 H(r), \\
	r_0 &= r_H \left( 1 + \left( \frac{r_H}{R} \right)^2 H(r_H) \right)^{1/2}, \\
	X_i &= H(r)^{1/3}/H_i(r), \\
	A_{i\mu} &= -\frac{e_i}{r^2 + q_i} \delta_{\mu,0}, \\
	e_i &= \sqrt{q_i(r_0^2 + q_i)}, \\
	\tilde \Delta &= \sum_{i=1}^3 X_i r_i^2, \\
	\sum_i r_i^2 &= 1.
}
In addition to the metric, we have the self-dual five-form $F^{(5)} = dc_4 = d\tilde c_4 + * d\tilde c_4$ with \cite{Yamada:2008em}
\SP{
	\tilde c_4 = \left[ \left( \frac{r}{R} \right)^4 \Delta -  \sum_i \frac{r_0^2 + (-r_H^2 + q_i)}{R^2} r_i^2 \right] dt \wedge \epsilon^{(3)} + \\ + \sum_i \left( \frac{e_i}{R^2} \right) r_i^2 (R d\phi_i)\wedge \epsilon ^{(3)},
}
where $\Delta \equiv H^{2/3} \tilde{\Delta}$, and $\epsilon ^{(3)}$ is the volume form with respect to $R^2 d\Omega_{3,1}$.

After going to the co-rotating frame
\EQ{
	\phi_i \rightarrow \phi_i - R^{-1} A_{i0}(r_H) t,
}
in which the horizon of the black hole is static, the only change in the metric \eqref{eq:metric} is
\EQ{
	A_{i0} \rightarrow \frac{e_i}{r_H^2 + q_i} - \frac{e_i}{r^2 + q_i}.
}
Also, the new expression for $d\tilde c_4$ is
\SP{
	\tilde c_4 = \left[ \left( \frac{r}{R} \right)^4 \Delta + \sum_i \frac{1}{R^2} \left\{ \frac{e_i^2}{r_H^2 + q_i} - (r_0^2 - r_H^2 + q_i) \right\} r_i^2 \right] dt \wedge \epsilon^{(3)} + \\ + \sum_i \left( \frac{e_i}{R^2} \right) r_i^2 (R d\phi_i)\wedge \epsilon ^{(3)}.
}
We can identify the chemical potentials of the field theory on the boundary as
\EQ{
	\mu_i = A_{i0}(\infty)/R = R^{-1} \frac{e_i}{r_H^2 + q_i}.
}

\subsection{TsT-Transformation}

The idea of Lunin and Maldacena \cite{Lunin:2005jy} was to obtain the Type IIB background describing the $\beta$-deformed theory by performing a TsT-trans-formation on the solution describing $\mathcal{N}=4$ SYM at finite temperature. The TsT-transformation is a solution generating technique which involves a T-duality, followed by a shift, and then another T-duality. The general rules for T-duality transformations are given in \cite{Bergshoeff:1995as}. We also found \cite{Frolov:2005dj} a useful reference for how to derive the action of a TsT-transformation on the metric $g$ and the NS 2-form $b$. T acts on $g$, $b$, and the dilaton $\phi$ as follows ($i,j > 1$):
\AL{
	g_{11} &\rightarrow \frac{1}{g_{11}} \\
	g_{ij} &\rightarrow g_{ij} - \frac{g_{1i} g_{1j} - b_{1i} b_{1j}}{g_{11}} \\
	g_{1i} &\rightarrow \frac{b_{1i}}{g_{11}} \\
	b_{ij} &\rightarrow b_{ij} - \frac{b_{1i} b_{1j} - b_{1i} g_{1j}}{g_{11}} \\
	b_{1i} &\rightarrow \frac{g_{1i}}{g_{11}} \\
	e^{2\phi} &\rightarrow \frac{e^{2\phi}}{g_{11}}
}
A shift s, given by
\EQ{
	\varphi_2 \rightarrow \varphi_2 + \gamma \varphi_1,
}
acts on $g$ as
\AL{
	g_{11} &\rightarrow g_{11} + \gamma^2 g_{22} + 2\gamma g_{12} \\
	g_{1i} &\rightarrow g_{1i} + \gamma g_{2i},
}
and on $b$ as
\EQ{
	b_{1i} \rightarrow b_{1i} + \gamma b_{2i}.
}
Starting with $b = 0$, a TsT-transformation gives for $i,j > 2$ ($G_{ij}$, $B_{ij}$ are the TsT-transformed fields):
\SP{
	G_{ij} = G g_{ij} + G \gamma^2 \bigg[ g_{ij}g_{22}g_{11} + g_{1i}g_{2j}g_{12} + g_{1j}g_{2i}g_{12} - \\ - g_{1i}g_{1j}g_{22} - g_{ij}g_{12}g_{12} - g_{2i}g_{2j}g_{11} \bigg],
}
where
\EQ{
	G \equiv \frac{1}{1 + \gamma^2 (g_{22}g_{11}-g_{12}^2)}.
}
For $i \leq 2$ or $j \leq 2$, we have
\EQ{
	G_{ij} = G g_{ij}.
}
For $b$, a TsT-transformation gives
\EQ{
	B_{ij} = G \gamma ( g_{1i}g_{2j} - g_{1j}g_{2i} ).
}
(Note that if $g_{1i}=g_{2i}=g_{1j}=g_{2j}=0$, then $G_{ij} = g_{ij}$ and $B_{ij} = b_{ij}$.) The dilaton transforms as
\EQ{
	e^{2\phi} \rightarrow G e^{2\phi},
}
and, finally, the $n$-forms transform as \cite{Imeroni:2008cr}
\EQ{
\label{eq:TsTnform}
    \sum_q C_q \wedge e^{-B} = \sum_q c_q \wedge e^{-b} + \gamma \left[ \sum_q c_q \wedge e^{-b}
\right]_{[\varphi^1][\varphi^2]},
}
where for a general $p$-form $\omega_p$ we have defined
\EQ{
	\omega_p = \bar \omega_p + \omega_{p[y]} \wedge dy,
}
where $\bar \omega_p$ does not contain any legs in $dy$.

\subsection{The $\beta$-deformed Solution}

Let us first take a look at the how the TsT-transformation was used in \cite{Lunin:2005jy} to obtain a Type IIB supergravity background that describes the $\beta$-deformed theory at finite temperature. Starting with the solution in $AdS_5 \times S^5$
\AL{
	\label{eq:beforeLM}
	ds^2 &= ds^2_{AdS_5} + R^2 \sum_{i=1}^3 (dr_i^2 + r_i^2d\phi_i^2), \\
	C_4 &= \omega_4 + 4 R^4 \omega_1 \wedge d\phi_1 \wedge \phi_2 \wedge \phi_3, \\
	e^{2\phi} &= e^{2\phi_0},
}
then going to coordinates
\AL{
	\phi_1 &= \varphi_3 - \varphi_2, \\
	\phi_2 &= \varphi_1 + \varphi_2 + \varphi_3, \\
	\phi_3 &= \varphi_3 - \varphi_1,
}
and performing a T-duality along $\varphi_1$, followed by a small shift $\varphi_2 \rightarrow \varphi_2 + \gamma \varphi_1$, ($\gamma \equiv \beta$) and then another T-duality along $\varphi_1$, one obtains the TsT-transformed Lunin-Maldacena solution
\ALlabel{
	ds^2 &= ds^2_{AdS_5} + R^2 \left[ \sum_{i=1}^3 (dr_i^2 + G r_i^2d\phi_i^2) + \hat \gamma^2 G r_1^2 r_2^2 r_3^2 \left( \sum_{i=1}^3 d\phi_i \right)^2 \right], \\
	B^{NS} &= R^2 \hat \gamma G (r_1^2 r_2^2 d\phi_1 \wedge d\phi_2 + r_2^2 r_3^2 d\phi_2 \wedge d\phi_3 + r_3^2 r_1^2 d\phi_3 \wedge d\phi_1), \\
	C_2 &= -4 R^2 \hat \gamma \omega_1 \wedge (d\phi_1 +d\phi_2 + d\phi_3), \\
	C_4 &= \omega_4 + 4 G R^4 \omega_1 \wedge d\phi_1 \wedge \phi_2 \wedge \phi_3, \\
	e^{2\phi} &= G e^{2\phi_0},
}{eq:LM}
where
\AL{
	G^{-1} &= 1 + \hat \gamma^2 (r_1^2 r_2^2 + r_2^2 r_3^2 +r_3^2 r_1^2), \\
	\hat \gamma &= R^2 \gamma \equiv R^2 \beta,
}
and
\AL{
	r_1 &= \cos \alpha, \\
	r_2 &= \sin \alpha \cos \theta, \\
	r_3 &= \sin \alpha \sin \theta, \\
	d\omega_1 &= \cos \alpha \sin^3 \alpha \sin \theta \cos \theta d\alpha \wedge d\theta, \\
	d\omega_4 &= \omega_{AdS_5}.
}

In order to obtain the correct background for the $\beta$-deformed theory at finite temperature and with chemical potentials, we should perform a TsT-transformation on the solution given in the previous section. First, we note that a coordinate change $\phi_i \rightarrow \phi_i + v t$ followed by TsT is the same as vice versa. This follows directly from the form of the transformation rules: as long as a coordinate transformation does not mix the two coordinates along which we T-dualize with each other, the TsT-transformed expressions behave as tensors. It is convenient to make the coordinate change
\EQ{
	\phi_i' = \phi_i + R^{-1} A_{i0} t,
}
after which, apart from a few scaling factors, the metric is the same as in \eqref{eq:beforeLM}. Only the components of the metric and $B^{NS}$ involving the coordinates $\phi_i$ are affected by the TsT-transformation. This means that we can take the LM solution \eqref{eq:LM} for the metric and $B^{NS}$ and simply make the following substitutions
\AL{
	R &\rightarrow \tilde{\Delta}^{-1/4} R, \\
	A^{(1)} &\rightarrow \tilde{\Delta}^{-1/4} A^{(1)}, \\
	r_i &\rightarrow X_i^{-1/2} r_i,
}
to obtain the correct form of the Type IIB supergravity solution describing $\beta$-deformed $\mathcal{N}=4$ SYM at finite temperature with chemical potentials:
\AL{
	ds^2_{10} &= \tilde{\Delta}^{1/2} ds_5^2 + R^2\tilde{\Delta}^{-1/2} \Bigg[ \sum_{i=1}^3 X_i^{-1} \left\{ dr_i^2 + G r_i^2 d\phi_i'^2 \right\} + \nonumber \\* & +
	\hat{\gamma}^2 G \frac{r_1^2 r_2^2 r_3^2}{X_1 X_2 X_3} \left( \sum_{i=1}^3 d\phi_i' \right)^2 \Bigg], \\
	B^{NS} &= \hat{\gamma} G R^2 \tilde{\Delta}^{-1/2} \bigg( \frac{r_1^2 r_2^2}{X_1 X_2} d\phi_1' \wedge d\phi_2' + \nonumber \\
	&+ \frac{r_2^2 r_3^2}{X_2 X_3} d\phi_2' \wedge d\phi_3' + \frac{r_3^2 r_1^2}{X_3 X_1} d\phi_3' \wedge d\phi_1' \bigg), \\
	e^{2\phi} &= G e^{2\phi_0},
}
where
\AL{
	G^{-1} &= 1+\hat{\gamma}^2 \left( \frac{r_1^2 r_2^2}{X_1 X_2} + \frac{r_2^2 r_3^2}{X_2 X_3} + \frac{r_3^2 r_1^2}{X_3 X_1}  \right), \\
	\hat{\gamma} &\equiv R^2 \tilde{\Delta}^{-1/2} \gamma.
}
Using \eqref{eq:TsTnform}, we have that
\ALlabel{
	C_0 &= 0, \\
	C_2 &= \gamma \left[ c_4 \right]_{[\varphi^1][\varphi^2]}, \\
	C_4 - C_2 \wedge B &= c_4, \\
	C_6 - C_4 \wedge B &= 0, \\
	C_8 &= 0,
}{eq:TsTnformexpl}
where we have used that $b=0$ and $B \wedge B = 0$.

\section{Probe-Brane Calculation}
\subsection{Coulomb Branch}
We will now perform a probe-brane calculation in the TsT-transformed background. The Coulomb branch of the theory is probed by a D3-brane, static in the co-rotating frame (in which the horizon of the black hole also is static), and extending in all the directions of $AdS_5$ except the radial direction \cite{Kraus:1998hv}. In the field theory, separating a D3-brane from the stack of $N$ branes at the origin, corresponds to turning on VEVs
\EQ{
	\phi_i = \textnormal{diag} (v_i,-\frac{v_i}{N-1}, \cdots, -\frac{v_i}{N-1}).
}
The action for a general D$p$-brane has the form
\SP{
	S_{\text{D}p} = - \tau_p \int d^{p+1}\sigma \ e^{-\phi}
		\sqrt{- \det \left(\hat{G}_{ab} + F_{ab} - \hat{B}_{ab}\right) }
		 - \\ - \tau_p \int_{\mathcal{M}_{p+1}} \sum_q \hat{C}_q \wedge e^{F-\hat{B}},
}
where $\tau_p = \frac{1}{(2\pi)^p g_s}$, and hats denote pullbacks onto the world-volume $\mathcal{M}_p$ of the brane. For the D3-brane, $F = \hat B = 0$, and the induced metric is given by
\SP{
	\tilde{\Delta}^{1/2} \left( - H(r)^{-2/3} f(r) dt^2 + H(r)^{1/3} r^2 d\Omega_{3,1}^2 \right)
	+ \\ + \tilde{\Delta}^{-1/2} \Bigg\{ \sum_{i=1}^3 X_i^{-1} G r_i^2  A_{i0}^2 + \hat{\gamma}^2 G \frac{r_1^2 r_2^2 r_3^2}{X_1 X_2 X_3} \left( \sum_{i=1}^3 A_{i0} \right)^2 \Bigg\} dt^2,
}
so that
\SP{
	e^{-\phi} \sqrt{- \det \left(\hat{G}_{ab}\right) } = e^{-\phi_0} r^3 \tilde{\Delta} \Bigg\{ G^{-1} H^{1/3} f - \\ - \tilde{\Delta}^{-1} H \Bigg[ \sum_{i=1}^3 X_i^{-1} r_i^2  A_{i0}^2 + \hat{\gamma}^2 \frac{r_1^2 r_2^2 r_3^2}{X_1 X_2 X_3} \left( \sum_{i=1}^3 A_{i0} \right)^2 \Bigg] \Bigg\}^{1/2}.
}
For the Wess-Zumino term, we have
\SP{
	\label{eq:WZD3}
	\int_{\mathcal{M}_4} ( \hat{C}_4 - \hat{C}_2 \wedge \hat{B} ) = \int_{\mathcal{M}_4} \hat{c}_4 = \\
	= \int_{\mathcal{M}_4} \left[ \left( \frac{r}{R} \right)^4 \Delta + \sum_i \frac{1}{R^2} \left\{ \frac{e_i^2}{r_H^2 + q_i} - (r_0^2 - r_H^2 + q_i) \right\} r_i^2 \right] dt \wedge \epsilon^{(3)},
}
which is the same as for the $\Ncal = 4$ case analyzed in \cite{Yamada:2008em}.

First we note that, at the horizon, the terms that are introduced by the deformation have no dependence on any of the coordinates parameterizing the $S^5$; this is because $f(r_H) = A_{i0}(r_H) = 0$. If we turn on just one chemical potential $\mu_2 = \mu_3 = 0$, then the probe-brane action is minimized for $r_1=1$, $r_2 = r_3 = 0$, in which case $G = 1$, and all $\gamma$-dependence disappears. Therefore, the analysis is exactly the same as for the undeformed case; for close to but above critical chemical potential, there will be a metastable state at $r = r_H$, which decays towards the run-away direction $r = \infty$ \cite{Yamada:2008em}. For two equal chemical potentials ($\mu_1 = \mu_2$, $\mu_3 = 0$), the undeformed probe-brane action is minimized for $r_3 = 0$, but has no dependence on $r_1$ or $r_2$. Since no such dependence is introduced at the horizon by the $\beta$-deformation, there is still a meta-stable state at $r = r_H$. For three equal critical chemical potentials, the probe-brane action has no dependence on either of the coordinates $r_i$ in the undeformed case. Again, at the horizon, no such dependence is introduced by the $\beta$-deformation. We note that a probe-brane at the black hole horizon $r = r_H$ should correspond to zero VEVs in the field theory.

\subsection{Higgs Branch}
\begin{figure}[t]
	\centering
		\includegraphics[width=9cm]{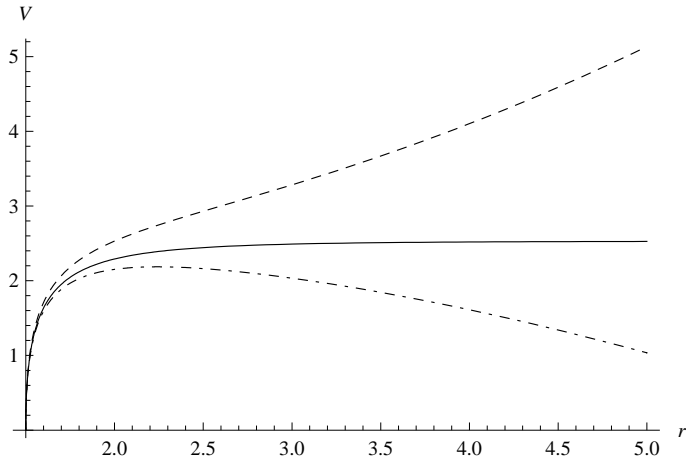}
	\caption{\footnotesize The D3 probe-brane action $V$ in the \textit{undeformed} background for chemical potentials $\mu_i = (q,q,q)$ as a function of the radius $r$. Up to a factor of $n$, this is the same as the action of a D5 probe-brane in the \textit{deformed} background. Everything is in units of $R$. The solid line corresponds to critical chemical potential $q = 1$, the dashed line corresponds to $q = 0.7$, and the dot-dashed line corresponds to $q = 1.2$. In all cases, we have put $r_H = 1.5$.}
	\label{fig:plot}
\end{figure}
The Higgs branch is probed by a D5-brane extending in the same directions as the D3-brane of the Coulomb branch, but in addition wrapping the torus formed by the two coordinates of the $S^5$ that are involved in the TsT-transformation \cite{Dorey:2003pp, Dorey:2004iq}. In the field theory, this corresponds to VEVs given by
\AL{
	\left\langle \phi_1 \right\rangle &= \Lambda^{(1)} \otimes U_{(n)}, \\
	\left\langle \phi_2 \right\rangle &= \Lambda^{(2)} \otimes V_{(n)}, \\
	\left\langle \phi_3 \right\rangle &= \Lambda^{(3)} \otimes V_{(n)}^\dagger U_{(n)}^\dagger,
}
with
\EQ{
	\Lambda^{(i)} = \textnormal{diag} (v^{(i)},-\frac{v^{(i)}}{m-1}, \cdots, -\frac{v^{(i)}}{m-1}).
}
Also, we will need to turn on a world-volume flux along the directions of the torus:
\EQ{
    F_{\varphi_1 \varphi_2} = \frac{1}{\gamma}.
}
One way of seeing this is that the D3-brane RR-charge of one D5-brane should be the same as that of $n = 1/\gamma$ D3-branes. Using \eqref{eq:TsTnformexpl} and $F \wedge F = B \wedge B = 0$, the Wess-Zumino term is
\SP{
	\int_{\mathcal{M}_6} \left( \hat{C}_6 + \hat{C}_4 \wedge (F - \hat{B}) + \frac{1}{2} \hat{C}_2 \wedge (F - \hat{B})^2 \right) = \int_{\mathcal{M}_6} \hat{c}_4 \wedge F,
}
which indeed is equal to $n$ times the corresponding expression \eqref{eq:WZD3} for a D3-brane, which in turn is the same as that for a D3-brane in the undeformed background corresponding to $\Ncal = 4$ studied in \cite{Yamada:2008em}.

For the world-volume part of the action, a more involved calculation gives
\SP{
	e^{-\phi} \sqrt{- \det \left(\hat{G}_{ab} + F_{ab} - \hat{B}_{ab}\right) } = \\ = e^{-\phi_0} \gamma^{-1} r^3 \tilde{\Delta} \left( H^{1/3} f - \tilde{\Delta}^{-1} H \sum_{i=1}^3 X_i^{-1} r_i^2  A_{i0}^2 \right)^{1/2},
}
which also is precisely equal to $n$ times the corresponding result for a D3-brane in the undeformed background. Therefore all the results of \cite{Yamada:2008em} apply in the TsT-transformed case. In particular, for nearly critical chemical potentials, there is a metastable state with a D5-brane situated at $r = r_H$, which will eventually be ``ejected'' towards infinite radius.

\section{Summary}
We have found the Type IIB supergravity background which describes $\beta$-deformed $\Ncal = 4$ SYM with chemical potentials at strong 't~Hooft coupling. At finite temperature, this solution describes a black hole rotating in the internal (deformed) $S^5$. The Coulomb branch is probed by a D3-brane, whereas the Higgs branch is probed by a D5-brane wrapping a torus. On both the Coulomb branch and the Higgs branch, for near (and above) critical chemical potentials there are metastable states in which the probe-branes reside at the black hole horizon and tunnel out towards infinite radius. This matches the weak coupling picture of the previous chapter.

\newpage

\chapter{Holographic Methods for Computing Spectra}
\label{ch:4}

The holographic prescription for computing the glueball spectrum is to study fluctuations around a particular background and look for solutions that satisfy correct boundary conditions in the IR and UV. Such solutions exist only for specific values of $K^2 = - M^2$, where $K$ is the four-momentum of the fluctuations. These $K^2$ correspond to poles of the correlator $\left\langle \mathcal{O} \mathcal{O} \right\rangle$ (where $\mathcal O$ is the operator in the dual field theory corresponding to the fluctuation in question), and give us the glueball spectrum of the dual field theory. In \cite{Berg:2005pd}, an explicitly gauge-invariant formalism was developed for studying fluctuations in five-dimensional non-linear sigma models consisting of a number of scalars coupled to gravity. The gauge-invariant formalism has the advantage that it allows one to study fluctuations of both the scalars and the metric degrees of freedom, while effectively decoupling them from each other. As we will see, the linearized equations of motion for the fluctuations can be solved algebraically for the gravitational degrees of freedom, and one ends up with a system of coupled differential equations that involve only the scalar fluctuations. Formulas for these linearized equations of motion for the fluctuations were given in \cite{Berg:2005pd} in terms of a superpotential $W$, from which the potential for the scalars could be derived.

This chapter closely follows \cite{Berg:2005pd}. However, we will derive the generalized versions of the formulas given therein, which hold for an arbitrary potential $V$ not necessarily obtainable from a superpotential. These methods have applications to both bottom-up approaches where the models are formulated in five dimensions, as well as top-down approaches where the five dimensional system originates from a consistent truncation of a higher-dimensional model in string theory or M-theory. In the following two chapters, we will apply them to compute the spectra of a few ten dimensional systems in Type IIB supergravity, for which there exist consistent truncations to five dimensional non-linear sigma models.

\section{The Model}

We start with a non-linear sigma model whose action is
\SP{
\label{eq:5daction}
	S = \int dr \int d^d x \sqrt{-g}\left[\frac{R}{4} - \frac{1}{2}G_{ab} g^{MN} \partial_M \Phi^a \partial_N \Phi^b - V(\Phi) \right],
}
where $G_{ab}(\Phi)$ is the non-linear sigma model metric and $V(\Phi)$ is a potential for the scalars. The cases we will be interested in have $d=4$, and the backgrounds will be of the form
\EQ{
	ds^2 = dr^2 + e^{2A}dx_{1,3}^2,
}
where $A(r)$ is a warp factor.

The equations of motion for the scalars following from the action \eqref{eq:5daction} read \cite{Berg:2005pd}
\SP{
\label{eq:zerothorderscalar}
	\nabla^2 \Phi^a + \mathcal{G}^a_{bc} g^{MN} (\partial_M \Phi^b) (\partial_N \Phi^c) - V^a = 0,
}
whereas Einstein's field equations read
\SP{
	- R_{MN} + 2 G_{ab} (\partial_M \Phi^a) (\partial_N \Phi^b) + \frac{4}{d-1} g_{MN} V = 0.
}
Here, we have defined $V_a \equiv \partial V / \partial \Phi^a$, and indices are lowered and raised using the non-linear sigma model metric $G_{ab}$ and its inverse. Furthermore, $\mathcal{G}^a_{\ bc}$ is the Christoffel symbol with respect to the non-linear sigma model metric
\EQ{
		\mathcal{G}^a_{\ bc} = \frac{1}{2} G^{ad} ( \partial_c G_{db} + \partial_b G_{dc} - \partial_d G_{bc} ).
}
For special cases, $V$ can be written in terms of a superpotential $W$ as follows:
\SP{
	V = \frac{1}{2} W^a W_a - \frac{d}{d-1} W^2.
}
When this is the case, and provided the background is assumed to depend only on the radial coordinate $r$, we obtain the first order equations of motion from the superpotential as
\SP{
\label{eq:Einsteinzerothorder}
	A' &= - \frac{2 W}{d-1}, \\
	\Phi'^a &= W^a,
}
where prime denotes the derivative with respect to $r$.

\section{Equations of Motion in the ADM formalism}

We will now generalize the results of \cite{Berg:2005pd} to cases where the potential $V$ can not necessarily be written in terms of a superpotential. The idea is to slice space-time along the radial coordinate and rewrite everything in terms of $d$-dimensional quantities in the ADM formalism.

We start by writing the metric on the form
\SP{
	g_{MN} =
	\left(
	\begin{array}{ll}
	 	\tilde g_{\mu\nu} & n_\nu \\
	 	n_\mu & n_\mu n^\mu + n^2
	\end{array}
\right),
}
where tilde is used to refer to $d$-dimensional quantities and the indices $\mu$ and $\nu$ run over the $d$-dimensional space-time.
The inverse metric is given by
\SP{
	g^{MN} =
	\frac{1}{n^2}
	\left(
	\begin{array}{ll}
	 	n^2 \tilde g^{\mu\nu} + n^\mu n^\nu & -n^\nu \\
	 	-n^\mu & 1
	\end{array}
\right).
}
The tangent vectors $X^M_\mu$ are given by $X^r_\mu = 0$ and $X^\nu_\mu = \delta^\nu_\mu$. We have a normal vector $N_M = (0,n)$, $N^M = n^{-1} (-n^\mu,1)$. The second fundamental form is
\SP{
	\mathcal{K}_{\mu\nu} = n \Gamma^r_{\mu\nu} = -\frac{1}{2n} (\partial_r g_{\mu\nu} - \nabla_\mu n_\nu - \nabla_\nu n_\mu),
}
and one can derive the following relations
\SP{
\label{gammarelations}
	\Gamma^\sigma_{\mu\nu} =& \tilde \Gamma^\sigma_{\mu\nu} - \frac{n^\sigma}{n} \mathcal{K}_{\mu\nu}, \\
	\Gamma^r_{\mu r} =& \frac{1}{n} \partial_\mu n + \frac{n^\nu}{n} \mathcal{K}_{\mu\nu}, \\
	\Gamma^\sigma_{\mu r} =& \nabla_\mu n^\sigma - \frac{n^\sigma}{n} \partial_\mu n - n \mathcal{K}_{\mu\nu} \left( g^{\nu\sigma} + \frac{n^\nu n^\sigma}{n^2} \right), \\
	\Gamma^r_{rr} =& \frac{1}{n} (\partial_r n + n^\nu \partial_\nu n + n^\mu n^\nu \mathcal{K}_{\mu\nu}), \\
	\Gamma^\sigma_{rr} =& \partial_r n^\sigma + n^\mu \nabla_\mu n^\sigma - n \nabla^\sigma n - 2n \mathcal{K}^\sigma_\mu n^\mu - n^\sigma \Gamma^r_{rr},
}
where $\tilde \Gamma^\sigma_{\mu\nu}$ is the $d$-dimensional Christoffel symbol corresponding to the metric $\tilde g_{\mu\nu}$.

Let us now write down the expressions for the equations of motion using the quantities defined above. The equation of motion for the scalars becomes
\SP{
\label{eq:scalars}
	\Big\{ \partial_r^2 - 2n^\mu \partial_\mu \partial_r + n^2 \nabla^2 + n^\mu n^\nu \nabla_\mu \partial_\nu - (n \mathcal{K}^\mu_\mu + \partial_r \ln n - n^\mu \partial_\mu \ln n) \partial_r + &\\
	\left[ n \nabla^\mu n - \partial_r n^\mu + n^\nu \nabla_\nu n^\mu + n^\mu (n \mathcal{K}^\nu_\nu + \partial_r \ln n - n^\nu \partial_\nu \ln n) \right] \partial_\mu \Big\} \Phi^a + &\\
	\mathcal{G}^a_{\ bc} \Big[ (\partial_r \Phi^b) (\partial_r \Phi^c) - 2n^\mu (\partial_\mu \Phi^b) (\partial_r \Phi^c) + &\\ (n^2 \tilde g^{\mu\nu} + n^\mu n^\nu) (\partial_\mu \Phi^b) (\partial_\nu \Phi^c) \Big] - n^2 G^{ab} \frac{\partial V }{\partial \Phi^b} = 0,
}
Einstein's equations separate into normal, mixed, and tangential components, obtained by projecting with $P^{MN} = N^M N^N - \tilde g^{\mu\nu} X^M_\mu X^N_\nu$, $P^{MN}_\mu = N^M X^N_\mu$, and $P^{MN}_{\mu\nu} = X^M_\mu X^N_\nu$, respectively. The normal component reads
\SP{
\label{eq:normal}
	(n \mathcal{K}^\mu_\nu) (n \mathcal{K}^\nu_\mu) - (n \mathcal{K}^\mu_\mu)^2 + n^2 \tilde R - 4n^2 V + 2G_{ab} \Big[ (\partial_r \Phi^a) (\partial_r \Phi^b) - &\\ 2n^\mu (\partial_\mu \Phi^a) (\partial_r \Phi^b) + (n^\mu n^\nu - n^2 \tilde g^{\mu\nu}) (\partial_\mu \Phi^a) (\partial_\nu \Phi^b) \Big] = 0.
}
The mixed component is given by
\SP{
\label{eq:mixed}
	\partial_\mu (n \mathcal{K}^\nu_\nu) - \nabla_\nu (n \mathcal{K}^\nu_\mu) - n \mathcal{K}^\nu_\nu \partial_\mu \ln n + n \mathcal{K}^\nu_\mu \partial_\nu \ln n - &\\ 2G_{ab} (\partial_r \Phi^a - n^\nu \partial_\nu \Phi^a) \partial_\mu \Phi^b = 0.
}
and the tangential component is
\SP{
\label{eq:tangential}
	- \partial_r (n \mathcal{K}^\mu_\nu) + n^\sigma \nabla_\sigma (n \mathcal{K}^\mu_\nu) + n \mathcal{K}^\mu_\nu (n \mathcal{K}^\sigma_\sigma + \partial_r \ln n - n^\sigma \partial_\sigma \ln n) + &\\ n \nabla^\mu \partial_\nu n + n \mathcal{K}^\mu_\sigma \nabla_\nu n^\sigma - n \mathcal{K}^\sigma_\nu \nabla_\sigma n^\mu - n^2 \tilde R^\mu_\nu + &\\ 2n^2 G_{ab} (\nabla^\mu \Phi^a) (\partial_\nu \Phi^b) + \frac{4 n^2 V}{d-1} \delta^\mu_\nu = 0.
}

\section{Linearized Equations of Motion}

We will now expand the equations of motion in fluctuations of the metric and the scalar fields to linear order. To this end, we expand the scalars as
\SP{
	\Phi^a \rightarrow \Phi^a + \varphi^a,
}
and the metric as
\SP{
	g_{\mu\nu} &\rightarrow e^{2A} (\eta_{\mu\nu} + h_{\mu\nu}), \\
	n_\mu &\rightarrow \nu_\mu, \\
	n &\rightarrow 1 + \nu,
}
with
\SP{
	h^\mu_\nu = {h^{TT}}^\mu_\nu + \partial^\mu \epsilon_\nu + \partial_\nu \epsilon^\mu + \frac{\partial^\mu \partial_\nu}{\Box} H + \frac{1}{d-1} \delta^\mu_\nu h,
}
where ${h^{TT}}^\mu_\nu$ is traceless and transverse, and $\epsilon^\mu$ is transverse. Altogether, we have the fluctuation variables $\{ \varphi, \nu, \nu^\mu, {h^{TT}}^\mu_\nu, h, H, \epsilon^\mu \}$. To first order, these transform under diffeomorphisms as
\SP{
	\delta \varphi^a &= \Phi'^a \xi^r, \
	\delta \nu = \partial_r \xi^r, \
	\delta \nu^\mu = \partial^\mu \xi^r + e^{2A} \partial_r \xi^\mu, \
	\delta {h^{TT}}^\mu_\nu = 0, \\
	\delta \epsilon^\mu &= \Pi^\mu_\nu \xi^\nu, \
	\delta H = 2 \partial_\mu \xi^\mu, \
	\delta h = 2 (d-1) A' \xi^r,
}
where
\SP{
	\Pi^\mu_\nu = \delta^\mu_\nu - \frac{\partial^\mu \partial_\nu}{\Box}
}
is the transverse projector.

In \cite{Berg:2005pd}, the following gauge invariant variables were defined
\SP{
\label{eq:gaugeinvariantvariables}
	\mathfrak a^a &= \varphi^a - \frac{\Phi'^a}{2(d-1) A'} h, \\
	\mathfrak b &= \nu - \frac{\partial_r (h/A')}{2(d-1)}, \\
	\mathfrak c &= e^{-2A} \partial_\mu \nu^\mu - \frac{e^{-2A} \Box h}{2(d-1) A'} - \frac{1}{2} \partial_r H, \\
	\mathfrak d^\mu &= e^{-2A} \Pi^\mu_\nu \nu^\nu - \partial_r \epsilon^\mu, \\
	\mathfrak e^\mu_\nu &= {h^{TT}}^\mu_\nu.
}
Next, the fluctuations were separated into two groups $X = \{ h, H, \epsilon^\mu \}$ and $Y = \{ \varphi, \nu, \nu^\mu, {h^{TT}}^\mu_\nu \}$. The variables $Y$ were rewritten in terms of the gauge invariant variables and $X$. Then Einstein's equations were expanded order by order, and it was shown that by performing a diffeomorphism, these could be written on a form such that at each order all the terms involving X cancelled, using the equations of motion at the previous order in the expansion. In this way, gauge invariant expressions for the linearized equations of motion were found. Indeed, these were the same as what is obtained by putting the variables $X$ to zero by hand everywhere, only keeping $Y$ in the expansion, then switching to the gauge invariant variables at the end of the calculation. This can be viewed as the gauge choice $X = 0$, but as pointed out in \cite{Berg:2005pd} really leads to gauge invariant expressions. Thus, we will expand around a background (assumed to be dependent only on $r$) to linear order using the following prescription:
\SP{
	\phi^a &\rightarrow \Phi^a + \mathfrak{a}^a, \\
	n &\rightarrow 1 + \mathfrak{b}, \\
	n^\mu &\rightarrow e^{2A} (\mathfrak{d}^\mu + \frac{\partial^\mu}{\Box} \mathfrak{c}), \\
	\tilde g_{\mu\nu} &\rightarrow e^{2A} (\eta_{\mu\nu} + \mathfrak{e}_{\mu\nu}).
}
It will be useful that
\SP{
	n \mathcal{K}^\mu_\nu \rightarrow - \partial_r A + \frac{1}{2} \left( \partial^\mu \mathfrak{d}_\nu + \partial_\nu \mathfrak{d}^\mu + 2 \frac{\partial^\mu \partial_\nu}{\Box} \mathfrak{c} - \partial_r \mathfrak{e}^\mu_\nu \right),
}
\SP{
	n \mathcal{K}^\mu_\mu \rightarrow - d \partial_r A + \mathfrak{c},
}
and
\SP{
	\tilde R^\mu_\nu = -\frac{1}{2} e^{-2A} \Box \mathfrak{e}^\mu_\nu.
}

Expanding the linearized equations of motion for the scalars \eqref{eq:scalars} to first order, we obtain
\SP{
\label{eq:scalarfirstorder}
	&\partial_r^2 \mathfrak{a}^a + e^{-2A} \Box \mathfrak{a}^a + d A' \partial_r \mathfrak{a}^a + 2 \mathcal{G}^a_{\ bc} \Phi'^b \partial_r \mathfrak{a}^c + \\& \partial_d \mathcal{G}^a_{\ bc} \Phi'^b \Phi'^c a^d - \frac{\partial V^a}{\partial \Phi^c} \mathfrak{a}^c - \Phi'^a (\mathfrak{c} + \partial_r \mathfrak{b}) - 2 V^a \mathfrak{b} = 0.
}
Defining a ``background covariant'' derivative as
\SP{
	D_r \varphi^a = \partial_r + \mathcal{G}^a_{\ bc} \Phi'^b \varphi^c,
}
and using \eqref{eq:zerothorderscalar}, we can write \eqref{eq:scalarfirstorder} as
\SP{
\label{eq:scalarfirstorder2}
	\Big[ D_r^2 + d A' D_r + e^{-2A} \Box \Big] \mathfrak{a}^a - &(V^a_{|c} - \mathcal{R}^a_{\ bcd} \Phi'^b \Phi'^d) \mathfrak{a}^c - \\& \Phi'^a (\mathfrak{c} + \partial_r \mathfrak{b}) - 2 V^a \mathfrak{b} = 0,
}
where $\mathcal R^a_{\ bcd}$ is the Riemann tensor with respect to the non-linear sigma model metric
\EQ{
	\mathcal R^a_{\ bcd} = \partial_c \mathcal G^a_{\ bd} - \partial_d \mathcal G^a_{\ bc} + \mathcal G^a_{\ ce} \mathcal G^e_{\ bd} - \mathcal G^a_{\ de} \mathcal G^e_{\ bc}.
}

Let us now expand Einstein's equations. At first order, the normal component of \eqref{eq:normal} gives
\SP{
\label{eq:normalfirstorder}
	2(d-1) A' \mathfrak{c} + 4 \Phi'_a (D_r \mathfrak{a}^a) - 4 V_a \mathfrak{a}^a - 8 V \mathfrak b = 0,
}
where we use the notation $\Phi'_a \equiv G_{ab} \Phi'^b$. The mixed components \eqref{eq:mixed} give
\SP{
\label{eq:mixedfirstorder}
	-\frac{1}{2} \Box \mathfrak{d}_\mu + (d-1) A' \partial_\mu \mathfrak{b} - 2 \Phi'_a \partial_\mu \mathfrak{a}^a = 0.
}
Here, we have used that $\mathfrak{e}^\mu_\nu$ and $\mathfrak{d}_\mu$ are transverse, and that $R = 0$ at first order. The tangential component of \eqref{eq:mixedfirstorder} implies that
\SP{
	\mathfrak{b} = \frac{2 \Phi'_a \mathfrak{a}^a}{(d-1) A'}.
}
Plugging into \eqref{eq:normalfirstorder} gives
\SP{
	\mathfrak{c} = \frac{8V \Phi'_a \mathfrak{a}^a}{(d-1)^2 A'^2} + \frac{2 V_a \mathfrak{a}^a}{(d-1) A'} - \frac{2 \Phi'_a D_r \mathfrak{a}^a}{(d-1) A'}.
}
Thus, we have obtained $\mathfrak b$ and $\mathfrak c$ algebraically in terms of $\mathfrak a^a$.
Using \eqref{eq:Einsteinzerothorder}, one can show that
\SP{
	\partial_r \mathfrak{b} = \Big[ - \frac{2d \Phi'_a}{d-1} + \frac{2 V_a }{(d-1) A'} + \frac{4 \Phi'_b \Phi'^b \Phi'_a}{(d-1)^2 A'^2} + \frac{2 \Phi'_a D_r}{(d-1) A'} \Big] \mathfrak{a}^a.
}
Plugging everything into \eqref{eq:scalarfirstorder2} finally gives us
\SP{
\label{eq:diffeq}
	&\Big[ D_r^2 + d A' D_r - e^{-2A} K^2 \Big] \mathfrak{a}^a - \\ &\Big[ V^a_{|c} - \mathcal{R}^a_{\ bcd} \Phi'^b \Phi'^d + \frac{4 (\Phi'^a V_c + V^a \Phi'_c )}{(d-1) A'} + \frac{16 V \Phi'^a \Phi'_c}{(d-1)^2 A'^2} \Big] \mathfrak{a}^c = 0.
}
This is the linearized equation of motion for the scalar fluctuations $\mathfrak a^a$ that we need to solve. As explained, it will in general only admit solutions with the correct IR and UV behaviour for special values of $-K^2 = M^2$, which in turn give us the spectrum. Let us point out that it is in general non-trivial to determine which boundary conditions are correct to impose.

In the special case where $V$ can be written in terms of a superpotential $W$, \eqref{eq:diffeq} agrees with the formula given in \cite{Berg:2005pd}:
\SP{
\label{eq:diffeqscalars}
		\Bigg[ \left( \delta^a_b D_z + W^a_{|b} - \frac{W^a W_b}{W} - \frac{2d}{d-1} W \delta^a_b \right) \left( \delta^b_c D_z - W^b_{|c} + \frac{W^b W_c}{W} \right) - \\ \delta^a_c e^{-2A} K^2 \Bigg] \mathfrak{a}^c = 0.
}

\section{Numerical Methods}

We will now describe how to set up the computation of the spectrum numerically. Suppose that we have a system of $n$ scalar fields satisfying a second order linear differential equation, and that the boundary conditions in the IR single out $p$ linearly independent solutions, whereas the boundary conditions in the UV single out $q$ solutions. A solution is completely characterized by evaluating it and and its derivative at a chosen point. Therefore, let us form vectors $({\mathfrak{a}_{IR}}_{(i)}, \partial_\rho {\mathfrak{a}_{IR}}_{(i)})$ where different $i$ denote different solutions in the IR, and we have suppressed the field index. These are $p$ column vectors of size $2n$. By evolving them numerically from the IR, we can evaluate them at any point we like, and therefore they are functions of $\rho$. Similarly, we form $q$ column vectors from the UV solutions, $({\mathfrak{a}_{UV}}_{(i)}, \partial_\rho {\mathfrak{a}_{UV}}_{(i)})$. The question that we need to answer is whether, for a particular value of $K^2$, we can find a solution that interpolates between the correct IR and UV behaviours. In other words, we want to know whether we can find a linear combination of the $p$ solutions in the IR and write it in terms of a linear combination of the $q$ solutions in the UV. For $p = q = n$, this is true if and only if the deterimant of the matrix formed by putting the IR and UV column vectors next to each other is equal to zero. It is convenient to evaluate this matrix at a point chosen between the IR and UV. In other words, the linearly independent solutions satisfying the boundary conditions in the IR and in the UV, respectively, are evolved numerically to a midpoint, where the determinant is evaluated. If it is zero for a particular value of $K^2$, there is a pole in the correlator. This is the midpoint determinant method described in \cite{Berg:2006xy}.

We would now like to generalize this method to include cases where $p$ and $q$ are not necessarily equal to $n$. In such cases, the matrix obtained by putting the IR and UV column vectors next to each other is not generally a square matrix, and therefore we can not answer the question of whether the vectors are linearly independent by evaluating a determinant. The method we will use instead is the following. First we normalize the vectors $({\mathfrak{a}_{IR}}_{(i)}, \partial_\rho {\mathfrak{a}_{IR}}_{(i)})$ and $({\mathfrak{a}_{UV}}_{(i)}, \partial_\rho {\mathfrak{a}_{UV}}_{(i)})$. Let us denote by $X^a_{\ i}$ ($i = 1, \ldots, p + q$, $a = 1, \ldots, 2n$) the matrix formed by putting these normalized column vectors next to each other. Then we construct an orthonormal basis $e^i_{\ a}$ ($i = 1, \ldots, p + q$, $a = 1, \ldots, 2n$) for the subspace spanned by these vectors. Finally, we project the normalized vectors onto the basis and form a matrix $Y^i_{\ j} = e^i_{\ a} X^a_{\ j}$. This is now the $(p+q) \times (p+q)$ matrix whose determinant we compute at a midpoint between the IR and UV. Again, if it is equal to zero, there is a pole in the correlator.

\section{Summary}

We have studied a generic five-dimensional non-linear sigma consisting of a number of scalars coupled to gravity. Expanding around a background and linearizing the equations of motion in fluctuations of the scalar fields and the metric, we found that the equations of motion for the fluctuation of the metric degrees of freedom could be solved algebraically in terms of the scalar fluctuations. In the end, the linearized equations of motion become a set of coupled differential equations for the scalar fluctuations, equation \eqref{eq:diffeq}. The holographic prescription for obtaining the spectrum is to solve this differential equation for different values of $-K^2 = M^2$ and impose that the fluctuations obey the correct behaviour in the IR and in the UV. It is only for special values of $M^2$ that well-behaved solutions exist, and these $M^2$ make up the spectrum. Finally, we have described numerical methods to be used for the practical implementations of these studies.

\newpage

\chapter{Glueball Spectra of SQCD-like Theories}
\label{ch:5}

In this chapter, we will study the spectrum of scalar glueballs in SQCD-like theories, whose gravity description is in terms of $N_c$ D5 branes wrapping an $S^2$ inside a CY3-fold, and $N_f$ backreacting D5 flavor branes wrapping a non-compact two-cycle inside the same CY3-fold \cite{Casero:2006pt}. The dual field theory is believed to be similar in the IR to $\mathcal N = 1$ SQCD with a quartic superpotential for the quark superfields. However, the full theory cannot be dual to SQCD for a number of reasons. It does not have an $SU(N_f) \times SU(N_f) \times U(1)_R$ global symmetry as SQCD does, but instead only one $SU(N_f)$ (broken further to $U(1)^{N_f}$ by smearing the flavor branes as will be discussed later). Also, for $N_f < N_c$, the Affleck-Dine-Seiberg superpotential \cite{Affleck:1983mk} tells us that SQCD does not have a vacuum, whereas for the systems we will study backgrounds exist with $N_f < N_c$.

Using the holographic techniques described in the previous chaper, we will find how the mass of the lightest scalar glueball in the spectrum depends on the number of flavors for a few different backgrounds. First we will show that a consistent truncation to a five-dimensional non-linear sigma model exists. This five-dimensional model contains four scalar fields coupled to gravity. In the gravity picture, Seiberg duality is realized for these theories as a diffeomorphism, i.e. just a change of variables \cite{Casero:2006pt, HoyosBadajoz:2008fw}. Therefore, the background itself does not change under Seiberg duality, but since we have changed variables, the dictionary interpretation of the dual field theory is changed. We show that the Lagrangian of the five-dimensional non-linear sigma model is invariant under a set of transformations of the scalar fields and $N_c \rightarrow N_f - N_c$. It follows that anything that can be computed within this framework will obey Seiberg duality.

The backgrounds correponding to the setup described above have been found to fall into two categories known as Type A and Type N \cite{HoyosBadajoz:2008fw, Casero:2007jj}. Type A backgrounds are special cases of Type N backgrounds for which the VEV of the gaugino condensate as well as the mesons are zero. We will study the spectrum of a few backgrounds of Type A for which the dilaton grows linearly in the UV. In the IR, there are different possible behaviours for the background (known as Type I, II and III \cite{Casero:2007jj}) corresponding to different vacua in the dual field theory. These backgrounds have a singularity in the IR which is ``good'' according to the criterion given in \cite{Maldacena:2000mw}, and are believed to capture the non-perturbative physics of the dual field theory. This criterion states that the $g_{00}$ component of the metric should not increase as we approach the singularity (the idea is that proper energy excitations should correspond to lower and lower energy excitations from the point of view of the field theory as one approaches the singularity in the IR).

The D5 flavor branes are smeared along the transverse angular coordinates, breaking the $SU(N_f)$ global symmetry to $U(1)^{N_f}$ (this procedure was first introduced in the context of flavor branes in \cite{Bigazzi:2005md}). The consistent truncation to five dimensions does not contain fluctuations of the gauge fields on the branes. However, it still contains fluctuations of the Ramond-Ramond 3-form $F_{(3)}$. Therefore, when $N_f \sim N_c$, the fluctuations that we consider mix glueballs and mesons. Since the fluctuations do not involve the gauge fields on the brane, the meson-glueballs whose spectrum we compute are $U(1)^{N_f}$-singlets.

Imposing the boundary condition on the fluctuations in the IR that their kinetic terms are regular, and in the UV that the fluctuations correspond to normalizable modes, we find that the mass of the lightest scalar glueball increases as the number of flavors is increased, until the point $N_f = 2 N_c$ is reached after which the opposite behaviour is observed. For a particular class of backgrounds that are Seiberg dual to themselves, we demonstrate explicitly that the spectrum obeys Seiberg duality.

There is by now a large literature on systems with back-reacting flavors. In the future, it would be interesting to apply the same techniques to study the glueball spectra of the various systems studied in \cite{HoyosBadajoz:2008fw, Casero:2007jj, Ramallo:2006et, Murthy:2006xt, Paredes:2006wb, Benini:2006hh, Casero:2007pz, Bertoldi:2007sf, Casero:2007ae, Benini:2007gx, Hirano:2007cj, Zeng:2007ta,  Burrington:2007qd, Benini:2007kg, Caceres:2007mu, Canoura:2008at, Cremonesi:2008zw, Bigazzi:2008gd, Bigazzi:2008zt, Bigazzi:2008ie, Bigazzi:2008cc, Arean:2008az, Gaillard:2008wt, Bigazzi:2008qq, Ramallo:2008ew, Bigazzi:2009gu, Caceres:2009bk, Bigazzi:2009bk, Nunez:2009da, Cotrone:2007qa}.

This chapter is organized as follows. In section~5.1, we describe the general setup and the backgrounds that we will study. In section~5.2, we derive the consistent truncation to the five-dimensional non-linear sigma model and discuss the Seiberg duality it obeys. Section~5.3 contains the computation of the spectra. Finally, we summarize our results in section~5.4.

\section{Gravity Duals of SQCD-like Theories}

The backgrounds we will be interested in are obtained from wrapping $N_c$ D5 color branes on an $S^2$ inside a CY3-fold, then adding $N_f$ back-reacting flavor branes that wrap a non-compact two-cycle inside the same CY3-fold. This is described in detail in \cite{Casero:2006pt}, where evidence is given for that the backgrounds obtained are dual to a field theory with similar behaviour in the IR as $\mathcal N = 1$ SQCD with a quartic superpotential for the quark superfields.

\subsection{Action and Equations of Motion}

We will now write the Type IIB supergravity action and the equations of motion that follow from it. The action (in Einstein frame) is given by
\SP{
\label{eq:ActionTypeIIBwithFlavors}
	S = \hat S_{IIB}+ S^{(flavors)},
}
where $\hat S_{IIB}$ describes Type IIB supergravity in the truncation to the metric, the dilaton, and the RR 3-form $(g_{\mu\nu}, \phi, F_{(3)})$, and $S^{(flavors)}$ is the action of the flavor branes. We have that
\SP{
	\hat S_{IIB} = \frac{1}{2 \kappa_{(10)}^2} \int d^{10}
	x \sqrt{-g} \left[ R - \frac{1}{2} \partial_\mu \phi 
	\partial^\mu \phi - \frac{1}{12} e^{\phi} F_{(3)}^2 \right],
}
where $\kappa_{10}$ is the 10d gravitational coupling constant. We will choose coordinates as $(x^\mu, \rho, \theta, \varphi, \tilde \theta, \tilde \varphi, \psi)$, where $\rho$ is the radial coordinate, the angles $0 \leq \theta \leq \pi$ and $0 \leq \varphi < 2 \pi$ parametrize an $S^2$, and the angles $0 \leq \tilde \theta \leq \pi$, $0 \leq \tilde \varphi < 2 \pi$, and $0 \leq \psi < 4 \pi$ parametrize an $S^3$. The flavor branes extend along the external coordinates $x^\mu$, the radial coordinate $\rho$, and the angular coordinate $\psi$. Their action is given by
\SP{
  S^{(flavors)} = T_{D5} \sum^{N_f} \Bigg[ - \int_{\mathcal{M}_6} d^6 x e^{\phi/2} \sqrt{- g_{(6)}} + \int_{\mathcal{M}_6} P[C_6] \Bigg],
}
where $T_{D5}$ is the D5-brane tension, $g_{(6)}$ is the determinant of the pullback of the metric to $\mathcal{M}_6$, the world volume of the flavor brane, and similarly $P[C_6]$ is the pullback of Ramond-Ramond $6$-form $C_6$. In order to simplify the analysis and avoid delta function sources in the equations of motion, we distribute the flavor branes evenly over the transverse angular coordinates $(\theta, \varphi, \tilde \theta, \tilde \varphi)$. This so-called smearing of the flavor branes breaks the global $SU(N_f)$ symmetry to $U(1)^{N_f}$. We obtain
\SP{
  S^{(flavors)} =
	\frac{T_{D5} N_f}{(4\pi)^2} \Bigg[ - \int 
	d^{10}x\sin\theta\sin\tilde{\theta} e^{\phi/2} \sqrt{-g_{(6)}} +
	\int C_6\wedge \Omega_4  \Bigg],
}
where
\SP{
	\Omega_4= \sin\theta \sin\tilde{\theta} 
	d\theta \wedge d\tilde{\theta} \wedge d\varphi\wedge d\tilde{\varphi}.
}

The equation of motion for the dilaton is
\SP{
	\frac{1}{\sqrt{-g}} \partial_\mu \left( g^{\mu \nu} \sqrt{-g} \partial_\nu \phi \right) - \frac{1}{12} e^\phi F_{(3)}^2 - \frac{N_f}{8} e^{\phi/2} \frac{\sqrt{-g_{(6)}}}{\sqrt{-g_{(10)}}} \sin \theta \sin \tilde \theta = 0,
}
while Einstein's equations read
\SP{
	R_{\mu\nu} - \frac{1}{2} g_{\mu\nu} R =& \frac{1}{2} \left( \partial_\mu \phi \partial_\nu \phi - \frac{1}{2} g_{\mu\nu} \partial_\lambda \phi \partial^\lambda \phi \right) + \\& \frac{e^\phi}{12} \left( 3 F_{\mu \sigma \lambda} F_\nu^{\ \sigma \lambda} - \frac{1}{2} g_{\mu\nu} F_{(3)}^2 \right) + T_{\mu\nu}^{(flavor)},
}
where
\SP{
	T^{\mu\nu}_{(flavor)} = - \frac{N_f}{8} \sin \theta \sin \tilde \theta e^{\phi/2} \delta^\mu_\alpha \delta^\mu_\beta g^{\alpha \beta}_{(6)} \frac{\sqrt{-g_{(6)}}}{\sqrt{-g_{(10)}}}.
}
Finally, Maxwell's equation for the $F_{(3)}$ is given by
\SP{
	\partial_\mu \left( \sqrt{-g} e^\phi F^{\mu\nu\lambda} \right) =0.
}

\subsection{Type A Backgrounds}
In this chapter, we will be interested in so-called Type A backgrounds. For these backgrounds the VEV of the gaugino condensate is zero, as are the VEVs of the meson matrix. Type A backgrounds can be obtained starting from the ansatz \cite{Casero:2006pt}
\SP{
	\label{eq:ansatzTypeA}
	ds^2 =& \mu^2 e^{2f} \Bigg[ \mu^{-2} dx_{1,3}^2 + e^{2k} d\rho^2 + e^{2h} (d\theta^2 + \sin^2 \theta d\varphi^2) + \\
	& \frac{e^{2\tilde g}}{4} (d\tilde\theta^2 + \sin^2 \tilde\theta d\tilde\varphi^2) + \frac{e^{2k}}{4} (d\psi + \cos \tilde\theta d\tilde\varphi + \cos \theta d\varphi)^2 \Bigg], \\
	F_{(3)} =& - \mu^2 \Big[ \frac{N_c}{4} \sin \tilde\theta d\tilde\theta \wedge d\tilde\varphi + \frac{N_f-N_c}{4} \sin \theta d\theta \wedge d\varphi \Big] \wedge \\& (d\psi + \cos \tilde\theta d\tilde\varphi + \cos \theta d\varphi),
}
with $\mu^2 = \alpha' g_s$. Here, $f$, $k$, $h$, and $\tilde g$ are taken to be functions of the radial coordinate $\rho$. For the backgrounds that we will be interested in the IR is at $\rho = 0$ and the UV at $\rho = \infty$.

Making a change of variables to $P$, $Q$, and $Y$ through
\SP{
	e^{2h} =& \frac{P + Q}{4}, \\
	e^{2\tilde g} =& P - Q \\
	e^{2k} =& 4Y,
}
the BPS equations can be solved as ($f = \phi / 4$) \cite{HoyosBadajoz:2008fw, Casero:2007jj}
\SP{
	Q =& Q_0 + (2 N_c - N_f) \rho, \\
	Y =& \frac{1}{8} (P' + N_f), \\
	e^{4 (\phi - \phi_0)} =& \frac{e^{4\rho}}{(P^2 - Q^2) Y},
}
where $\phi_0$ and $Q_0$ are integration constants, and $P$ satisfies a second order differential equation given by
\SP{
	P'' + (P' + N_f) \left( \frac{P' + Q' + 2 N_f}{P - Q} + \frac{P' - Q' + 2N_f}{P + Q} - 4 \right) = 0.
}

\subsection{IR and UV Expansions}

\begin{figure}[p]
\centering
	\includegraphics[width=9cm]{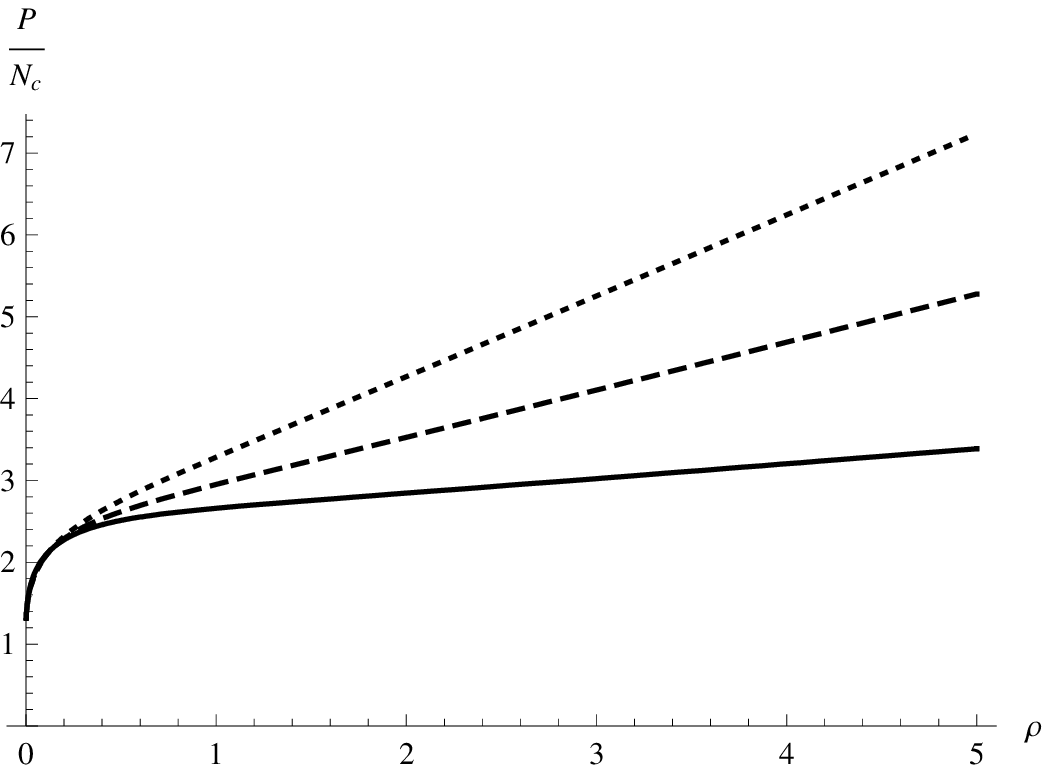}
	\caption{$P$ as a function of $\rho$ for the Type A background with Type II IR behaviour and $Q_0 = 1.2$. The different lines correspond to different number of flavors: dotted is $N_f = N_c$, dashed is $N_f = 1.4 N_c$, and solid is $N_f = 1.8 N_c$.}
	\label{fig:PplotTypeII_1}
\end{figure}

\begin{figure}[p]
\centering
	\includegraphics[width=9cm]{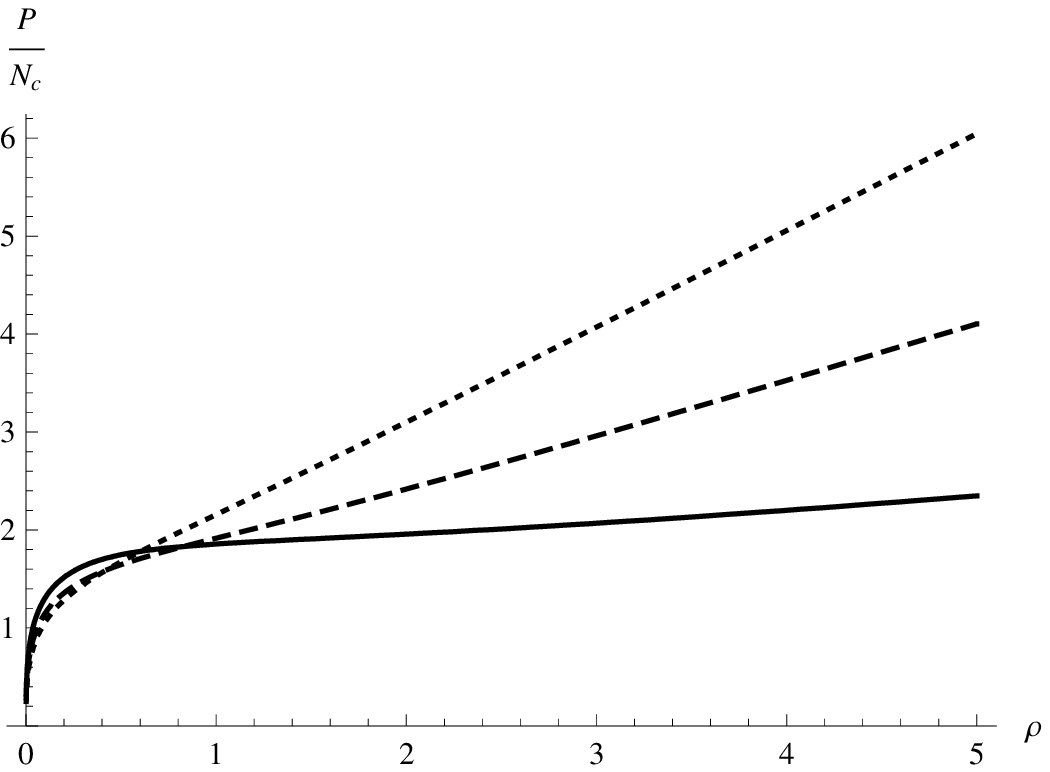}
	\caption{$P$ as a function of $\rho$ for the Type A background with Type III IR behaviour. The different lines correspond to different number of flavors: dotted is $N_f = N_c$, dashed is $N_f = 1.4 N_c$, and solid is $N_f = 1.8 N_c$.}
	\label{fig:PplotTypeIII_1}
\end{figure}

\begin{figure}[p]
\centering
	\includegraphics[width=9cm]{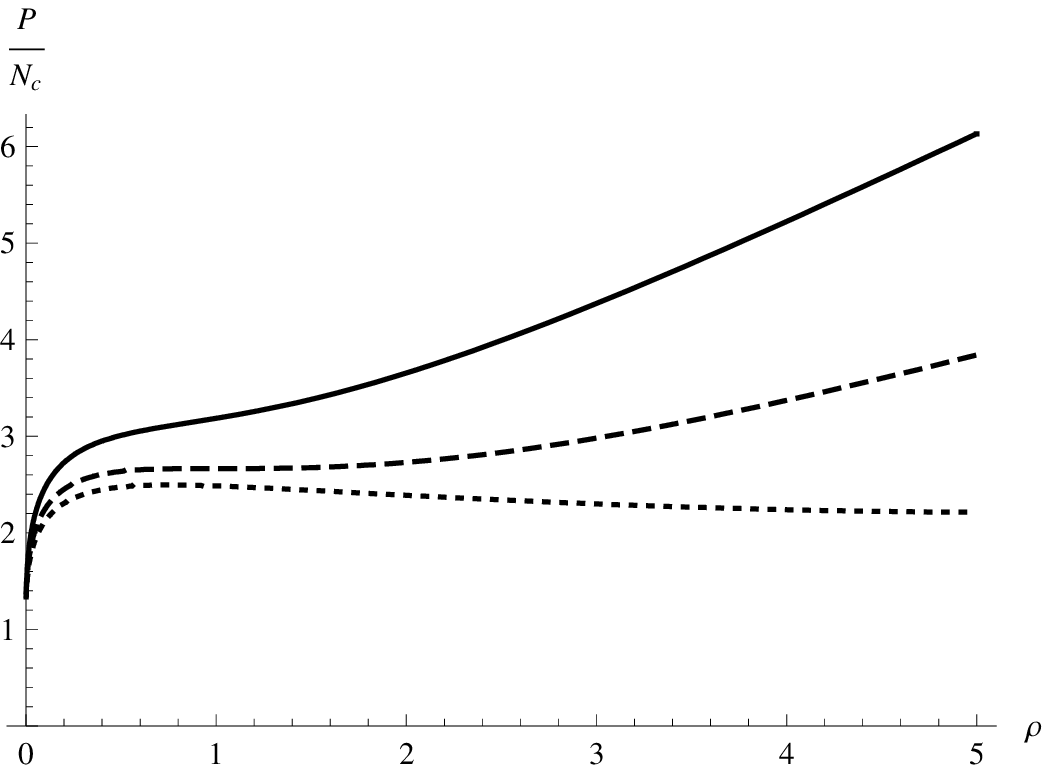}
	\caption{$P$ as a function of $\rho$ for the Type A background with Type II IR behaviour and $Q_0 = 1.2$. The different lines correspond to different number of flavors: dotted is $N_f = 2.2 N_c$, dashed is $N_f = 2.6 N_c$, and solid is $N_f = 3 N_c$.}
	\label{fig:PplotTypeII_2}
\end{figure}

\begin{figure}[p]
\centering
	\includegraphics[width=9cm]{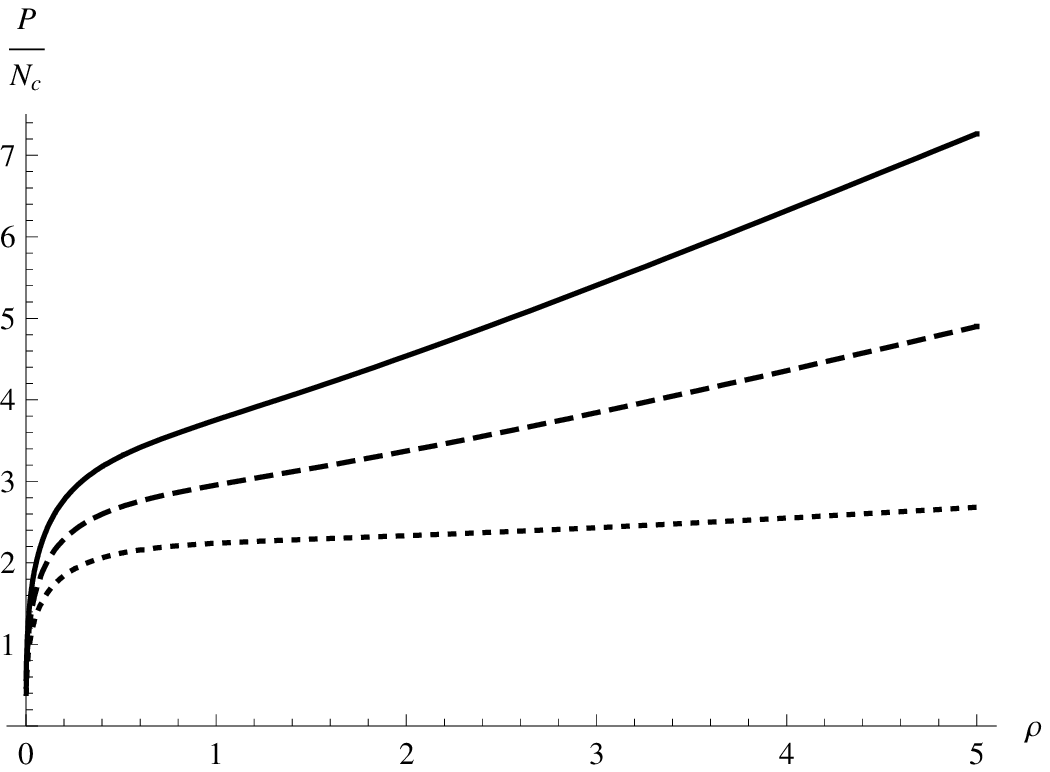}
	\caption{$P$ as a function of $\rho$ for the Type A background with Type III IR behaviour. The different lines correspond to different number of flavors: dotted is $N_f = 2.2 N_c$, dashed is $N_f = 2.6 N_c$, and solid is $N_f = 3 N_c$.}
	\label{fig:PplotTypeIII_2}
\end{figure}

We will be interested in backgrounds for which $P$ grows linearly in the UV.\footnote{The alternative is that the $P$ grows exponentially in the UV, in which case the spectrum only contains a continuum.} For $N_f < 2 N_c$, these have the UV expansion (around $\rho = \infty$) given by
\EQ{
\label{eq:UVexpansion}
	P_{UV} = (2 N_c - N_f) \rho + (N_c + Q_0) + \frac{N_f N_c}{4(2 N_c - N_f)} \rho^{-1} + \mathcal{O}\left(\rho^{-2}\right),
}
whereas if $N_f > 2 N_c$
\EQ{
	P_{UV} = - (2 N_c - N_f) \rho - ((1 + Q_0) N_c - N_f) - \frac{N_f (N_f - N_c)}{4(2 N_c - N_f)} \rho^{-1} + \mathcal{O}\left(\rho^{-2}\right).
}

In the IR, there are several different possible behaviours for the background. Here, we will focus on two different ones, that of Type II and Type III \cite{Casero:2007jj}. For Type II and $Q_0 > 0$ we have that \cite{Casero:2006pt}
\EQ{
	P_{IR}^{(II)} = Q_0 + 4 h_1 \sqrt{\rho} - \left( 2 N_c + \frac{8 h_1^2}{3 Q_0} + N_f \right) \rho + \mathcal{O}\left(\rho^{3/2}\right).
}
There are two integration constants: $Q_0$ and $h_1$. Solutions exist that interpolate smoothly between the Type II IR and the linear dilaton behaviour in the UV \cite{Casero:2006pt}. In order to obtain such solutions, one must dial the integration constants $Q_0$ and $h_1$, essentially making $h_1$ a function of $Q_0$. This leaves us with one free parameter $Q_0$ for the Type II solutions.

In the case of Type III, $Q_0 = 0$ and $P$ has the following behaviour in the IR:
\EQ{
	P_{IR}^{(III)} = 4 h_1 \rho^{1/3} - \frac{9 N_f}{5} \rho + \frac{8 h_1}{3} \rho^{4/3} + \mathcal{O}\left(\rho^{5/3}\right).
}
Now, requiring that the solution has the UV asymptotics of $P_{UV}$ completely fixes the one parameter $h_1$. Both the Type II and Type III backgrounds have a singularity in the IR, which satisfies the criterion for being a ``good'' singularity given in \cite{Maldacena:2000mw}.

Figure~\ref{fig:PplotTypeII_1} shows $P$ as a function of $\rho$ for the Type A background with Type II IR behaviour and $Q_0 = 1.2$ for a few different number of flavors $N_f < 2 N_c$. Figure~\ref{fig:PplotTypeIII_1} shows the corresponding plots for the Type A background with Type III IR behaviour. Figures~\ref{fig:PplotTypeII_2}~and~\ref{fig:PplotTypeIII_2} are the same as Figures~\ref{fig:PplotTypeII_1}~and~\ref{fig:PplotTypeIII_1} but for flavors $N_f > 2 N_c$.

\subsection{Dual Field Theory}

We will now describe in more detail some aspects of the field theory conjectured to be dual to the backgrounds described above. First, consider the case of no flavors. Then the dual field theory is a four-dimensional $\mathcal N = 1$ supersymmetric field theory, obtained from a twisted compactification of six-dimensional SYM on $S^2$ where the twisting is such that it preserves four supercharges \cite{Andrews:2005cv, Andrews:2006aw}. At weak coupling, this field theory consists of a massless vector multiplet, $V$, as well as a Kaluza-Klein tower of massive chiral and vector multiplets, $\Phi_k$ and $V_k$. The infinite number of KK modes reflects the fact that the UV completion is not given by a quantum field theory (in fact, it is given by a Little String Theory). If it were possible to separate the scale set by the size of the $S^2$ from the scale $\Lambda$ at which the theory becomes strongly coupled, we would have a gravity dual of $\mathcal N = 1$ SYM. Unfortunately, this is not the case. The Lagrangian of the field theory without flavors has the generic form
\SP{
	\mathcal L =& \int d^4\theta \sum_k \left( \Phi^\dagger_k e^V \Phi_k + m_k \left| V_k \right|^2 \right) + \\&
	\int d^2\theta \left[ \mathcal W_\alpha \mathcal W^\alpha + \sum_k \left( W_{k,\alpha} W_k^\alpha + \mu_k \left| \Phi_k \right|^2 + \mathcal W (\Phi_k, V_k) \right) \right],
}
where $\mathcal W_a$ and $W_{k,\alpha}$ are the curvatures of $V$ and $V_k$, and $m_k$ and $\mu_k$ are the masses of the massive vector and chiral multiplets comprising the Kaluza-Klein tower. The superpotential $\mathcal W(\Phi_k, V_k)$, governing the interactions between the KK chiral and vector multiplets, is given by
\SP{
	\mathcal W(\Phi_k, V_k) = \sum_{i,j,k} z_{ijk} \Phi_i \Phi_j \Phi_k + \sum_k \hat f(\Phi_k) W_{k,\alpha} W_k^\alpha.
}

With the introduction of flavors, we also have to add the terms \cite{Casero:2006pt}
\SP{
	\int d^4 \theta \left( \bar Q^\dagger e^V \bar Q + Q^\dagger e^{-V} Q \right) + \int d^2 \theta \sum_{p,i,j,a,b} \kappa^{ij}_p \bar Q^{a,i} \Phi^{ab}_p Q^{b,j}
}
to the Lagrangian. Here, $a,b = 1, \ldots, N_c$ are indices of the fundamental and anti-fundamental representations of $SU(N_c)$ and $i,j = 1, \ldots, N_f$ are indices of the fundamental and anti-fundamental representations of $SU(N_f)$. Since the smearing procedure described above breaks $SU(N_f)$ to $U(1)^{N_f}$, the $\kappa^{ij}_p$ must serve the role of breaking this symmetry in the field theory, however its exact form is not known. In principle, we could have also considered the more general case where there is a superpotential for the flavors too. Again, the form of such a superpotential is not known.

As mentioned above, it is not possible to separate the scale set by the size of the $S^2$ from the scale $\Lambda$ at which the field theory becomes strongly coupled. Nevertheless, we can imagine integrating out at least some of the KK modes, if not all the way down to $\Lambda$. This then gives rise to an effective superpotential $W_{eff}$ containing quartic terms
\SP{
	W_{eff} \sim \sum_{p,i,j} \frac{\kappa^2_{p,i,j}}{2 \mu^2_p} (\bar Q_i Q_j)^2.
}
This means that in the IR the field theory is similar to $\mathcal N = 1$ SQCD with a quartic superpotential (bearing in mind that not all the KK modes can be integrated out in this fashion).

\section{5d Formalism}

\subsection{5d Effective Action}
\label{sec:5deffaction}

We will now derive the 5d effective action of the non-linear sigma model which is a consistent truncation of the 10d system discussed in the previous section. We start with the ansatz given in \eqref{eq:ansatzTypeA}, and plug it into the Type IIB supergravity action given by \eqref{eq:ActionTypeIIBwithFlavors}. We will assume that the background functions $(f, \tilde g, h, k, \phi)$ only depend on the coordinates $(x^\mu, \rho)$, and integrate over the angular coordinates. In fact, because of Lorentz invariance it is sufficient to first consider the case where the background functions only depend on the radial coordinate $\rho$, then generalize to the case when they can also depend on the external coordinates $x^\mu$. Performing the integration over the angular coordinates yields
\EQ{
	S_{10d} = \frac{4 \mu^4 N_c^2 (4\pi)^3}{2 \kappa_{(10)}^2} \int d\rho \int d^4 x e^{8f+2\tilde g+2h+2k} \left( T - V \right),
}
where
\SP{
	T = e^{-2k} \Bigg[& \frac{9 \left(f'\right)^2}{4}+h' f'+\frac{k' f'}{2}+\tilde{g}'
   f'+\frac{\left(h'\right)^2}{16}- \\& \frac{\left(\phi '\right)^2}{64}+\frac{1}{16}
   \left(\tilde{g}'\right)^2+\frac{h' k'}{8}+\frac{1}{4} h' \tilde{g}'+\frac{1}{8} k'
   \tilde{g}' \Bigg],
}
and
\SP{
	V &= \frac{1}{256} e^{-2 \left(2 (f+h)+k+2 \tilde{g}\right)} \times \\& \Bigg[16 e^{4 h+\phi } N_c^2+e^{\phi
   +4 \tilde{g}} \left(N_c-N_f\right){}^2 + 8 e^{\frac{1}{2} \left(4
   (f+h+k)+\phi +4 \tilde{g}\right)} N_f + \\& e^{4 f+2 k} \left(e^{4 \tilde{g}} \left(-16 e^{2
   h}+e^{2 k}\right)+16 e^{4 h+2 k}-64 e^{4 h+2 \tilde{g}}\right) \Bigg].
}
Notice  that the Wess-Zumino term, whose only effect is to change the Bianchi identity of $F_3$, does not appear in this action, from which the Einstein, dilaton and Maxwell equations are derived.

Let us change coordinates to
\SP{
& f = A + p - \frac{x}{2}, \;\;
\tilde g = -A - \frac{g}{2} + \log 2  - p + x, \\
& h = -A + \frac{g}{2} - p + x, \;\;\; k = -A + \log 2  - 4 p, 
}
with inverse
\SP{
	A &=\frac{1}{3} \left( 8f +2\tilde g +2h +k \right) - \log 2, \;\;\;
	g = -\tilde g + h + \log 2, \\
	p &= - \frac{1}{6} \left( 4 f + \tilde g + h +2k \right) + \frac{1}{2} \log 2, \;\;\;
	x = 2 f + \tilde g + h - \log 2,
}
and also change the radial coordinate as $dr = e^{A+k} d\rho$. This leads to
\EQ{
	S_{10d} = \frac{4 \mu^4 N_c^2 (4\pi)^3}{2 \kappa_{(10)}^2} \int dr \int d^4 x e^{4A} \left( T - V \right),
}
with
\EQ{
	T = 3 A'^2-\frac{g'^2}{4}-3p'^2-\frac{x'^2}{2}-\frac{\phi'^2}{8},
}
and
\SP{
\label{eq:V5dTypeA}
	V =& \frac{1}{128} e^{-2 (g+2 (p+x))} \Bigg[ 16 \left(-4 e^{g+6 p+2 x} \left(1+e^{2 g}\right)+e^{4
   g}+1\right)+ \\& e^{12 p+2 x+\phi } \left(e^{4 g} N_c^2+\left(N_c-N_f\right){}^2\right)+8 e^{2
   g+6 p+x+\frac{\phi }{2}} N_f\Bigg]
}
Recognizing that for a metric given by $ds_5^2 = dr^2 + e^{2A} dx_{1,3}^2$ (where the function $A$ is the warp factor) the Ricci scalar is (up to partial integrations) equal to $R = -12A'^2$, we can write this as the action of a 5d non-linear sigma model
\SP{
	S_{5d} = \int dr \int d^4 x \sqrt{-g} \Bigg[ \frac{R}{4} - \frac{1}{2} G_{ab} g^{MN} \partial_M \Phi^a \partial_N \Phi^b - V(\Phi) \Bigg],
}
where $\Phi = [g, p, x, \phi]$, and the non-linear sigma model metric is diagonal with entries $G_{gg} = \frac{1}{2}$, $G_{pp} = 6$, $G_{xx} = 1$, and $G_{\phi \phi} = \frac{1}{4}$. One can verify that every solution to the 5d equations of motion following from this action also solves the full 10d Type IIB supergravity equations of motion. This shows that the five-dimensional non-linear sigma model is a consistent truncation of the 10d Type IIB supergravity system. Finally, let us point out that by studying the five-dimensional system, we cannot see excitations of KK modes in the $S^2 \times S^3$. Therefore, only part of the spectrum is accessible for us to study using the methods of Chapter~\ref{ch:4}.

\subsection{Superpotential from the BPS Equations}

Using the BPS equations, it is possible to find a superpotential $W$, in terms of which the potential $V$ can be written as
\EQ{
\label{eq:VfromW}
	V = \frac{1}{2} W^a W_a - \frac{4}{3} W^2.
}
In order to derive the BPS equations, we will consider spinors in Type IIB SUSY variations that satisfy \cite{Casero:2006pt}
\SP{
\epsilon=i\epsilon^*,\;\;\ 
\Gamma_{\theta\varphi}\epsilon=\Gamma_{\tilde{\theta}\tilde{\varphi}}\epsilon,\;\;\; 
\Gamma_{r\tilde{\theta}\tilde{\varphi}\psi}\epsilon= \epsilon.
} 
The gravitino 
variation $\delta \psi_x = 0$ gives
\SP{
	f' = \frac{e^{-2 f-2 g-2 h+\frac{\phi }{2}}}{16} \left(e^{2 g} \
N_f-\left(e^{2 g}-4 e^{2 h}\right) N_c\right)
}
where prime denotes differentiation with respect to $\rho$.
These equations are the same as the ones coming from the dilatino 
variations with $\phi = 4f$. 
Further, $\delta \psi_\theta = 0$ gives
\SP{
	h' = \frac{1}{4} e^{-2 (f+h)} \left(e^{\phi /2} N_c+e^{2 (f+k)}-e^{\phi \
/2} N_f\right),
}
while $\delta \psi_{\tilde \theta} = 0$ gives
\SP{
	\tilde g' = e^{-2 \left(f+\tilde{g}\right)} \left(e^{2 (f+k)}-e^{\phi /2} N_c\right),
}
and, finally, $\delta \psi_{\psi} = 0$ gives
\SP{
	k' = \frac{1}{4} e^{-2 \left( f+h+\tilde{g}\right)} \Big[& -e^{\phi /2} \
\left(4 e^{2 h}-e^{2 \tilde{g}}\right) N_c-4 e^{2 (f+h+k)}+ \\& 8 e^{2 \
\left(f+h+\tilde{g}\right)}-e^{2 \left(f+k+\tilde{g}\right)}-e^{\frac{\phi }{2}+2 \tilde{g}} N_f\Big].
}
The equation of motion for $A$ gives us an expression for the superpotential
\EQ{
	W = - \frac{3}{2} \frac{\partial A}{\partial r} = -\frac{e^{-A-k}}{2} \left( 8f' +2g' +2h' +k' \right),
}
where (as above) prime denotes differentiation with respect to $\rho$.
Using the above BPS equations, one arrives at
\SP{
	W = \frac{1}{16} e^{-g-2 (p+x)} \Big[ e^{6 p+x+\frac{\phi }{2}}
   \left(\left(-1+e^{2 g}\right) N_c+N_f\right)- \\ 4
   \left(1+e^{2 g}+2 e^{g+6 p+2 x}\right)\Big].
}
Using \eqref{eq:VfromW}, one can check that this superpotential $W$ reproduces the Type A 5d potential $V$ given in \eqref{eq:V5dTypeA}. Also, the equations of motion derived from the superpotential are precisely the BPS equations given above.

\subsection{Seiberg Duality}

For the models considered in this chapter, Seiberg duality is realized on the gravity side as a diffeomorphism. While the background does not change, the change of variables means that the dictionary describing quantities in the dual quantum field theory changes.

In terms of the variables ($P$, $Q$, $Y$, $\phi$), Seiberg duality transforms \cite{Casero:2006pt}
\SP{
	Q &\rightarrow -Q, \\
	N_c &\rightarrow N_f - N_c,
}
leaving $P$, $Y$, and $\phi$ unchanged. Using the relations
\SP{
	e^{3A} =& \frac{e^{2 \phi} (P^2 - Q^2) \sqrt{Y}}{16}, \\
	e^{2g} =& \frac{P + Q}{P - Q}, \\
	e^{6p} =& \frac{4 e^{- \phi}}{\sqrt{P^2 - Q^2} Y}, \\
	e^{2x} =& \frac{e^{\phi} (P^2 - Q^2)}{16}
}
we see that in terms of the 5d variables, a Seiberg duality simply takes the form $g \rightarrow -g$ (and, as usual, $N_c \rightarrow N_f - N_c$). It is straightforward to see that both the potential $V$ and the non-linear sigma model metric $G_{ij}$ are invariant under this transformation. It follows that the whole 5d theory exhibits Seiberg duality, and therefore anything that we can compute within this framework, including the spectrum, will manifest Seiberg duality.

Considering that Seiberg duality is normally only a duality in the IR, it may seem odd that the whole 5d Lagrangian is invariant under the Seiberg duality transformations. However, in \cite{Strassler:2005qs} it is argued that $\mathcal N = 1$ SQCD with a quartic superpotential for the quark superfields can satisfy an {\it exact} Seiberg duality, where not only the IR of two different field theories are the same, but they in fact represent two different descriptions of the {\it same} renormalization group flow. The fact that in our setup Seiberg duality corresponds to diffeomorphisms supports the view that Seiberg duality is exact for these backgrounds. However, it is not clear how this happens from a field theory perspective. One can make the following schematic argument in the IR. Starting with a superpotential $W = h \mu^{-1} (\tilde Q Q)^2$, we first Seiberg dualize, obtaining $\hat W = h \mu M^2 + \tilde q M q$, where $M$ is the meson field and $q$ and $\tilde q$ are the dual quark and anti-quark fields. For $N_f < 2 N_c$, $h$ is a relevant operator, so we can integrate out $M$, solving 0 = $\partial \hat W / \partial M = 2 h \mu M + \tilde q q$. Plugging $M$ back into $\hat W$ gives us an effective superpotential $\tilde W = - h^{-1} \mu^{-1} (\tilde q q)^2 / 2$. As can be seen, this superpotential is of the same form as the one that we started with, but describing a theory with $\tilde N_c = N_f - N_c$ colors and $\tilde N_f$ flavors. Furthermore, the coupling $h$ has been inverted, consistent with that for $\tilde N_f > \tilde 2 N_c$ the coupling appearing in front of the quartic term of the superpotential is irrelevant. It is, however, clear that this argument only works in the IR. In order to argue from the field theory that Seiberg duality is exact, one would have to take into account the KK modes that become important in the UV. From this point of view, it still remains somewhat mysterious why the backgrounds we are considering seem to have an exact Seiberg duality and what it means in terms of the dual field theory.

\section{Scalar Spectra}

In this section, we will study some different Type A backgrounds, and compute the mass of the lightest scalar glueball as a function of the number of flavors. In Chapter~\ref{ch:4}, a system of coupled differential equations for the fluctuations were derived, equation \eqref{eq:diffeq}
\SP{
	&\Big[ D_r^2 + d A' D_r - e^{-2A} K^2 \Big] \mathfrak{a}^a - \\ &\Big[ V^a_{|c} - \mathcal{R}^a_{\ bcd} \Phi'^b \Phi'^d + \frac{4 (\Phi'^a V_c + V^a \Phi'_c )}{(d-1) A'} + \frac{16 V \Phi'^a \Phi'_c}{(d-1)^2 A'^2} \Big] \mathfrak{a}^c = 0.
}
Changing the radial coordinate as $dr = e^{A+k} d\rho$, \eqref{eq:diffeq} becomes
\SP{
\label{eq:flucdiffrho}
	\Big[ \delta^a_b \partial_\rho^2 + S^a_b \partial_\rho + T^a_b - \delta^a_b e^{2k} K^2 \Big] \mathfrak{a}^b = 0,
}
with
\SP{
	S^a_b =& 2 \mathcal{G}^a_{\ bc} \partial_\rho \Phi^c + 4 \left(\partial_\rho p + \partial_\rho A \right) \delta^a_b, \\
	T^a_b =& \partial_b \mathcal{G}^a_{\ cd} \partial_\rho \Phi^c \partial_\rho \Phi^d - \\& 4 e^{-8p} \Bigg[ \left( \frac{4 (V^a \partial_\rho \Phi^c + V^c \partial_\rho \Phi^a)}{3 \partial_\rho A} + \frac{16 V \partial_\rho \Phi^a \partial_\rho \Phi^c}{9 (\partial_\rho A)^2} \right) G_{cb} + \partial_b V^a \Bigg].
}
This is the second order linear differential equation for the scalar fluctuations that we need to solve for different values of $K^2$ imposing the certain boundary behaviour in the IR and UV. In the IR, we will require that the kinetic terms for the fluctuations are regular. In the UV, we will require that the fluctuations are normalizable. This gives us the spectrum. In the following, we will put $x_f = N_f/N_c$, and rescale $P \rightarrow N_c P$ and similarly for $Q$ and $Y$. All masses are given in units of $\sqrt{\alpha' g_s N_c}$.

\subsection{Boundary Conditions in the UV}
\subsection{$N_f < 2 N_c$}
We will now expand the differential equations for the scalars \eqref{eq:flucdiffrho} in the UV. For $N_f < 2 N_c$, the background is given by \eqref{eq:UVexpansion}. We obtain that
\SP{
	S^a_b =& 2 \delta^a_b + \mathcal{O}\left(\rho^{-1}\right), \\
	T =&
\left(
\begin{array}{llll}
 -4 & 0 & 4 & -2 \\
 0 & -6 & -1 & -\frac{1}{2} \\
 2 & -6 & -3 & \frac{1}{2} \\
 -4 & -12 & 2 & -3
\end{array}
\right) + \mathcal{O}\left(\rho^{-1}\right).
}
A basis that diagonalizes $T$ to leading order is given by
\SP{
	B =
\left(
\begin{array}{llll}
 -1 & \frac{2}{3} & 0 & 1 \\
 \frac{1}{2} & \frac{1}{6} & 1 & 0 \\
 1 & -\frac{1}{6} & -3 & \frac{1}{2} \\
 0 & 1 & -6 & -1
\end{array}
\right),
}
such that $B^{-1} T B$ is diagonal. To leading order, this diagonalizes the differential equations for the fluctuations, and we obtain four independent differential equations
\SP{
	\partial_\rho^2 \mathfrak{a}^1 + 2 \partial_\rho \mathfrak{a}^1 - (8 + K^2) \mathfrak{a}^1 =& 0 \\
	\partial_\rho^2 \mathfrak{a}^2 + 2 \partial_\rho \mathfrak{a}^2 - (8 + K^2) \mathfrak{a}^2 =& 0 \\
	\partial_\rho^2 \mathfrak{a}^3 + 2 \partial_\rho \mathfrak{a}^3 - K^2 \mathfrak{a}^3 =& 0 \\
	\partial_\rho^2 \mathfrak{a}^4 + 2 \partial_\rho \mathfrak{a}^4 - K^2 \mathfrak{a}^4 =& 0
}
with solutions $\mathfrak{a}^{1,2} \sim e^{(-1 \pm \sqrt{9 + K^2}) \rho}$ and $\mathfrak{a}^{3,4} \sim e^{(-1 \pm \sqrt{1 + K^2}) \rho}$. Note that if we imagine expanding the fluctuations as $\mathfrak{a}^a = \lambda^a(\rho) e^{ \left(-1 \pm \sqrt{C^a - M^2} \right) \rho}$ with $\lambda^a(\rho) = \sum_n \mathfrak{a}^a_{n} \rho^{b_{a,n}}$, the above analysis captures the $C^a$ but not the function $\lambda^a(\rho)$. Therefore, the exponential factors are in general multiplied by powers of $\rho$.\footnote{The validity of the expansion in powers of $\rho^{-1}$ hinges on that it is possible to find a basis in which the components with different exponential behaviour do not mix. In the case of \cite{Elander:2009pk}, this can be checked explicitly. However, in that case, $P$ is exponentially close to $P = 2 N_c \rho$ in the UV, so that we can always work with analytical expressions. In the present case, only the UV expansion of $P$ in powers of $\rho^{-1}$ is known. While we cannot verify that there exists a basis in which the different exponential behaviours do not mix, the fact that we can go to reasonably high cut-offs in the UV (around $\rho = 15$) before the numerics break down suggests that such a basis exists.} However, the exponential behaviour is all we need for setting up the boundary conditions in the numerics. We are interested in the subleading behaviour so we pick the minus signs. In \cite{Berg:2006xy}, a normalizability condition for the fluctuations was given:
\EQ{
		\int dz e^{2A} G_{ab} \psi^a \psi^b = \int d\rho e^{3A + k} G_{ab} \psi^a \psi^b < \infty.
}
In our case, we have in the UV that ($x_f = N_f / N_c$)
\EQ{
		e^{3A + k} = e^{2\rho + 2 \phi_0} \left[ \frac{\sqrt{1 - x_f}}{8} \rho^{1/2} + \mathcal{O}(\rho^{-1/2}) \right],
}
so that the subdominant fluctuations are always normalizable, while the dominant ones are not. Let us also point out that for $M^2 > 1$ or $M^2 > 9$, we start getting oscillatory behaviour for fluctuations in the UV, signalling the start of a continuum.

\subsection{$N_f > 2 N_c$}

Similar considerations as in the last section (but with a different $B$) now lead to solutions of the form $\mathfrak{a}^{1,2} \sim e^{(-1 \pm \sqrt{9 + (x_f - 1) K^2}) \rho}$ and $\mathfrak{a}^{3,4} \sim e^{(-1 \pm \sqrt{1 + (x_f - 1) K^2}) \rho}$ (times powers of $\rho$). Note that the appearance of $x_f$ is Seiberg duality at work (restoring units, it takes $g_s \alpha' N_c K^2 \rightarrow g_s \alpha' N_c (x_f - 1) K^2$).

\subsection{Boundary Conditions for Type II in the IR}
For Type II backgrounds, it is natural to expand the fluctuations in the IR as
\EQ{
\label{eq:flucansatzTypeII}
	\mathfrak{a}^a = \sum_{n=0}^\infty \mathfrak{a}^a_n \rho^{n/2}.
}
We will choose boundary conditions such that the kinetic terms of the scalars do not blow up, i.e. that the derivative $\partial_\rho \mathfrak{a}^a$ does not blow up in the IR. This fixes $\mathfrak{a}^a_1 = 0$ and, after plugging in the ansatz \eqref{eq:flucansatzTypeII} into the differential equations for the fluctuations \eqref{eq:flucdiffrho}, leads to four linearly independent solutions
\SP{
	\mathfrak{a}_{(1)} =
	\begin{pmatrix} 1 \\ 0 \\ 0 \\ 0 \end{pmatrix} + 
	\begin{pmatrix} -2 +\frac{2 x_f}{Q_0} + \frac{16 h_1^2}{Q_0^2} \\ \frac{2 + 2 Q_0 - x_f}{2 Q_0} \\ \frac{2 + 4 Q_0 - x_f}{2 Q_0} \\ \frac{2 - x_f}{Q_0} \end{pmatrix} \rho +
	\mathcal{O}(\rho^{3/2}),
}
\SP{
	\mathfrak{a}_{(2)} =
	\begin{pmatrix} 0 \\ 1 \\ 0 \\ 0 \end{pmatrix} + 
	\begin{pmatrix} 12 + \frac{12}{Q_0} - \frac{6 x_f}{Q_0} \\ -6 + \frac{3x_f}{2 Q_0} + \frac{3}{h_1^2} \\ -12 + \frac{3x_f}{2 Q_0} + \frac{3}{h_1^2} \\ \frac{3x_f}{Q_0} + \frac{6}{h_1^2} \end{pmatrix} \rho +
	\mathcal{O}(\rho^{3/2}),
}
\SP{
	\mathfrak{a}_{(3)} =
	\begin{pmatrix} 0 \\ 0 \\ 1 \\ 0 \end{pmatrix} + 
	\begin{pmatrix} 4 + \frac{2}{Q_0} - \frac{x_f}{Q_0} \\ -2 + \frac{x_f}{4 Q_0} + \frac{1}{2h_1^2} \\ -4 + \frac{x_f}{4 Q_0} + \frac{1}{2h_1^2} \\ \frac{x_f}{2 Q_0} + \frac{1}{h_1^2} \end{pmatrix} \rho +
	\mathcal{O}(\rho^{3/2}),
}
\SP{
	\mathfrak{a}_{(4)} =
	\begin{pmatrix} 0 \\ 0 \\ 0 \\ 1 \end{pmatrix} + 
	\begin{pmatrix} \frac{2-x_f}{2 Q_0} \\ \frac{x_f}{8 Q_0} + \frac{1}{4h_1^2} \\ \frac{x_f}{8 Q_0} + \frac{1}{4h_1^2} \\ \frac{x_f}{4 Q_0} + \frac{1}{2 h_1^2} \end{pmatrix} \rho +
	\mathcal{O}(\rho^{3/2}).
}
This fixes our boundary conditions in the IR.

\subsection{Boundary Conditions for Type III in the IR}
For Type III backgrounds it is natural to expand the fluctuations as
\EQ{
	\mathfrak{a}^a = \sum_{n=0}^\infty \mathfrak{a}^a_n \rho^{n/3}.
}
We see that the requirement that the derivatives of the fluctuations do not blow up in the IR now leads to $\mathfrak{a}^a_1 = \mathfrak{a}^a_2 = 0$, which is a stronger requirement than for Type II, and consequently leads to fewer than four allowed linearly independent solutions in the IR:
\SP{
	\mathfrak{a}_{(1)} =&
	\begin{pmatrix} 0 \\ -\frac{1}{6} \\ 1 \\ 0 \end{pmatrix} + 
	\begin{pmatrix} 0 \\ -1 \\ -2 \\ 0 \end{pmatrix} \rho +
	\begin{pmatrix} 0 \\ -\frac{h_1 K^2}{12} \\ \frac{h_1 K^2}{2} \\ 0 \end{pmatrix} \rho^{4/3} + 
	\mathcal{O}(\rho^{5/3}), \\
	\mathfrak{a}_{(2)} =&
	\begin{pmatrix} 0 \\ -\frac{1}{12} \\ 0 \\ 1 \end{pmatrix} + 
	\begin{pmatrix} 0 \\ \frac{1}{2} \\ 1 \\ 0 \end{pmatrix} \rho +
	\begin{pmatrix} 0 \\ -\frac{h_1 K^2}{24} \\ 0 \\ -\frac{h_1 K^2}{2} \end{pmatrix} \rho^{4/3} + 
	\mathcal{O}(\rho^{5/3}).
}

Now that we have fixed the boundary conditions, singling out a number of allowed linearly independent solutions in the IR and UV, the question becomes whether for a particular value of $K^2$ it is possible to find linear combinations of the allowed solutions in the IR which when evolved towards the UV can be written as linear combinations of the allowed solutions in the UV. The numerical methods used for determining this are outlined in section~4.4.

\subsection{Results}

\begin{figure}[t]
\centering
	\includegraphics[width=9cm]{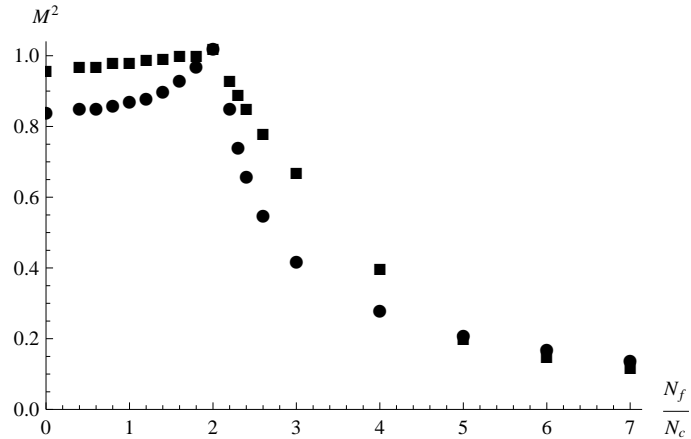}
	\caption{The mass squared of the lightest scalar glueball as a function of the number of flavors for a couple of Type A backgrounds with Type II IR behaviour: $Q_0=20$ (squares) and $Q_0=1.2$ (dots).}
	\label{fig:AbelianFlavorsTypeII}
\end{figure}

Figure~\ref{fig:AbelianFlavorsTypeII} shows the mass squared of the lightest scalar glueball as a function of $x_f = N_f/N_c$ for a couple of Type A backgrounds with Type II IR behaviour. As can be seen, the mass increases with the number of flavors, until the special point $N_f = 2 N_c$ (where the theory has certain peculiar properties) where it reaches the start of the continuum, $M^2 = 1$, and after that decreases as a function of the number of flavors.

\begin{figure}[p]
\centering
	\includegraphics[width=9cm]{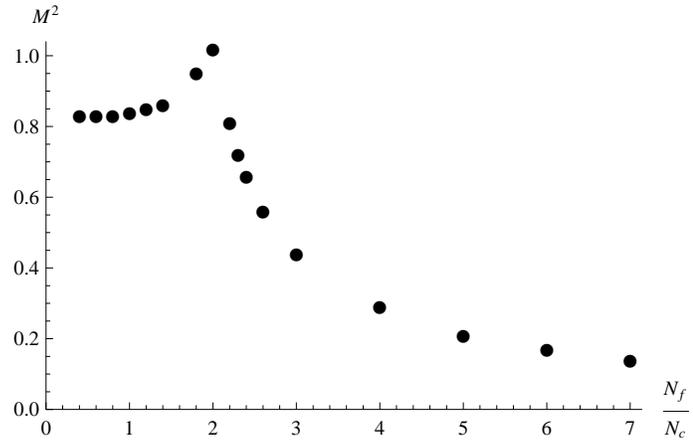}
	\caption{The mass squared of the lightest scalar glueball as a function of the number of flavors for the Type A background with Type III IR behaviour.}
	\label{fig:AbelianFlavorsTypeIII_1}
\end{figure}

\begin{figure}[p]
\centering
	\includegraphics[width=9cm]{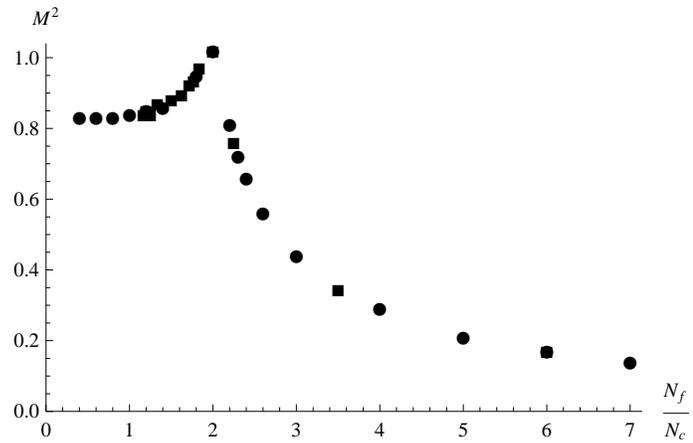}
	\caption{The mass squared of the lightest scalar glueball as a function of the number of flavors for the Type A background with Type III IR behaviour with the Seiberg dualized spectrum superimposed.}
	\label{fig:AbelianFlavorsTypeIII_2}
\end{figure}

Figure~\ref{fig:AbelianFlavorsTypeIII_1} shows the mass squared of the lightest scalar glueball as a function of $x_f = N_f / N_c$ for the Type A background with Type III IR behaviour. Again the same pattern can be seen.

Under Seiberg duality, the integration constant $Q_0 \rightarrow - Q_0$. Since for a Type A background with Type III IR behaviour $Q_0 = 0$, such backgrounds are Seiberg dual to themselves ($Q = (2 N_c - N_f) \rho \rightarrow (2 N_c - N_f) \rho$). In Figure~\ref{fig:AbelianFlavorsTypeIII_2}, the dots are the same as in Figure~\ref{fig:AbelianFlavorsTypeIII_1}, and the squares are what is obtained under Seiberg duality, mapping points as $x_f \rightarrow \frac{x_f}{x_f - 1}$. Also, under Seiberg duality, $M^2 \rightarrow (x_f - 1) M^2$, since we are working in units $\alpha' g_s N_c \rightarrow \alpha' g_s (N_f - N_c)$. As can be seen, the Seiberg dualized spectrum falls on the same trajectory.

\section{Summary}

We have been able to find a consistent truncation of the ten-dimensional Type IIB supergravity system describing $N_c$ D5 color branes and $N_f$ backreacting D5 flavor branes to five dimensions. The five-dimensional system is a non-linear sigma model coupled to gravity. In this model, Seiberg duality is realized at the level of the Lagrangian, i.e. any quantity that we can compute will automatically obey Seiberg duality.

We have computed the mass squared of the lightest scalar glueball for a few different Type A backgrounds, and found that the mass increases with the number of flavors for $N_f < 2 N_c$, but shows the opposite behaviour for $N_f > 2 N_c$. For a class of backgrounds that are Seiberg dual to themselves, we have seen explicitly how Seiberg duality is realized for the spectrum.

In the future, it would be interesting to apply the same techniques in order to compute the spectra of different systems with back-reacting flavors. For example, gravity duals that exhibit walking behaviour were found in \cite{Nunez:2008wi, Gurdogan:2009jd, Elander:2009pk}, and in particular one could imagine adding flavors to the walking backgrounds of \cite{Elander:2009pk} (for which $P$ grows linearly in the UV) studied in more detail in the next chapter and find out how their spectra are affected. It would be interesting to know what the effect of flavors is on the light scalar present for these backgrounds. We leave these questions for a future study.

\newpage

\chapter{Walking Dynamics from Gauge-Gravity Duality}
\label{ch:6}

In this chapter, we will study backgrounds in Type IIB supergravity, which exhibit walking behaviour, i.e. a suitably defined gauge coupling stays nearly constant in an intermediate energy regime. The backgrounds are obtained from the same kind of setup as those of the previous chapter, i.e. $N_c$ number of D5-branes wrapping an internal $S^2$, but we will now allow the solutions to have a more general form. In other words, the backgrounds we will study in this chapter will be of so-called Type N, which means that the VEV of the gaugino condensate is non-zero. Although, the walking backgrounds that are the main topic of this chapter do not have flavors, we will initially keep them in the analysis, so that we may generalize the results about Seiberg duality of the previous chapter to Type N systems.

The walking theories of this chapter share qualitative features of a certain class of phenomenological models known as Walking Technicolor. Technicolor models are gauge theories that become strongly coupled at the TeV scale \cite{Technicolor}. In these theories, electroweak symmetry would be dynamically broken much like chiral symmetry is broken by the chiral condensate in QCD. Furthermore, the large hierarchy between the electroweak scale and the Planck scale would no longer be a problem, for the same reason that there is no hierarchy problem associated with the smallness of $\Lambda_{\text{QCD}}$ relative to the Planck scale, i.e. dimensional transmutation. Technicolor models with walking dynamics \cite{WTC, Luty:2004ye} are viable candidates for physics beyond the Standard Model. However, finding explicit examples of such theories has proved difficult due to their strongly coupled nature. Recently, there has been a resurgence of interest in the lattice community, with many studies investigating whether field theories with walking behavior can be found \cite{Lattice}.

While plotting the gauge coupling as a function of energy scale gives an indication that one is dealing with a walking theory, it does not prove this is the case conclusively. The reason is that such a plot depends on what regularization scheme one uses. In the holographic picture, this is simply the fact that one could just as well have chosen a different radial coordinate corresponding to the energy scale. The lattice studies mentioned above are subject to analogous problems. Therefore, it is important to find well-defined physical questions to ask about the theory.

One such question is whether a light scalar exists in the spectrum. The existence of such a light scalar is conjectured to be due to the breaking of the approximate scale invariance of the walking region. It would then be the pseudo-Goldstone boson associated with dilatations, the dilaton.\footnote{Not to be confused with the dilaton of string theory.} It is an open question whether the dilaton is a generic feature of walking theories. From a phenomenological viewpoint, its existence would have significant consequences. Not only would its mass be lower than the dynamical scale set by where the walking theory becomes strongly coupled. To first order, it would also couple to the Standard Model fields in the same way as the Higgs does. Using the techniques of Chapter~\ref{ch:4}, we find the existence of a light state in the spectrum of the walking theories that we study. Its mass is suppressed by the length of the walking region, suggesting that it might be interpreted as a dilaton.

The structure of this chapter is as follows. In section~6.1, we review Type IIB supergravity backgrounds known as Type N. Even though the walking models which are the main topic of this chapter have no flavors, we keep them in the analysis for now. We find a consistent truncation to a five-dimensional non-linear sigma model, and generalize the results regarding Seiberg duality of the previous chapter. Next, in section~6.2, we describe the walking backgrounds whose spectra we will study. In section~6.3, we present our the results, and finally, we summarize our findings in section~6.4.

\section{Type N Backgrounds}

We will now describe Type N backgrounds. For now, we keep the flavor degrees of freedom, although eventually we will be interested in walking backgrounds, for which $N_f = 0$. Since Type N backgrounds are solutions of Type IIB supergravity coming from the same setup as in the previous chapter, i.e. $N_c$ D5 color branes wrapped on an $S^2$ inside a CY3-fold, and $N_f$ backreacting D5 flavor branes wrapped on a non-compact two-cycle inside the same CY3-fold, the formulas for the action and the equations of motion are the same as in section~5.1.1. The difference is that the ansatz for Type N is more general than the one for the Type A backgrounds of the previous chapter.

\subsection{Ansatz and BPS Equations}

The Type N ansatz is given by
\SP{
	\label{eq:ansatz}
	ds^2 =& \mu^2 e^{2f} \Bigg[ \mu^{-2} dx_{1,3}^2 + e^{2k} d\rho^2 + e^{2h} (d\theta^2 + \sin^2 \theta d\varphi^2) + \\
	& \frac{e^{2\tilde g}}{4} \left( (\tilde\omega_1 + a d\theta)^2 + (\tilde\omega_2 - a \sin \theta d\varphi)^2 \right) + \frac{e^{2k}}{4} (\tilde\omega_3 + \cos \theta d\varphi)^2 \Bigg], \\
	F_{(3)} =& \frac{\mu^2 N_c}{4} \Bigg[-(\tilde\omega_1 + b d\theta) \wedge (\tilde\omega_2 - b \sin \theta d\varphi) \wedge (\tilde\omega_3 + \cos \theta d\varphi) + \\
	& dy^\mu \wedge (\partial_\mu b (-d\theta \wedge \tilde\omega_1 + \sin \theta d\varphi \wedge \tilde\omega_2)) + (1 - b^2) \sin \theta d\theta \wedge d\varphi \wedge \tilde\omega_3 \Bigg] - \\
	& \frac{\mu^2 N_f}{4} \sin{\theta} d\theta \wedge d\varphi \wedge (d\psi +\cos{\tilde\theta} d\tilde\varphi),
}
where $\mu^2 = \alpha' g_s$, and
\SP{
	\tilde\omega_1 =& \cos \psi d\tilde\theta + \sin \psi \sin \tilde\theta d\tilde\varphi, \\
	\tilde\omega_2 =& -\sin \psi d\tilde\theta + \cos \psi \sin \tilde\theta d\tilde\varphi,\\
	\tilde\omega_3 =& d\psi + \cos \tilde\theta d\tilde\varphi.
}
As can be seen, the Type A ansatz corresponds to the special case $a = b = 0$.

Let us assume that the background functions $a$, $b$, $f$, $\tilde g$, $h$, and $k$ only depend on $\rho$. We will derive the BPS equations considering spinors in Type IIB with SUSY variations that satisfy
\SP{
	\epsilon = i \epsilon^*, \ \ \Gamma_{\theta\varphi} \epsilon = \Gamma_{12} \epsilon, \ \ \Gamma_{\rho 123} \epsilon = ( \mathcal A + \mathcal B \Gamma_{\varphi 2} ) \epsilon.
}
The gravitino variation $\delta \psi_x = 0$ gives
\SP{
	f' =&
	\frac{\mathcal{A} e^{-2 f-2 \tilde g-2 h+\frac{\phi }{2}}}{16} \Big[4 e^{2 h} \ N_c +e^{2 \tilde g} \left(N_f-\left(a^2-2 b a+1\right) N_c\right)\Big] + \\& \frac{\mathcal{B} e^{-2 f-\tilde g-h+\frac{\phi }{2}} N_c (b-a)}{4}
}
and
\SP{
	b' = 2 \mathcal{A} (a-b) + \frac{\mathcal{B} e^{-\tilde g-h}}{2 N_c} \Big[ 4 e^{2 h} N_c +e^{2 \tilde g} \ \left(N_f-\left(a^2-2 b a+1\right) N_c\right)\Big],
}
where prime denotes differentiation with respect to $\rho$.
These equations are the same as the ones coming from the dilatino variations with $\phi = 4f$. Further, $\delta \psi_\theta = 0$ gives
\SP{
	h' =& \frac{\mathcal{A} e^{-2 f-2 h}}{4} \Big[e^{\frac{\phi}{2}} \left(\left(a^2-2 b a+1\right) N_c-N_f\right)-\left(a^2-1\right) e^{2 (f+k)} \Big] + \\& \frac{\mathcal{B} e^{-2 f-\tilde g-h}}{2} \Big[a \left(-e^{2 (f+\tilde g)}-e^{2 (f+k)} +e^{\phi /2} N_c\right)-b e^{\phi /2} N_c\Big]
}
and
\SP{
	a' =& \mathcal{A} \left(-2 -2 e^{-2 \tilde g+2k}\right) a + \frac{\mathcal{B} e^{-2 f-3 \tilde g-h}}{2} \Big[ 2 \left(a^2-1\right) e^{2 (f+\tilde g+k)} - \\& 4 e^{2 h+\frac{\phi }{2}} N_c+ e^{2 \tilde g+\frac{\phi }{2}} \left(N_f-\left(a^2-2 b a+1\right) N_c\right)\Big].
}
$\delta \psi_{\tilde \theta} = 0$ gives
\SP{
	\tilde g' =& \mathcal{A} \left(e^{-2 \tilde g+2k}-e^{-2 f-2 \tilde g+\frac{\phi }{2}} N_c\right)
	\\&+\frac{\mathcal{B} e^{-2 f-\tilde g-h}}{2} \Big[a \left(e^{2 (f+\tilde g)} -e^{2 (f+k)} +e^{\phi /2} N_c\right)-b e^{\phi /2} N_c\Big]
}
and the constraint
\SP{
\label{eq:constraint1}
	0 = 4 \mathcal{A} a e^{-f-\tilde g-h+k} + \mathcal{B} e^{-f-2 \tilde g-2 h+k} \left(4 e^{2
   h}-\left(a^2-1\right) e^{2 \tilde g}\right).
}
Finally, $\delta \psi_{\psi} = 0$ gives
\SP{
	k' =& \frac{\mathcal{A} e^{-2 f-2 \tilde g-2 h}}{4} \Big[8 e^{2 (f+\tilde g+h)} +\left(a^2-1\right) e^{2 (f+\tilde g+k)} 
	\\&-4 e^{2 (f+h+k)} -4 e^{2 h+\frac{\phi }{2}} N_c+e^{2 \tilde g+\frac{\phi }{2}} \left(\left(a^2-2 b a+1\right) N_c-N_f\right)\Big]
	\\&+ \mathcal{B} e^{-2 f-\tilde g-h} \Big[a \left(-e^{2 (f+\tilde g)} +e^{2 (f+k)} +e^{\phi /2} N_c\right)-b e^{\phi /2} N_c\Big]
}
and the constraint
\SP{
\label{eq:constraint2}
	0 =& 4 \mathcal{A} e^{-3 f-\tilde g-h-k} \Big[b e^{\phi
   /2} N_c-a \left(-e^{2 (f+\tilde g)} +e^{2 (f+k)} +e^{\phi /2}
   N_c\right)\Big]
   \\& + \mathcal{B} e^{-3 f-2 \tilde g-2 h-k} \Big[8 e^{2 (f+\tilde g+h)} +\left(a^2-1\right) e^{2 (f+\tilde g+k)} 
   \\&-4 e^{2 (f+h+k)} -4
   e^{2 h+\frac{\phi }{2}} N_c+e^{2 \tilde g+\frac{\phi }{2}} \left(N_c a^2-2 N_c
   a b -N_f+N_c\right)\Big].
}
Solving for $\mathcal{A}$ in the first constraint and plugging into the second one leads to
\SP{
	0 =&
   e^{2 \tilde g} \left(e^{2 (f+\tilde g)} -4 e^{2
   (f+h)} +e^{\phi /2} (N_f-2 N_c)\right) a
   \\&+b e^{\phi /2} \left(e^{2g} +4 e^{2 h}\right) N_c
	-e^{2 f+4 \tilde g}  a^3
	+e^{2 \tilde g+\frac{\phi }{2}} a^2 b N_c .
}
Using that
\EQ{
	\mathcal{A}^2 + \mathcal{B}^2 = 1,
}
we obtain
\SP{
\label{eq:BPSA}
	\mathcal{A} = \frac{4 e^{2 h}-\left(a^2-1\right) e^{2 \tilde g}}{\sqrt{e^{4 \tilde g}
   \left(a^2-1\right)^2+16 e^{4 h}+8 \left(a^2+1\right) e^{2 (\tilde g+h)}}},
}
and
\SP{
\label{eq:BPSB}
	\mathcal{B} = -\frac{4 a e^{\tilde g+h}}{\sqrt{e^{4 \tilde g} \left(a^2-1\right)^2+16 e^{4 h}+8
   \left(a^2+1\right) e^{2 (\tilde g+h)}}}.
}
We would like to point out that due to the presence of constraints, it is non-trivial to find a superpotential $W$ that generates the 5d potential $V$. If we simply follow section~5.2.2 and use the equation of motion for $A$, i.e. $W = - \frac{3}{2} \partial A/\partial r$ (where $dr = e^{A+k} d\rho$), we would end up with expressions containing $\mathcal A$ and $\mathcal B$. In order to rewrite these as functions of the scalar fields, we would have to use one of the constraints (as we did in writing \eqref{eq:BPSA} and \eqref{eq:BPSB}). However, there is an ambiguity due to the fact that we there are two different constraints we can use, \eqref{eq:constraint1} and \eqref{eq:constraint2}. In either case, we end up with a superpotential that is only related to the 5d potential through $V = \frac{1}{2} W^a W_a - \frac{4}{3} W^2$ if evaluated on the classical solution. In other words, in order for this identity to be valid, one needs to use the constraint. This is enough when only considering the background, but since we will study fluctuations around the background in order to compute spectra, we need the more general formalism of Chapter~\ref{ch:4} that is valid for any 5d potential $V$.

In order to solve the BPS equations, it is convenient to go to the variables $P$, $Q$, $Y$, $\tau$, and $\sigma$ defined through \cite{HoyosBadajoz:2008fw}
\SP{
	e^{2h} =& \frac{1}{4} \left( \frac{P^2 - Q^2 }{P \cosh \tau - Q} \right), \\
	e^{2g} =& P \cosh \tau - Q , \\
	e^{2k} =& 4 Y, \\
	a =& \frac{P \sinh \tau}{P \cosh \tau - Q}, \\
	b =& \frac{\sigma}{N_c}.
}
In terms of these variables, the BPS equations can be written as a single second order differential equation for $P$:
\SP{
\label{eq:masterequationforP}
	P'' + (P' + N_f) \left( \frac{P' + Q' + 2 N_f}{P - Q} + \frac{P' Q' + 2 N_f}{P + Q} - 4 \cosh \tau \right),
}
where $Q$ and $\tau$ are given by
\SP{
	Q = \left( Q_0 + \frac{2 N_c - N_f}{2} \right) \cosh \tau + \frac{2 N_c - N_f}{2} (2 \rho \cosh \tau - 1)
}
and
\SP{
	\sinh \tau = \frac{1}{\sinh (2 (\rho - \rho_0))}.
}
The dilaton and $Y$ are given by
\SP{
	e^{4(\phi - \phi_0)} = \frac{1}{(P^2 - Q^2) Y \sinh^2 \tau}
}
and
\SP{
	Y = \frac{1}{8} (P' + N_f).
}
Here $Q_0$, $\phi_0$, and $\rho_0$ are integration constants. Without loss of generality, we put $\rho_0 = 0$ in the following.

\subsection{5d Effective Action}

We will now derive a consistent truncation to five dimensions of the ten-dimensional model under consideration. The derivation is analogous to that of section~\ref{sec:5deffaction}. Plugging the ansatz \eqref{eq:ansatz} into the Type IIB action given by \eqref{eq:ActionTypeIIBwithFlavors} and performing the integration over the angular coordinates yields
\EQ{
	S_{10d} = \frac{4 \mu^4 (4\pi)^3}{2 \kappa_{(10)}^2} \int d\rho \int d^4 x e^{8f+2\tilde g+2h+2k} \left( T - V \right),
}
where
\SP{
	T = e^{-2k} \Bigg[& -\frac{e^{2 \tilde{g}-2 h}}{128} a'^2-\frac{N_c^2 e^{-4 f-2 h+\phi -2 \tilde{g}}}{128}
   b'^2+\frac{9f'^2}{4}+\frac{h'^2}{16}- \\& \frac{\phi'^2}{64}+\frac{\tilde{g}'^2}{16} +f'
   h'+\frac{f' k'}{2}+\frac{h' k'}{8}+f' \tilde{g}'+\frac{1}{4} h'
   \tilde{g}'+\frac{1}{8} k' \tilde{g}' \Bigg],
}
and
\SP{
	V =& \frac{e^{-2 \left(2 (f+h)+k+2 \tilde{g}\right)}}{256} \times \\
   \Bigg[ & 8 e^{2 \left(2 f+h+2 k+\tilde{g}\right)} a^2+8
   e^{4 f+2 h+6 \tilde{g}} a^2+ 16 e^{4
   (f+h+k)}+ \\& \left(a^2-1\right)^2 e^{4
   \left(f+k+\tilde{g}\right)}- 64 e^{2 \left(2
   (f+h)+k+\tilde{g}\right)}- \\& 16 \left(a^2+1\right) e^{2
   \left(2 f+h+k+2 \tilde{g}\right)}+ 16 e^{4 h+\phi }
   N_c^2+ 8 (a-b)^2 e^{2 h+\phi +2 \tilde{g}}
   N_c^2+ \\& e^{\phi +4 \tilde{g}}
   \left(N_f- \left(a^2-2 b a+1\right)
   N_c\right)^2+ 8 e^{\frac{1}{2} \left(4
   (f+h+k)+\phi +4 \tilde{g}\right)} N_f \Bigg].
}

Changing to variables
\SP{
& f = A + p - \frac{x}{2}, \;\;
\tilde g = -A - \frac{g}{2} + \log 2  - p + x, \\
& h = -A + \frac{g}{2} - p + x, \;\;\; k = -A + \log 2  - 4 p, 
}
with inverse
\SP{
	A &=\frac{1}{3} \left( 8f +2\tilde g +2h +k \right) - \log 2, \;\;\;
	g = -\tilde g + h + \log 2, \\
	p &= - \frac{1}{6} \left( 4 f + \tilde g + h +2k \right) + \frac{1}{2} \log 2, \;\;\;
	x = 2 f + \tilde g + h - \log 2,
}
and also changing the radial coordinate as $dr = e^{A+k} d\rho$, we obtain
\EQ{
	S_{10d} = \frac{4 \mu^4 N_c^2 (4\pi)^3}{2 \kappa_{(10)}^2} \int dr \int d^4 x e^{4A} \left( T - V \right),
}
with
\EQ{
	T = 3A'^2-\frac{1}{4} e^{-2 g} a'^2-\frac{N_c^2 e^{\phi -2 x}}{64} b'^2-\frac{g'^2}{4}-3p'^2-\frac{x'^2}{2}-\frac{\phi'^2}{8},
}
and
\SP{
	V =& \frac{e^{-2 (g+2 (p+x))}}{128} \times \\ \Bigg[& e^{12 p+2 x+\phi }
   \left(2 e^{2 g} (a-b)^2+e^{4 g}+\left(a^2-2 b
   a+1\right)^2\right) N_c^2- \\& 2 \left(a^2-2 b a+1\right)
   e^{12 p+2 x+\phi } N_f N_c+e^{12 p+2 x+\phi } N_f^2+ 8 e^{2 g+6
   p+x+\frac{\phi }{2}} N_f + \\& 16
   \left(a^4+2 \left(\left(e^g-e^{6 p+2
   x}\right)^2-1\right) a^2+e^{4 g}-4 e^{g+6 p+2 x}
   \left(1+e^{2 g}\right)+1\right) \Bigg]
}
Recognizing that for a metric given by $ds_5^2 = dr^2 + e^{2A} dx_{1,3}^2$ the Ricci scalar is (up to partial integrations) equal to $R = -12A'^2$, we can write this as the action of a 5d non-linear sigma model
\SP{
	S_{5d} = \int dr \int d^4 x \sqrt{-g} \Bigg[ \frac{R}{4} - \frac{1}{2} G_{ab} g^{MN} \partial_M \Phi^a \partial_N \Phi^b - V(\Phi) \Bigg],
}
where $\Phi = [g, p, x, \phi, a, b]$, and the non-linear sigma model metric is diagonal with entries $G_{gg} = \frac{1}{2}$, $G_{pp} = 6$, $G_{xx} = 1$, $G_{\phi \phi} = \frac{1}{4}$, $G_{aa} = \frac{e^{-2g}}{2}$, and $G_{bb} = \frac{N_c^2 e^{-2x+\phi}}{32}$. It can be shown that any solution to the equations of motion following from this 5d action also satisfy the equations of motion of the original 10d system. Therefore, the 5d non-linear sigma model derived in this section is a consistent truncation.

\subsection{Seiberg Duality for Type N}

We can generalize the arguments of section~5.2.3 to Type N. For these backgrounds, Seiberg duality corresponds to the transformation \cite{HoyosBadajoz:2008fw}
\SP{
	Q &\rightarrow -Q, \\
	\sigma &\rightarrow -\sigma, \\
	N_c &\rightarrow N_f - N_c,
}
leaving $P$, $Y$, $\phi$, and $\tau$ unchanged.
Using the relations
\SP{
	e^{3A} =& \frac{e^{2 \phi} (P^2 - Q^2) \sqrt{Y}}{16}, \\
	e^{2g} =& \frac{P^2 - Q^2}{(\cosh \tau P- Q)^2}, \\
	e^{6p} =& \frac{4 e^{- \phi}}{\sqrt{P^2 - Q^2} Y}, \\
	e^{2x} =& \frac{e^{\phi} (P^2 - Q^2)}{16}
}
we see that in terms of the 5d variables, a Seiberg duality takes the form
\SP{
	e^g &\rightarrow \frac{e^g}{e^{2g} + a^2}, \\
	a &\rightarrow \frac{a}{e^{2g} + a^2}, \\
	b &\rightarrow \frac{N_c}{N_c-N_f} b, \\
	N_c &\rightarrow N_f - N_c.
}
Again, it is straightforward to see that both the non-linear sigma model metric $G_{ab}$ and the potential $V$ of the previous section are invariant under these transformations. It follows that the whole 5d theory obeys Seiberg duality.

\section{Walking Backgrounds}

In the remainder of this chapter, we will focus on a particular class of Type N solutions that have walking behaviour. These have $N_f = 0$ and $Q_0 = - N_c$, and can be thought of as deformations of the background known as non-singular Maldacena-Nunez \cite{Maldacena:2000yy}. In the IR, they behave as the walking solutions found in \cite{Nunez:2008wi}. Non-singular Maldacena-Nunez corresponds to $\hat P = 2 N_c \rho$. The reason that we want our solutions to asymptote to Maldacena-Nunez in the UV ($\rho \rightarrow \infty$) is that we then obtain a non-trivial (discrete) spectrum.

\begin{figure}[t]
\centering
	\includegraphics[width=9cm]{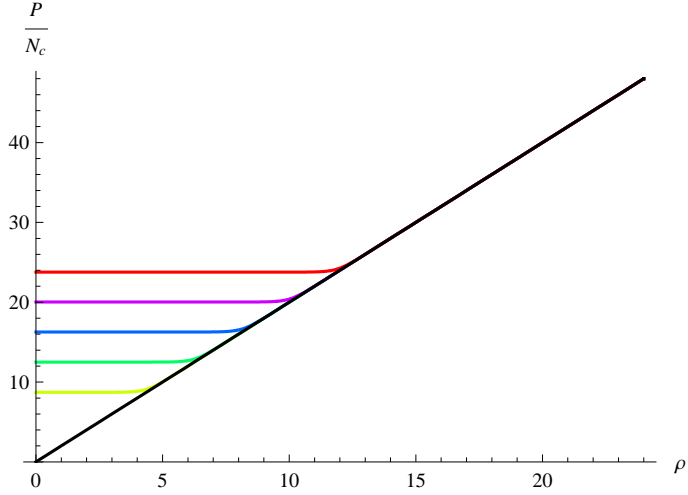}
	\caption{A few backgrounds with walking behaviour compared to non-singular Maldacena-Nunez (black line).}
	\label{fig:plotPWalking}
\end{figure}

Consider a small perturbation around $\hat P$, so that
\SP{
	P(\rho) = \hat P(\rho) + \epsilon p(\rho).
}
Linearizing the second order differential equation for $P$ that determines the background, equation \eqref{eq:masterequationforP}, we find two possible behaviours in the UV ($\rho \rightarrow \infty$): $p(\rho) \sim e^{-4\rho}$ and $p(\rho) \sim e^{2\rho}$ (up to factors that are powers of $\rho$ and are irrelevant for the purpose of setting up the numerics). In order for the perturbative expansion in $\epsilon$ to be consistent, we need to pick the first behaviour which is decaying exponentially. Solving \eqref{eq:masterequationforP}, we set up the boundary conditions in the UV corresponding to the small deformations around $\hat P$ so that
\SP{
	P = \hat P + N_c e^{4 (\rho_* - \rho)},
}
and evolve numerically towards the IR. $\rho_*$ sets the scale at which, going from the UV to the IR, the solutions start to deviate from non-singular Maldacena-Nunez. Another scale is set by the VEV of the gaugino condensate which is of order $\rho \sim 1$. A few examples of walking backgrounds are depicted in Figure~\ref{fig:plotPWalking}.

In the IR ($\rho \rightarrow 0$), $P$ becomes nearly constant for these backgrounds. More precisely, they fall into the class of Type N backgrounds that have an IR expansion known as Type I:
\SP{
\label{eq:IRexpansionP}
	P = P_0 + \frac{4}{3} c_+^3 P_0^2 \rho^3 + \frac{16}{15} P_0^2 c_+^3 \rho^5 + \mathcal O(\rho^6),
}
where $P_0$ and $c_+$ are integration constants. Due to the fact that we want the backgrounds to asymptote to non-singular Maldacena-Nunez in the UV, $c_+$ and $P_0$ are not independent, but need to dialed in such a way that the correct UV behaviour is obtained. Thus, we have a one-parameter family of solutions parameterized by $P_0 \approx 2 N_c \rho_*$. 

\begin{figure}[t]
\centering
	\includegraphics[width=9cm]{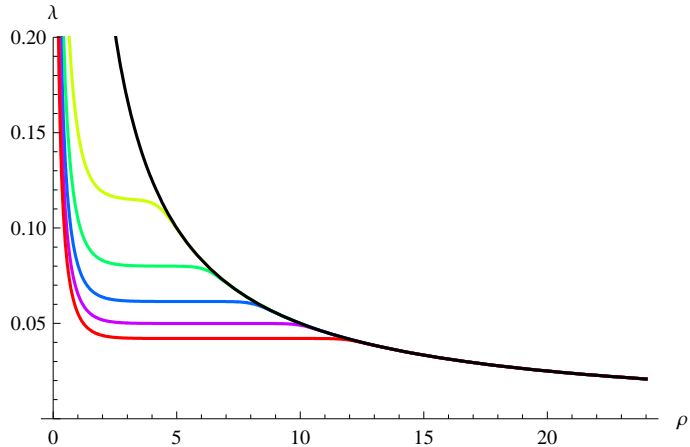}
	\caption{The four-dimensional gauge coupling $\lambda$ as a function of the radial coordinate $\rho$ for the same backgrounds as those in Figure~\ref{fig:plotPWalking}.}
	\label{fig:couplingconstant}
\end{figure}

A four-dimensional gauge coupling $\lambda$ can be defined which is essentially the inverse of the size of the $S^2$. It is given by
\SP{
	\lambda = \frac{g^2 N_c}{8 \pi^2} = \frac{N_c \coth \rho}{P}.
}
In Figure~\ref{fig:couplingconstant}, we plot this gauge coupling as a function of the radial coordinate $\rho$ for the same backgrounds as those in Figure~\ref{fig:plotPWalking}. As can be seen, there is a scale $\rho \sim 1$ set by the gaugino condensate below which the gauge coupling diverges. In an intermediate region, we can obtain walking behaviour. This behaviour continues until the scale set by $\rho_*$ after which all the backgrounds behave as non-singular Maldacena-Nunez towards the UV.

While Figure~\ref{fig:couplingconstant} certainly suggests that we are dealing with a walking theory, it does not prove this is the case conclusively. The reason is that the plot is regularization-scheme dependent. In the holographic picture, this corresponds to the fact that we can always rescale the radial coordinate $\rho$. In other words, we need to compute something that is actually physical, such as the spectrum.

\section{Scalar Spectra}

In this section, we will apply the methods developed in Chapter~\ref{ch:4} in order to compute the spectra of the walking backgrounds of the previous section. Let us remind the reader that the differential equation for the scalar fluctuations $\mathfrak a^a$ that we need to solve is given by \eqref{eq:flucdiffrho}
\SP{
\label{eq:flucdiffrho2}
	\Big[ \delta^a_b \partial_\rho^2 + S^a_b \partial_\rho + T^a_b - \delta^a_b e^{2k} K^2 \Big] \mathfrak{a}^b = 0,
}
with
\SP{
	S^a_b =& 2 \mathcal{G}^a_{\ bc} \partial_\rho \Phi^c + 4 \left(\partial_\rho p + \partial_\rho A \right) \delta^a_b, \\
	T^a_b =& \partial_b \mathcal{G}^a_{\ cd} \partial_\rho \Phi^c \partial_\rho \Phi^d - \\& 4 e^{-8p} \Bigg[ \left( \frac{4 (V^a \partial_\rho \Phi^c + V^c \partial_\rho \Phi^a)}{3 \partial_\rho A} + \frac{16 V \partial_\rho \Phi^a \partial_\rho \Phi^c}{9 (\partial_\rho A)^2} \right) G_{cb} + \partial_b V^a \Bigg].
}
Before doing so, we need to discuss which boundary conditions to impose on the scalar fluctuations in the IR and UV.

In the following, we will put $\mu = 1$, and rescale $P \rightarrow N_c P$, and similarly for $Q$ and $Y$. The masses that we will compute will be in units of $\sqrt{\alpha' g_s N_c}$.

\subsection{Boundary Conditions in the UV}

In the UV, the background is exponentially close to non-singular Maldacena-Nunez. Let us go to a basis in which the matrices $S$ and $T$ become diagonal in the UV to leading order in $1/\rho$ (ignoring exponentially suppressed terms). Such a basis is given by
\EQ{
	\mathfrak a^a \rightarrow B^a_b \mathfrak a^b,
}
where
\EQ{
    B =
    \left(
    \begin{array}{llllll}
        -1 & \frac{2}{3} & 0 & 1 & 0 & 0 \\
         \frac{1}{2} & \frac{1}{6} & 1 & 0 & 0 & 0 \\
         1 & -\frac{1}{6} & -3 & \frac{1}{2} & 0 & 0 \\
         0 & 1 & -6 & -1 & 0 & 0 \\
         0 & 0 & 0 & 0 & 1 & -1 \\
         0 & 0 & 0 & 0 & 1 & 1
    \end{array}
\right).
}
The matrices $S$ and $T$ take the following form in the UV:
\SP{
    S =&
\left(
\begin{array}{llllll}
 \frac{8 \rho}{4 \rho-1} & 0 & 0 & 0 & 0 & 0 \\
 0 & \frac{8 \rho}{4 \rho-1} & 0 & 0 & 0 & 0 \\
 0 & 0 & \frac{8 \rho}{4 \rho-1} & 0 & 0 & 0 \\
 0 & 0 & 0 & \frac{8 \rho}{4 \rho-1} & 0 & 0 \\
 0 & 0 & 0 & 0 & \frac{4-8 \rho}{1-4 \rho} & 0 \\
 0 & 0 & 0 & 0 & 0 & \frac{4-8 \rho}{1-4 \rho}
\end{array}
\right),
}
\SP{
    T =&
\left(
\begin{array}{llllll}
 -\frac{8 \left(48 \rho^2-24 \rho+5\right)}{3 (1-4 \rho)^2} & -\frac{16}{9 (1-4 \rho)^2} & 0 & 0 & 0 & 0 \\
 -\frac{8}{(1-4 \rho)^2} & -\frac{32 \left(12 \rho^2-6 \rho+1\right)}{3 (1-4 \rho)^2} & 0 & 0 & 0 & 0 \\
 0 & 0 & 0 & 0 & 0 & 0 \\
 0 & 0 & 0 & \frac{2 (1-2 \rho)^2}{\rho^2 (4 \rho-1)} & 0 & 0 \\
 0 & 0 & 0 & 0 & \frac{4}{4 \rho-1} & 0 \\
 0 & 0 & 0 & 0 & 0 & \frac{32 \rho-4}{1-4 \rho}
\end{array}
\right).
}
Expanding the fluctuations as
\EQ{
    \mathfrak a^a = e^{C_a \rho} \sum_n \mathfrak a^a_n \rho^{b_{a,n}},
}
where the exponents $b_{a,n}$ can take on non-integer values, plugging into the differential equation \eqref{eq:flucdiffrho2}, and expanding in powers of $1/\rho$, we find\footnote{When setting up the boundary conditions in the UV, the exponential behaviour is the most important, since up to an overall factor that affects both $\mathfrak a^a$ and its derivative, the effect of the $b_{a,n}$ is suppressed by $1/\rho$. Therefore, $C_a$ is all we need for running the numerics.}
\SP{
    C_{1,2,6} = -1 \pm \sqrt{9 + K^2}, \\
    C_{3,4,5} = -1 \pm \sqrt{1 + K^2}.
}
We are interested in the subdominant behaviour, so we pick the minus signs in these expressions. In \cite{Berg:2006xy}, a normalizability condition for the fluctuations was given:
\EQ{
		\int dr e^{2A} G_{ab} \mathfrak a^a \mathfrak a^b = \int d\rho e^{3A + k} G_{ab} \mathfrak a^a \mathfrak a^b < \infty.
}
In our case, we have in the UV that
\EQ{
		e^{3A + k} = e^{2\rho + \phi_0/2} \left[ \frac{\rho^{1/2}}{16} + \mathcal{O}(\rho^{-1/2}) \right],
}
so that the subdominant fluctuations are always normalizable, while the dominant ones are not. Let us also point out that the presence of the square roots in the exponentials signal the start of a continuum at $M^2 = 1$ and at $M^2 = 9$ above which the fluctuations start to exhibit oscillatory behaviour in the UV. Note that the consistent truncation used in \cite{Berg:2006xy}, in order to compute the spectrum of non-singular Maldacena-Nunez, corresponds to keeping only the fluctuations that fall off as $e^{(-1 - \sqrt{1 + K^2}) \rho}$ in the UV, i.~e. $(\mathfrak a^3,\mathfrak a^4,\mathfrak a^5)$.

In conclusion, we now have six linearly independent solutions in the UV with the subdominant behaviour. We will evolve these numerically from the UV to a midpoint where we will compare them to solutions evolved from the IR. These IR solutions will be found in the next section.

\subsection{Boundary Conditions in the IR}

Using \eqref{eq:IRexpansionP}, we can expand the differential equations for the scalar fluctuations \eqref{eq:flucdiffrho2} in the IR ($\rho \rightarrow 0$). Writing the fluctuations as
\SP{
	\mathfrak a^a = \sum_{n=-\infty}^\infty \mathfrak a^a_n \rho^n
}
and plugging into \eqref{eq:flucdiffrho2}, we obtain
\SP{
\begin{cases}
	\mathfrak{a}^1 = \mathfrak{a}^1_0 + \mathfrak{a}^1_1 \rho + 4 (\mathfrak{a}^5_0 - \mathfrak{a}^1_0) \rho^2 + \mathcal O \left( \rho^3 \right), \\
	\mathfrak{a}^2 = \mathfrak{a}^2_0 + \mathfrak{a}^2_1 \rho + \mathcal O \left( \rho^3 \right), \\
	\mathfrak{a}^3 = \mathfrak{a}^3_0 + \mathfrak{a}^3_1 \rho + \mathcal O \left( \rho^3 \right), \\
	\mathfrak{a}^4 = \mathfrak{a}^4_0 + \mathfrak{a}^4_1 \rho + \mathcal O \left( \rho^3 \right), \\
	\mathfrak{a}^5 = \mathfrak{a}^5_0 + (-4 \mathfrak{a}^1_0 + 2 \mathfrak{a}^5_0) \rho^2 + \mathfrak{a}^5_3 \rho^3 + \mathcal O \left( \rho^4 \right), \\
	\mathfrak{a}^6 = \mathfrak{a}^6_{-1} \rho^{-1} -\frac{2}{3} \mathfrak{a}^6_{-1} \rho + \mathfrak{a}^6_2 \rho^2 + \mathcal O \left( \rho^3 \right),
\end{cases}
}
The solutions are determined by the 12 integration constants
$\mathfrak{a}^1_0$, $\mathfrak{a}^1_1$, $\mathfrak{a}^2_0$, $\mathfrak{a}^2_1$, $\mathfrak{a}^3_0$, $\mathfrak{a}^3_1$, $\mathfrak{a}^4_0$, $\mathfrak{a}^4_1$, $\mathfrak{a}^6_{-1}$, $\mathfrak{a}^6_2$, $\mathfrak{a}^5_0$, and $\mathfrak{a}^5_3$. Suppose we want to impose boundary conditions such that the kinetic terms of the action do not blow up in the IR. The kinetic term of the action is
\EQ{
    \int dr \sqrt{-g} G_{ab} g^{rr} \partial_r \Phi^a \partial_r \Phi^b =
    \int d\rho e^{3A-k} G_{ab} \partial_\rho \Phi^a \partial_\rho \Phi^b.
}
We have that
\EQ{
    e^{3A-k} = \frac{1}{2} e^{4(A+p)} = \frac{e^{\phi_0/2}}{8 \sqrt{2} c_+^{3/2}} + \mathcal{O}(\rho^3).
}
All components of the non-linear sigma model metric are of order 1, except
\EQ{
    \label{eq:BC}
    G_{55} = \frac{1}{8\rho^2} + \mathcal{O}(\rho^0).
}
This means that we must set $\mathfrak a^5_0 = \mathfrak a^6_{-1} = 0$ in the IR. It is not clear which IR boundary conditions to impose on the first four fields. For definiteness, we will in the following choose $a^a_0 = 0$, i.e. Dirichlet, for all the scalar fluctuations. Six parameters remain: $a^1_1$, $a^2_1$, $a^3_1$, $a^4_1$, $a^5_3$, and $a^6_2$. Again, by alternately setting all but one of these parameters equal to zero, we obtain six linearly independent solutions in the IR.

\subsection{Results}

\begin{figure}[t]
\centering
	\includegraphics[width=9cm]{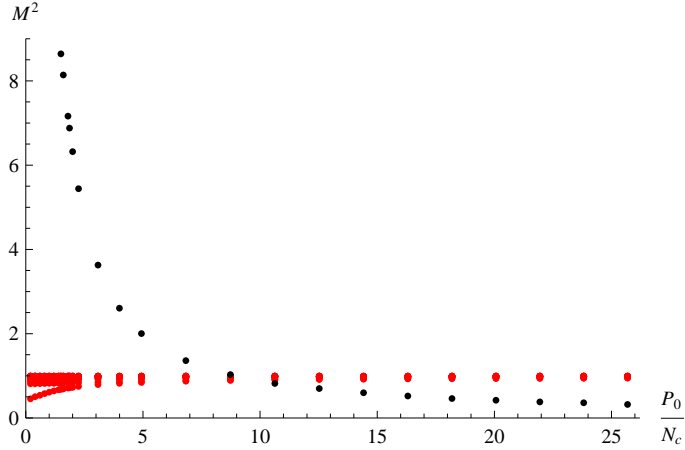}
	\caption{The spectrum as a function of $P_0/N_c \approx 2 \rho_*$.}
	\label{fig:spectrum}
\end{figure}

The numerical results are plotted in Figure~\ref{fig:spectrum}, where the spectrum as a function $P_0/N_c \approx 2 \rho_*$ is shown. As can be seen, the spectrum contains a state which becomes lighter as $P_0$ (or alternatively the length of the walking region $\rho_*$) is increased. When $\rho_*$ is of the same order as the scale set by the gaugino condensate, i.e. $\rho \sim 1$, this state disappears from the spectrum.

\begin{figure}[t]
\centering
	\includegraphics[width=9cm]{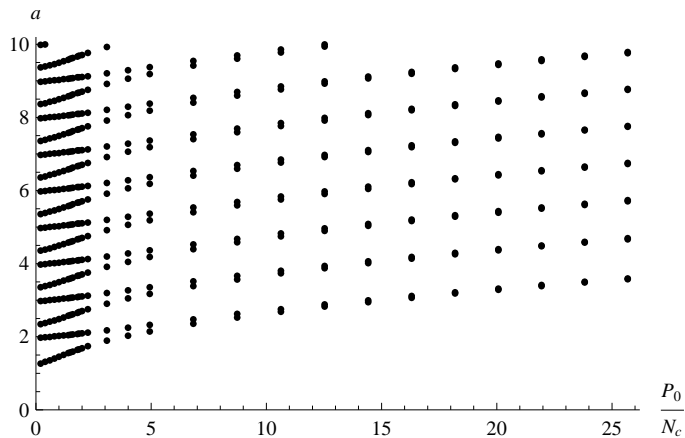}
	\caption{The two towers of states in terms of the variable $a$ defined in \eqref{eq:definitionofa}.}
	\label{fig:towers}
\end{figure}

Furthermore, the spectrum contains two towers of states converging on $M^2 = 1$. As the length of the walking region is increased, these become heavier, so that the discrete spectrum effectively disappears into the continuum. Going to a variable $a$ defined through
\EQ{
\label{eq:definitionofa}
	\sqrt{1 - M^2} = \frac{3}{4 a - 1},
}
we can more clearly see how the two towers behave. The result is plotted in Figure~\ref{fig:towers}. As $\rho_* \rightarrow 0$, the spectrum agrees with that of non-singular Maldacena-Nunez computed in \cite{Berg:2006xy}.

\section{Summary}

In this chapter, we studied ten-dimensional systems which can be thought of as $N_c$ D5-branes wrapping an internal $S^2$. At first, we included flavor degrees of freedom, obtained from $N_f$ back-reacting flavor branes. We derived a consistent truncation to a five dimensional non-linear sigma model consisting of six scalars coupled to gravity, and showed how Seiberg duality is realized from the five-dimensional point of view, generalizing the results of the previous chapter to apply to Type N systems.

We then turned our attention to a particular class of backgrounds that exhibit walking behaviour. We would like to emphasize that these are not Walking Technicolor models, since they do not yield a mechanism for electro-weak symmetry breaking. However, the set of results collected in this chapter supports the idea that this system is a very interesting laboratory, in which walking  can be studied systematically, and in which dynamical questions can be addressed in a calculable form, providing a guidance for  model building.

The class of solutions we found yields the four-dimensional gauge coupling 
of a walking theory (the Lagrangian of which, for present purposes, we do not need to know), in the sense that there is an intermediate region where the gauge coupling is approximately constant. While the interpretation in terms of the dual field theory is at this point not well understood, the very fact that we observe a particle in the spectrum with a mass much lower than the dynamical scale of the theory suggests that its existence is due to the spontaneous breaking of an approximate symmetry. If this symmetry is scale invariance, then the light scalar would be interpreted as the dilaton, the pseudo-Goldstone boson of dilatations. From the gravity point of view, it is clear that scale invariance is broken in the IR by the gaugino condensate, and in the UV at the scale set by $\rho_*$.

\end{document}